\documentclass[twocolumn,prd,aps,superscriptaddress,preprintnumbers,tightenlines,showpacs,nofootinbib,eqsecnum,amsfonts,amsmath]{revtex4}
\usepackage{thumbpdf} 
\usepackage{enumerate}
\usepackage{graphics,epsfig,placeins,wrapfig}
\usepackage{braket}
\usepackage{hhline}
\usepackage{placeins}
\usepackage{color}
\usepackage{calc}
\usepackage{amsmath,amssymb,graphicx,mathrsfs}
\usepackage{tensor}
\usepackage{bm}
\usepackage{times}
\usepackage[varg]{txfonts}
\usepackage[colorlinks, pdfborder={0 0 0}]{hyperref}
\usepackage[capitalise]{cleveref}
\usepackage{float}
\usepackage{dcolumn}
\usepackage{physics}
\usepackage[nolist,nohyperlinks]{acronym}
\usepackage{xspace}
\usepackage[abs]{overpic}
\usepackage{pict2e}
\usepackage{enumitem}
\usepackage[usenames,dvipsnames]{xcolor}
\usepackage[utf8]{inputenc}
\usepackage{acronym}
\usepackage{gensymb}
\usepackage{romannum}
\usepackage{subfigure}
\usepackage{cleveref}
\usepackage[normalem]{ulem}
\usepackage{longtable}
\usepackage{mathtools}
\usepackage{graphicx}
\usepackage{enumitem,amssymb}
\usepackage{cprotect}
\newlist{todolist}{itemize}{2}
\setlist[todolist]{label=$\square$}
\usepackage{pifont}

\Crefname{figure}{Fig.}{Figs.}

\newcommand{\AEI}{\affiliation{Max Planck Institute for Gravitational Physics (Albert Einstein Institute), Am M\"uhlenberg 1, Potsdam 14476, Germany}}

\newcommand{\Cornell}{\affiliation{Cornell Center for Astrophysics and Planetary Science, Cornell University, Ithaca, New York, 14853, USA}}
\newcommand{\Caltech}{\affiliation{Theoretical Astrophysics 350-17, California Institute of Technology, Pasadena, CA 91125, USA}}
\newcommand{\APC}{\affiliation{APC, Univ Paris Diderot, CNRS/IN2P3, CEA/Irfu, Obs de Paris, Sorbonne Paris Cité, France}}
\newcommand{\NCSA}{\affiliation{NCSA, University of Illinois at Urbana-Champaign, Urbana, Illinois, 61801, USA}}
\newcommand{\Potsdam}{\affiliation{Institute for Physics and Astronomy, University of Potsdam, Karl-Liebknecht-Str. 24/25, 14476 Potsdam, Germany}}
\newcommand{\CITA}{\affiliation{Canadian Institute for Theoretical Astrophysics, 60 St. George Street, University of Toronto, Toronto, ON M5S 3H8, Canada}}
\newcommand{\UofT}{\affiliation{Department of Physics 60 St. George Street, University of Toronto, Toronto, ON M5S 3H8, Canada}}
\newcommand{\JPL}{\affiliation{Jet Propulsion Laboratory, California Institute of Technology, Pasadena, CA 91109, USA}}
\newcommand{\Manchester}{\affiliation{University of Manchester, Sackville Building, Granby Row, Manchester M1 3BU, UK}}
\newcommand{\MoscowInst}{\affiliation{Moscow Institute of Physics and Technology, Dolgoprudny, Moscow region, Russia}}

\def\be{\begin{equation}}
\def\ee{\end{equation}}
\def\bea{\begin{eqnarray}}
\def\eea{\end{eqnarray}}
\newcommand{\bes}{\begin{subequations}}
\newcommand{\ees}{\end{subequations}}

\newcommand{\vS}{\mbox{\boldmath${S}$}}
\newcommand{\vR}{\mbox{\boldmath${r}$}}
\newcommand{\vP}{\mbox{\boldmath${p}$}}
\newcommand{\vL}{\mbox{\boldmath${L}$}}
\newcommand{\vN}{\mbox{\boldmath${N}$}}
\newcommand{\vLhat}{\mbox{\boldmath${\hat{L}}$}}

\newcommand{\vJ}{\mbox{\boldmath${J}$}}
\newcommand{\vE}{\mbox{\boldmath${\hat{e}}$}}

\newcommand{\vchi}{\mbox{\boldmath$\chi$}}

\newcommand{\aconfigs}{1523} % Total number of configurations analyzed
\newcommand{\EOBbthree}{$94\%$}
\newcommand{\EOBbone}{$57\%$}
\newcommand{\Phenombthree}{$83\%$}
\newcommand{\Phenombone}{$20\%$}

\allowdisplaybreaks

\graphicspath{{./Figures/}}

\begin{document}

\pagenumbering{arabic}

\title{Multipolar Effective-One-Body Waveforms for Precessing Binary Black Holes:\\ Construction and Validation}

\author{Serguei Ossokine}
\AEI
\author{Alessandra Buonanno}
\AEI
\affiliation{Department of Physics, University of Maryland, College Park, MD 20742, USA}
\author{Sylvain Marsat}
\APC
\AEI
\author{Roberto Cotesta}
\AEI
\author{Stanislav Babak}
\APC
\MoscowInst
\AEI
\author{\\Tim Dietrich}
\Potsdam
\AEI
\author{Roland Haas}
\NCSA
\AEI
\author{Ian Hinder}
\Manchester
\AEI
\author{Harald P. Pfeiffer}
\AEI
\author{Michael P\"urrer}
\AEI
\author{Charles J. Woodford}
\UofT
\CITA
\author{Michael Boyle}
\Cornell
\author{Lawrence E. Kidder}
\Cornell
\author{Mark  A. Scheel}
\Caltech
\author{B\'ela Szil\'agyi}
\JPL
\Caltech

\date{\today}

\begin{abstract}
As gravitational-wave detectors become more sensitive and broaden their frequency bandwidth, 
we will access a greater variety of signals emitted by compact binary systems, shedding light on their 
astrophysical origin and environment. A key physical effect that can distinguish among different 
formation scenarios is the misalignment of the spins with the orbital angular momentum, causing the spins 
and the binary's orbital plane to precess. To accurately model such precessing signals, especially 
when masses and spins vary in the wide astrophysical range, it is crucial to include multipoles 
beyond the dominant quadrupole. Here, we develop the first {\it multipolar} precessing waveform model 
in the effective-one-body (EOB) formalism for the entire coalescence stage (i.e., inspiral, 
merger and ringdown) of binary black holes: \verb+SEOBNRv4PHM+. In the nonprecessing limit, the model reduces to 
\verb+SEOBNRv4HM+, which was calibrated to numerical-relativity (NR) simulations, and 
waveforms from black-hole perturbation theory. 
We validate the \verb+SEOBNRv4PHM+ by comparing it to the public catalog of 1405 precessing NR waveforms 
of the Simulating eXtreme Spacetimes (SXS) collaboration, and also to 118 SXS precessing NR waveforms, 
produced as part of this project, which span mass ratios 1-4 and (dimensionless) black-hole's spins 
up to 0.9. We stress that \verb+SEOBNRv4PHM+ is not calibrated to NR simulations in the 
precessing sector. We compute the unfaithfulness against 
the \aconfigs \ SXS precessing NR waveforms, and find that, for \EOBbthree\ (\EOBbone\ ) of the cases, the maximum value, in the 
total mass range $20\mbox{--}200 M_\odot$, is below $3\%$ ($1\%$). Those numbers change to \Phenombthree\ (\Phenombone\ ) 
when using the inspiral-merger-ringdown, multipolar, precessing phenomenological model \verb+IMRPhenomPv3HM+. 
We investigate the impact of such unfaithfulness values with two Bayesian, parameter-estimation studies on synthetic signals.
We also compute  the unfaithfulness between 
those waveform models as a function of the mass and spin parameters to identify in which part of the parameter 
space they differ the most. We validate them also against the multipolar, precessing NR surrogate model \verb+NRSur7dq4+, 
and find that the \verb+SEOBNRv4PHM+ model outperforms \verb+IMRPhenomPv3HM+.
\end{abstract}


\maketitle

\section{Introduction}
\label{sec:Intro}

Since the Laser Interferometer Gravitational wave Observatory (LIGO) 
detected a gravitational wave (GWs) from a binary--black-hole (BBH) 
in 2015~\cite{Abbott:2016blz}, multiple observations of GWs from BBHs have been made with 
the LIGO~\cite{TheLIGOScientific:2014jea} and Virgo~\cite{TheVirgo:2014hva} detectors~\cite{TheLIGOScientific:2016pea,LIGOScientific:2018mvr,Zackay:2019tzo,Venumadhav:2019lyq,Nitz:2019hdf}.  
Two binary neutron star (BNSs) systems have been observed~\cite{TheLIGOScientific:2017qsa,Abbott:2020uma}, one of them both 
in gravitational and electromagnetic radiation~\cite{GBM:2017lvd,Monitor:2017mdv}, opening the exciting new chapter 
of multi-messenger GW astronomy. Mergers of compact-object binaries are expected to be detected at an
even higher rate with LIGO and Virgo ongoing and future, observing runs~\cite{Aasi:2013wya}, and with 
subsequent third-generation detectors on the ground, such as the Einstein Telescope and 
Cosmic Explorer, and the Laser Interferometer Space Antenna (LISA). In order to 
extract the maximum amount of astrophysical and cosmological information, the accurate modeling of 
GWs from binary systems is more critical than ever. Great progress has been made in this direction, 
both through the development of analytical methods to solve the two-body problem in General
Relativity (GR), and by ever-more expansive numerical-relativity (NR) simulations. 

One of the key areas of interest is to improve the modeling of systems where the misalignment of
the spins with the orbital angular momentum causes the spins and the orbital
plane to precess~\cite{Apostolatos:1994mx}. Moreover, when the binary's component masses are asymmetric, 
gravitational radiation is no longer dominated by the quadrupole moment, and higher multipoles need to 
be accurately modeled~\cite{Blanchet:2013haa}. Precession and higher multipoles lead to  very rich dynamics, 
which in turn is imprinted on the GW signal. Their measurements will be able to shed light on  
the formation mechanism of the observed systems, probe the astrophysical environment, break degeneracy 
among parameters, allowing  more accurate measurements of cosmological parameters, masses and spins, 
and more sophisticated tests of GR. 

Faithful waveform models for precessing compact-object binaries have been developed within the 
effective-one-body (EOB) formalism~\cite{Taracchini:2013wfa,Pan:2013rra,Babak:2016tgq}, 
and the phenomenological approach~\cite{Hannam:2013oca,Khan:2015jqa,Khan:2018fmp,Pratten:2020fqn,Garcia-Quiros:2020qpx} through calibration 
to NR simulations. Recently, an inspiral-merger-ringdown phenomenological waveform model that tracks precession and 
includes higher modes was constructed in Ref.~\cite{Khan:2019kot} (henceforth, \verb+IMRPhenomPv3HM+)~\footnote{During the final preparation of this work, a new frequency-domain phenomenological model with precession and higher modes ({\tt IMRPhenomXPHM}~\cite{Pratten:2020ceb}), and a time-domain phenomenological precessing model with the  dominant mode ({\tt IMRPhenomTP}~\cite{Estelles:2020c}) were developed. We leave the comparison to these models for future work.} 
The model describes the six spin degrees of freedom in the inspiral phase, but not in the late-inspiral, merger and ringdown stages.  
In the co-precessing frame~\cite{Buonanno:2002fy,Schmidt:2010it,Boyle:2011gg,O'Shaughnessy:2011fx,Schmidt:2012rh},  
in which the BBH is viewed face-on at all times and the GW radiation resembles the 
nonprecessing one, it includes the modes $(\ell, m)= (2,\pm 2), (2,\pm 1), (3,\pm 3), (3,\pm 2), 
(4,\pm 4)$ and $(4,\pm 3)$. Here, building on the multipolar 
aligned-spin EOB waveform model of Ref.~\cite{Bohe:2016gbl,Cotesta:2018fcv}, 
which was calibrated to 157 NR simulations~\cite{Mroue:2013xna,Chu:2015kft}, and 13 waveforms from 
BH perturbation theory for the (plunge-)merger and ringdown~\cite{Barausse:2011kb}, we develop the first EOB waveform model that 
includes both spin-precession and higher modes (henceforth, \verb+SEOBNRv4PHM+). The model describes 
the six  spin degrees of freedom throughout the BBH coalescence. It differs from the one 
of Refs.~\cite{Pan:2013rra,Babak:2016tgq}, not only because it includes  in the co-precessing 
frame the $(3,\pm 3)$, $(4,\pm 4)$ and $(5,\pm 5)$ 
modes, beyond the $(2,\pm 2)$ and $(2,\pm  1)$ modes, but also because it uses an improved description 
of the two-body dynamics, having been calibrated~~\cite{Bohe:2016gbl} to a large set of NR waveforms~\cite{Mroue:2013xna}. We note that \verb+IMRPhenomPv3HM+ and \verb+SEOBNRv4PHM+ are not completely independent 
because the former is constructed fitting (in frequency domain) hybridized waveforms obtained by stitching together EOB and NR waveforms.  
We stress that both \verb+SEOBNRv4HM+ and  \verb+IMRPhenomPv3HM+ are not calibrated to NR simulations in the 
precessing sector. Finally, the surrogate approach, which  interpolates NR waveforms, has been used to construct several waveform models that include higher modes~\cite{Varma:2018mmi} and precession~\cite{Blackman:2017pcm}. In this paper, we  consider the state-of-the-art surrogate waveform model with full spin precession and higher modes~\cite{Varma:2019csw} (henceforth, {\tt NRSur7dq4}), 
developed for binaries with mass ratios 1-4, (dimensionless) BH's spins up to $0.8$ and binary's 
total masses larger than $\sim 60 M_\odot$. It includes in the co-precessing frame all modes up to  $\ell =4$. 
Table~\ref{tbl:wf_models} summarizes the waveform models used in this paper. 

\begin{table}
	\begin{ruledtabular}
	\begin{tabular}{lll}
		Model name & Modes in the co-precessing frame & Reference\\
		\hline
		{\tt SEOBNRv3P} &  $(2,\pm2)$, $(2,\pm1)$ & \cite{Pan:2013rra,Babak:2016tgq}\\
		{\tt SEOBNRv4P} & $(2,\pm2)$, $(2,\pm1)$ &  this work \\
		{\tt SEOBNRv4PHM} & $(2,\pm2)$, $(2,\pm1)$, $(3,\pm3)$, $(4,\pm4)$\\
		& $(5,\pm5)$ & this work \\
		{\tt IMRPhenomPv2} & $(2,\pm2)$ & \cite{Hannam:2013oca}\\
		{\tt IMRPhenomPv3} & $(2,\pm2)$  & \cite{Khan:2018fmp}\\
		{\tt IMRPhenomPv3HM} & $(2,\pm2)$, $(2,\pm1)$, $(3,\pm3)$, $(3,\pm2)$,\\
		& $(4,\pm 4)$, $(4,\pm3)$&\cite{Khan:2019kot} \\
		{\tt NRSur7dq4} & all with $\ell\leq4$ &\cite{Varma:2019csw} \\
	\end{tabular}
	\end{ruledtabular}
	\caption{The waveform models used in this paper. We also specify which modes are included in the co-precessing frame}
	\label{tbl:wf_models}
\end{table}

The best tool at our disposal to validate waveform models built from approximate solutions of the 
Einstein equations, such as the ones obtained from post-Newtonian (PN) theory, BH perturbation theory 
and the EOB approach, is their comparison to NR waveforms. So far, NR simulations of  BBHs have been mostly limited to mass ratio $\leq 4$ and (dimensionless) spins $\leq 0.8$, and length   
of $15\mbox{--}20$ orbital cycles before merger~\cite{Jani:2016wkt,Healy:2017psd,Healy:2019jyf,Huerta:2019oxn,Boyle:2019kee} (however, see Ref.~\cite{Hinder:2018fsy}
where simulations with larger spins and mass ratios were obtained through a synergistic use of NR codes).
Here, to test our newly constructed EOB precessing waveform model, we enhance the 
NR parameter-space coverage, while maintaining a manageable computational cost, and perform $118$ 
new NR simulations with the pseudo spectral Einstein code (SpEC) of the Simulating eXtreme 
Spacetimes (SXS) collaboration.  The new NR simulations span BBHs with  
mass ratios  $1\mbox{--}4$, and dimensionless spins in the range $0.3\mbox{--}0.9$, and different 
spins' orientations.  To assess the accuracy of the different precessing waveform models, 
we compare them to the NR waveforms of the public SXS catalogue~\cite{Boyle:2019kee}, 
and to the new $118$ NR waveforms produced for this paper.

The paper is organized as follows. In Sec.~\ref{sec:NR} we discuss the new NR simulations of BBHs, 
and assess their numerical error. In Sec.~\ref{sec:multiEOB} we develop the multipolar EOB waveform 
model for spin-precessing BBHs, \verb+SEOBNRv4PHM+, and highlight the improvements with respect to the 
previous version~\cite{Pan:2013rra,Babak:2016tgq}, which was used in LIGO and Virgo 
inference analyses~\cite{Abbott:2016izl,Abbott:2017vtc,LIGOScientific:2018mvr}. 
In Sec.~\ref{sec:compEOBNR} we validate the accuracy of the multipolar precessing EOB model 
by comparing it to NR waveforms. We also compare the performance of \verb+SEOBNRv4PHM+ 
against the one of \verb+IMRPhenomPv3HM+, and study in which region of the parameter space 
those models differ the most from NR simulations, and also from each other. In Sec.~\ref{sec:peEOBNR} 
we use Bayesian analysis to explore the impact of the accuracy of the precessing waveform models 
when extracting astrophysical information and perform two synthetic NR injections in zero noise. 
In Sec.~\ref{sec:concl} we summarize our main conclusions and discuss future  work. Finally, 
in Appendix~\ref{sec:comparisonNRSurr} we compare the precessing waveform models to the 
NR surrogate {\tt NRSur7dq4}, and in Appendix~\ref{sec:NRparam} we list the parameters 
of the 118 NR simulations done for this paper.
 
In what follows, we use geometric units $G=1=c$ unless otherwise specified.

\section{New numerical-relativity simulations of spinning, precessing binary black holes}
\label{sec:NR}

Henceforth, we denote with $m_{1,2}$ the two BH masses (with $m_1 \geq m_2$), $\vS_{1,2} \equiv
m_{1,2}^2\,\vchi_{1,2}$ their spins, $q = m_1/m_2$ the mass ratio, $M = m_1+m_2$ the 
binary's total mass, $\mu= m_1m_2/M$ the reduced mass, and $\nu = \mu/M$ the 
symmetric mass ratio. We indicate with $\vJ = \vL + \vS $ the total angular momentum, 
where $\vL$ and $\vS = \vS_1 +\vS_2$, are the orbital angular momentum and the 
total spin, respectively

\subsection{New $\mathbf{118}$ precessing numerical-relativity waveforms}

\begin{figure}
\includegraphics[width=\linewidth]{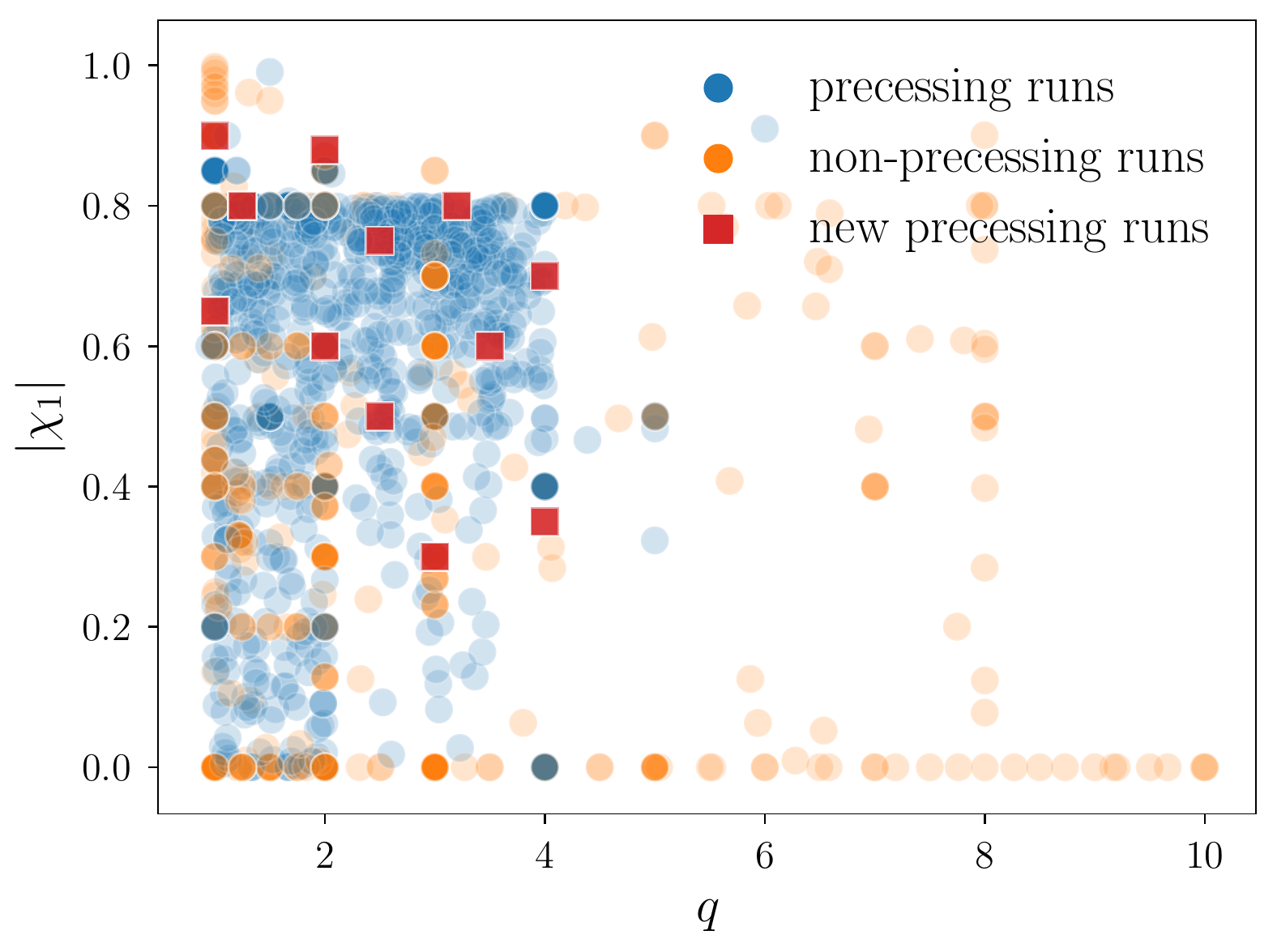}
\caption{Parameter space coverage in $q\mbox{--}\chi_{1}$ space for {\tt SpEC} waveforms. For runs
  from the {\tt SpEC} catalog~\cite{Boyle:2019kee} the opacity was changed so that runs with similar
  parameters are clearly visible. We indicate with squares precessing BBH runs performed as part of this paper. 
\label{fig:Param_space}}
\end{figure}

The spectral Einstein code ({\tt SpEC}) \footnote{\url{www.black-holes.org}} of the Simulating eXtreme Spacetimes
(SXS) collaboration is a multi-domain collocation code designed for the solution 
of partial differential equations on domains with simple topologies.
It has been used extensively to study the mergers of compact-object
binaries composed of BH~\cite{Scheel:2014ina,Lovelace:2014twa,Szilagyi:2015rwa,Lovelace:2016uwp,Afle:2018slw,Boyle:2019kee} and NSs~\cite{Foucart:2015gaa,Haas:2016cop,Foucart:2018lhe,Vincent:2019kor}, including in theories 
beyond GR~\cite{Okounkova:2018abo,Okounkova:2019dfo,Okounkova:2019zjf,Okounkova:2020rqw}. 
SpEC employs a first-order symmetric-hyperbolic formulation of Einstein's equations~\cite{Lindblom:2005qh} in the damped harmonic 
gauge~\cite{Lindblom:2009tu,Szilagyi:2009qz}. Dynamically controlled
excision boundaries are used to treat spacetime singularities~\cite{Hemberger:2012jz,Scheel:2014ina} 
(see Ref.~\cite{Boyle:2019kee} for a recent, detailed overview).

Significant progress has been made in recent years by several NR groups to improve the coverage of the BBH parameter 
space~\cite{Jani:2016wkt,Healy:2017psd,Healy:2019jyf,Huerta:2019oxn,Boyle:2019kee,Hinder:2018fsy}, 
mainly motivated by the calibration of analytical waveform models and surrogate models 
used in LIGO and Virgo data analysis.  While large strides have been made for aligned-spin cases, the exploration of
precessing waveforms has been mostly limited to $q\leq 4, \chi_{1,2} \equiv 
|\vchi_{1,2}| \leq 0.8$, typically $15\mbox{--}20$ orbital cycles before merger, 
and a large region of parameter space remains to be explored. Simulations with high mass ratio ($q\ge4$) and high spin ($|\vchi_{1}|>0.5$) are challenging, primarily due to the need
to resolve the disparate length scales in the binary system, which increases the computational cost for a given level of accuracy. Furthermore, for high spin, the apparent horizons can be  dramatically smaller, which makes it more difficult to control the excision boundaries, further increasing the computational burden.

Here, we want to improve the parameter-space coverage of the SXS catalog~\cite{Boyle:2019kee}, while 
maintaining a manageable computational cost, thus we restrict to  simulations in 
the range of mass ratios $q=1\mbox{--}4$ 
and (dimensionless) spins $\chi_{1,2}=0.3\mbox{--}0.9$, with the spin magnitudes decreasing as the mass ratio
increases. In Fig.~\ref{fig:Param_space} we display, in the $q\mbox{--} \chi_1$ parameter space,  
the precessing and non-precessing waveforms from the published SXS catalogue~\cite{Boyle:2019kee}, 
and the new precessing waveforms produced as part of this work. 

We choose to start all the simulations at the same (angular) orbital frequency, $M\Omega_0 \approx 0.0157$, where the value 
is not exact as it was modified slightly during the eccentricity-reduction procedure in {\tt SpEC}~\cite{Buonanno:2010yk}. 
This corresponds to a physical GW starting frequency of $20$ Hz at $50 M_{\odot}$ and 
results in the number of orbits up to merger varying between $15$ and $30$  in 
our new catalog.

We parametrize the directions of the spins by three angles, the angles $\theta_{1,2}$ between the spins 
and the unit vector along the Newtonian orbital angular momentum, $\vLhat_{\rm N}$, and the angle
$\Delta \phi$ between the projections of the spins in the orbital plane. Explicitly, 
\begin{subequations}
\begin{align}
\theta_{i}  &=\arccos (\vchi_{i} \cdot \vLhat_{\rm N})\,, \\
\Delta \phi &= \arccos \left(\frac{\cos\theta_{12}-\cos\theta_1\cos\theta_2}{\sin\theta_1\sin\theta_2}\right)\,,
\end{align}
\end{subequations}
where $\cos\theta_{12}=\vchi_{1}\cdot \vchi_{2}$. We make the choice
that ${\mathbf \chi}_{1}$ lies in the $\vLhat_{\rm N}-\mathbf{n}$ plane, where $\mathbf{n}$  
is the unit vector along the binary's radial separation, at the start of the
simulation.  The angles are chosen to be $\theta_{i,0}\in \{60^{\circ},\theta_{\rm
max}\}$, and $\Delta\phi_0 \in \{0,90^{\circ}\}$. Here $\theta_{max}$ is the angle that maximizes the opening angle of $\vL_{\rm N}$ around the total angular momentum
$\vJ$ and is computed assuming that the two spins are co-linear, giving
\begin{equation}
\cos \theta_{\rm max} = -\frac{|\mathbf{S}|}{|\vL_{\rm N}|} =
-\frac{m_{1}^{2}\,\chi_{1}+m_{2}^{2}\,\chi_{2}}{|\vL_{\rm N}|}\,,
\end{equation}
with $|\vL_{\rm N}|=\mu M^{2/3}\Omega^{-1/3}$ for circular orbit, 
being $\Omega$ the orbital angular frequency. For each choice
of $\{q,\chi\}$ we choose 10 different configurations divided into two 
categories: i) $\chi_{1}=\chi_{2}=\chi,\theta_{i,0}\in\{60^{\circ},\theta_{\rm
  max}\},\ \Delta \phi_0\in\{0,90^{\circ}\}$ giving eight runs, and ii)
$\chi_{1} = \chi, \chi_{2}=0, \theta_{1,0} \in \{60^{\circ},\theta_{\rm
  max}\}$ giving two runs. The detailed parameters can be found in Appendix~\ref{sec:NRparam}.

These choices of the spin directions allow us to test the multipolar 
precessing model \verb+SEOBNRv4PHM+ in many different regimes, including where the effects of 
precession are maximized, and where spin-spin effects are significant or diminished.

\subsection{Unfaithfulness for spinning, precessing waveforms}

The gravitational signal emitted by non-eccentric BBH systems 
and observed by a detector depends on 15 parameters: the component masses $m_1$ and
$m_2$ (or equivalently the mass ratio $q = m_1/m_2 \geq 1$ and
the total mass $M = m_1 + m_2$), the dimensionless spins
$\vchi_1(t)$ and $\vchi_2(t)$, the direction to observer from the
source described by the angles $(\iota,\varphi_0)$, the luminosity
distance $d_L$, the polarization $\psi$, the location in the sky
$(\theta,\phi)$ and the time of arrival $t_c$. The gravitational strain 
can be written as:
\begin{align}
\label{eq:det_strain}
h(t) \equiv & F_+(\theta,\phi,\psi) \ h_+(\iota,\varphi_0, d_L, \boldsymbol{\xi},t_{\mathrm{c}};t) \nonumber \\
&+ F_\times(\theta,\phi,\psi)\ h_\times(\iota,\varphi_0, d_L, \boldsymbol{\xi},t_\mathrm{c};t)\,,
\end{align}
where to simplify the notation we introduce the function $\boldsymbol{\xi} \equiv \left(q, M, \vchi_{1}(t), \vchi_{2}(t)\right)$.
The functions $F_+(\theta,\phi,\psi)$ and $F_\times(\theta,\phi,\psi)$ are the antenna 
patterns~\cite{Sathyaprakash:1991mt,Finn:1992xs}:
\begin{subequations}
\begin{align}
F_+(\theta,\phi,\psi) &= \frac{1+ \cos^2(\theta)}{2} \ \cos(2\phi) \ \cos(2\psi)+\\ \nonumber
& -\cos(\theta) \ \sin(2\phi)\ \sin(2\psi),\\ 
F_\times(\theta,\phi,\psi) &= \frac{1+ \cos^2(\theta)}{2} \ \cos(2\phi) \ \sin(2\psi)+\\  \nonumber 
&+ \cos(\theta) \ \sin(2\phi)\ \cos(2\psi).
\end{align}
\end{subequations}
Equation \eqref{eq:det_strain} can be rewritten as:
\begin{align}
\label{eq:det_strain_kappa}
h(t) \equiv & \mathcal{A}(\theta,\phi)\big[\cos\kappa(\theta,\phi,\psi) \ h_+(\iota, \varphi_0, d_L, \boldsymbol{\xi}, t_{\mathrm{c}};t) \nonumber \\
&+ \sin\kappa(\theta,\phi,\psi) \ h_\times (\iota, \phi, d_L, \boldsymbol{\xi},t_{\mathrm{c}};t) \big],
\end{align}
where $\kappa(\theta,\phi,\psi)$ is the \textit{effective polarization}~\cite{Capano:2013raa} defined as:
\begin{equation}
e^{i \kappa(\theta,\phi,\psi)} = \frac{F_+(\theta,\phi,\psi) + i F_\times(\theta,\phi,\psi)}{\sqrt{F_+^2(\theta,\phi,\psi) + F_\times^2(\theta,\phi,\psi)}},
\end{equation}
which has support in the region $[0, 2\pi)$, while $\mathcal{A}(\theta,\phi)$ reads:
\begin{equation}
\mathcal{A}(\theta,\phi) = \sqrt{F_+^2(\theta,\phi,\psi) + F_\times^2(\theta,\phi,\psi)}\,.
\end{equation}
Henceforth, to ease the notation we suppress the dependence on $(\theta,\phi,\psi)$ in $\kappa$.

Let us introduce the inner product between two waveforms $a$ and $b$~\cite{Sathyaprakash:1991mt,Finn:1992xs}:
\begin{equation}
\left( a, b\right) \equiv 4\ \textrm{Re}\int_{f_{\rm in}}^{f_{\rm max}} df\,\frac{\tilde{a}(f) \ \tilde{b}^*(f)}{S_n(f)},
\end{equation}
where a tilde indicates the Fourier transform, a star the complex
conjugate and $S_n(f)$ is the one-sided power spectral density (PSD)
of the detector noise. We employ as PSD the Advanced LIGO’s “zero-detuned
high-power” design sensitivity curve~\cite{Barsotti:2018}. Here we use $f_{\rm in} = 10 {\rm Hz}$  
and $f_{\rm max} = 2 {\rm kHz}$, when both waveforms fill the band. For cases where this is 
not the case (e.g the NR waveforms) we set $f_{\rm in}=1.05f_{\rm start}$, where $f_{\rm start}$ is the starting frequency of the waveform.

To assess the closeness between two waveforms $s$ (e.g., the signal) and $\tau$ (e.g., the template), 
as observed by a detector, we define the following \textit{faithfulness} function~\cite{Cotesta:2018fcv}:
\begin{equation}
\label{eq:faith}
\mathcal{F}(M_{\textrm{s}},\iota_{\textrm{s}},{\varphi_0}_{\textrm{s}},\kappa_{\textrm{s}}) \equiv  \max_{t_c, {\varphi_0}_{\tau}, \kappa_{\tau}} \left[\left . \frac{ \left( s,\,\tau \right)}{\sqrt{ \left( s,\,s \right) \left( \tau,\,\tau \right)}}\right \vert_{\substack{\iota_{\mathrm{s}} = \iota_{\tau} \\\boldsymbol{\xi}_{\mathrm{s}}(t_{\mathrm{s}} = t_{0_\mathrm{s}}) = \boldsymbol{\xi}_{\tau}(t_\tau = t_{0_\mathrm{\tau}})}} \right ].
\end{equation}
While in the equation above we set the inclination angle $\iota$ of signal and template waveforms to  be the same, the  coalescence time $t_c$
and the  angles ${\varphi_0}_{\tau}$ and  $ \kappa_{\tau}$ of  the template
waveform are adjusted to maximize  the faithfulness. This is a typical
choice motivated by the fact these quantities are not interesting from
an  astrophysical  perspective.  The   maximizations  over  $t_c$  and
${\varphi_0}_{\tau}$ are performed numerically, while the maximization
over   $\kappa_{\tau}$  is   done  analytically   following  the
procedure described  in Ref. ~\cite{Capano:2013raa} (see Appendix  A therein).

The condition $\boldsymbol{\xi}_{\mathrm{s}}(t_{\mathrm{s}} =
t_{0_\mathrm{s}}) = \boldsymbol{\xi}_{\tau}(t_\tau =
t_{0_\mathrm{\tau}})$ in Eq.~\eqref{eq:faith} enforces that the mass
ratio $q$, the total mass $M$ and the spins $\vchi_{1,2}$ of the
template waveform at $t = t_0$ (i.e., at the beginning of the
waveform) are set to have the same values of the ones in the signal
waveform at its $t_0$. When computing the faithfulness between NR
waveforms with different resolutions this condition is trivially
satisfied by the fact that they are generated using the same initial
data. In the case of the faithfulness between NR and any model from the   \verb+SEOBNR+  family, it
is first required to ensure that $t_0$ has the same physical meaning
for both waveforms. Ideally $t = t_{0_\mathrm{\tau}}$ in the \verb+SEOBNR+ 
waveform should be fixed by requesting that the frequency of the \verb+SEOBNR+
$(2,2)$ mode at $t_{0_\mathrm{\tau}}$ coincides with the NR (2,2) mode
frequency at $t_{0_\mathrm{s}}$. This is in practice not possible
because the NR (2,2) mode frequency may display small oscillations
caused by different effects --- for example the persistence of the
junk radiation, some residual orbital eccentricity or spin-spin
couplings~\cite{Buonanno:2010yk}.  Thus, the frequency of the \verb+SEOBNR+ 
$(2,2)$ mode at $t = t_{0_\mathrm{\tau}}$ is chosen to guarantee the
same time-domain length of the NR waveform. \footnote{The difference
  between the NR (2,2) mode frequency and the {\tt SEOBNRv4PHM} (2,2) frequency
  chosen at $t = t_0$ is never larger than $5\%$.}.
In practice, we
require that the peak of $\sum_{\ell, m} |h_{\ell m}|^2$, as elapsed
respectively from $t_{0_\mathrm{s}}$ and $t_{0_\mathrm{\tau}}$, occurs
at the same time in NR and \verb+SEOBNR+. 
For waveforms from the \verb+IMRPhenom+ family we adopt a different approach, and following 
the method outlined in Ref.~\cite{Khan:2018fmp}, also optimize the faithfulness numerically over the
reference frequency of the waveform.

The faithfulness defined in Eq.~\eqref{eq:faith} is  still a function of 4
parameters (i.e., $M_{\textrm{s}},\iota_{\textrm{s}},{\varphi_0}_{\textrm{s}},\kappa_{\textrm{s}}$), 
therefore it does not allow to describe the agreement between waveforms in a compact form. For  this  purpose  we define  the \textit{sky-and-polarization-averaged faithfulness}~\cite{Babak:2016tgq} as:
\begin{equation}
\overline{\mathcal{F}}(M_\mathrm{s}, \iota_\mathrm{s}) \equiv  \frac{1}{8\pi^2}\int_{0}^{2\pi} d\kappa_{\mathrm{s}} \int_{0}^{2\pi} d{\varphi_0}_{\mathrm{s}} \ \mathcal{F}(M_\mathrm{s},\iota_{\textrm{s}},{\varphi_0}_{\textrm{s}},\kappa_{\textrm{s}}).
\label{eq:avg_faith}
\end{equation}
Despite the apparent difference, the sky-and-polarization-averaged
faithfulness $\overline{\mathcal{F}}$ defined above is equivalent to the one given in Eqs.~(9) 
and (B15) of Ref.~\cite{Babak:2016tgq}. The definition in 
Eq.~\eqref{eq:avg_faith} is less computationally expensive because,
thanks to the parametrization of the waveforms in
Eq.~\eqref{eq:det_strain_kappa}, it allows one to write the sky-and-polarization-averaged 
faithfulness as a double integral instead of the
triple integral in Eq.~(B15) of Ref.~\cite{Babak:2016tgq}.
We also define the sky-and-polarization-averaged, signal-to-noise (SNR)-{\it weighted} faithfulness as:
\newpage\begin{widetext}
\begin{equation}
\overline{\mathcal{F}}_{\mathrm{SNR}}(M_\mathrm{s},\iota_{\mathrm{s}}) \equiv \sqrt[3]{\frac{\int_{0}^{2\pi} d\kappa_ {\mathrm{s}} \int_{0}^{2\pi} d{\varphi_0}_{\mathrm{s}} \ \mathcal{F}^{3}(M_{\textrm{s}},\iota_{\textrm{s}},{\varphi_0}_{\textrm{s}},\kappa_{\textrm{s}}) \ \mathrm{SNR}^3(\iota_{\textrm{s}},{\varphi_0}_{\textrm{s}},\kappa_{\textrm{s}})}{\int_{0}^{2\pi} d\kappa_{\mathrm{s}} \int_{0}^{2\pi} d{\varphi_0}_{\mathrm{s}} \ \mathrm{SNR}^3(\iota_{\textrm{s}},{\varphi_0}_{\textrm{s}},\kappa_{\textrm{s}})}}.
\end{equation}
\end{widetext}
where the $\mathrm{SNR}(\iota_{\textrm{s}},{\varphi_0}_{\textrm{s}},\theta_\textrm{s}, \phi_\textrm{s},\kappa_{\textrm{s}},{D_{\mathrm{L}}}_{\mathrm{s}},
\boldsymbol{\xi}_\mathrm{s},{t_c}_\mathrm{s})$ is defined as:
\begin{equation}
\mathrm{SNR}(\iota_{\textrm{s}},{\varphi_0}_{\textrm{s}},\theta_\textrm{s}, \phi_\textrm{s}, \kappa_{\textrm{s}},{D_{\mathrm{L}}}_{\mathrm{s}},\boldsymbol{\xi}_\mathrm{s},{t_c}_\mathrm{s}) \equiv \sqrt{\left(h_{\mathrm{s}},h_{\mathrm{s}}\right)}.
\end{equation}
This is also an interesting metric because weighting the faithfulness
with the SNR takes into account that, at fixed distance, 
the SNR of the signal depends on its phase and on the effective
polarization (i.e., a combination of waveform polarization and
sky-position). Since the SNR scales with the luminosity distance, the
number of detectable sources scale with the $\mathrm{SNR}^3$,
therefore signals with a smaller SNR are less likely to be
observed. Finally, we define the unfaithfulness (or
mismatch) as
\begin{equation}
\overline{\mathcal M} = 1 -  \overline{\mathcal{F}}\,.
\label{mismatch}
\end{equation}

\subsection{Accuracy of new numerical-relativity waveforms}

To assess the accuracy of the new NR waveforms, we compute the
sky-and-polarization-averaged unfaithfulness defined in Eq.~(\ref{eq:avg_faith})
between the highest and second highest resolutions in the NR simulation. 

\begin{figure}
\includegraphics[width=\linewidth]{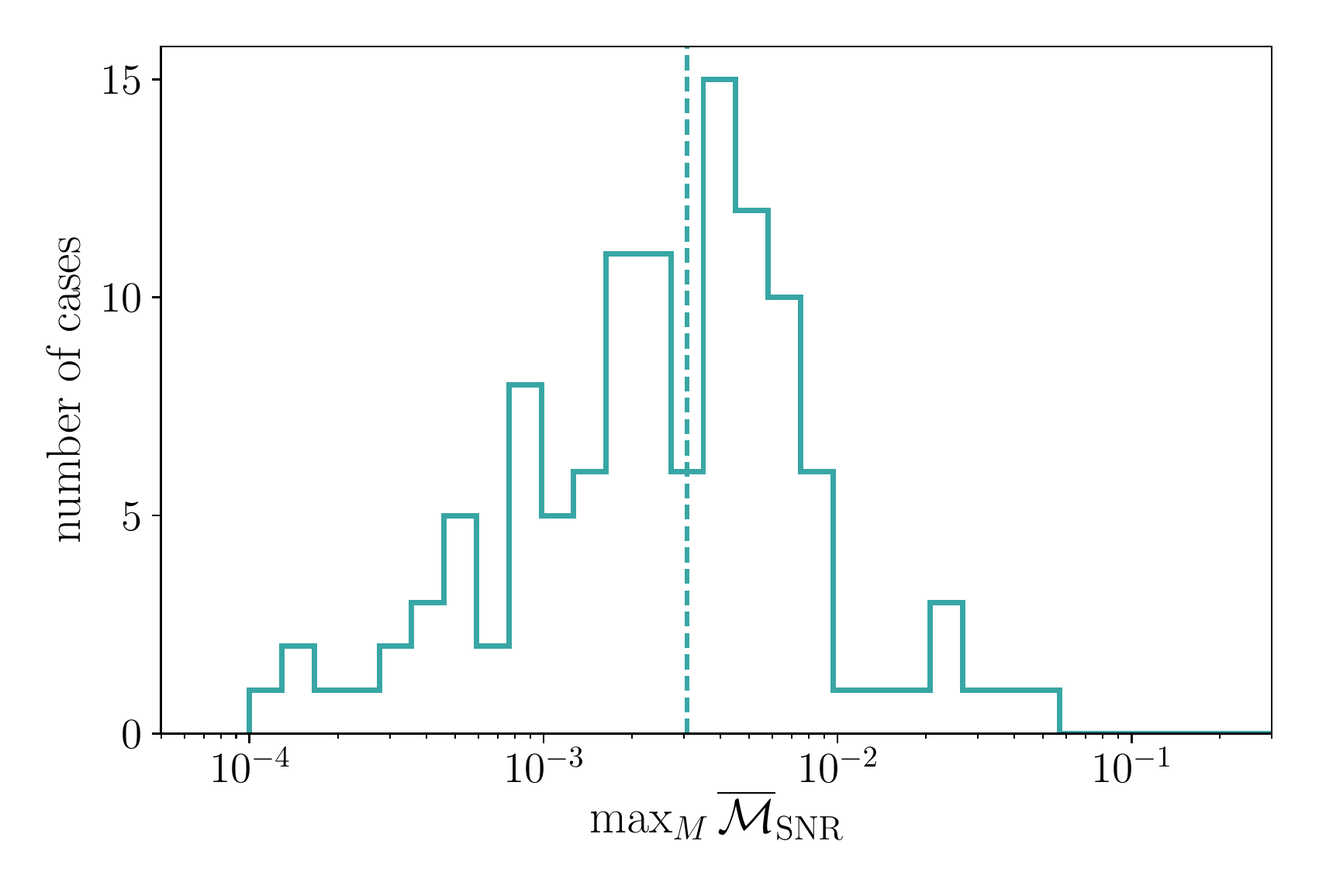}
 \caption{The sky-and-polarization-averaged unfaithfulness between 
the highest and second highest resolutions in the NR simulation 
maximized over the total mass for the new 
118 NR precessing waveforms. The inclination used is $\pi/3$. The vertical dashed line shows the median.}
  \label{fig:unfaith_NRNR}
\end{figure}

Figure~\ref{fig:unfaith_NRNR} shows a histogram of the unfaithfulness,
evaluated at $\iota_{s}={\pi}/{3}$ maximized over the total mass, between 20 and
200 $M_\odot$. It is apparent that the unfaithfulness is  below $1 \%$ for most
cases, but there are several cases with much higher unfaithfulness. 
This tail to high unfaithfulness has been observed previously, when evaluating 
the accuracy of SXS simulations in Ref.~\cite{Varma:2019csw}. Therein, it was 
established that, when the non-astrophysical junk radiation perturbs the parameters of
the simulation sufficiently, the different resolutions actually correspond to
different physical systems. Thus, taking the difference between adjacent resolutions 
is no longer an appropriate estimate of the error.

We also find that the largest unfaithfulness occurs when the difference in parameters is
largest, thus confirming that it is the difference in parameters that dominates
the unfaithfulness.

\subsection{Effect of mode asymmetries in numerical-relativity waveforms}
\label{sec:mode_asymm}

The gravitational polarizations at time $t$ and location $(\varphi_0,\iota)$ on the coordinate sphere from the binary can be decomposed in $-2$--spin-weighted spherical harmonics, as follows
\begin{equation}
h_{+}(\varphi_0,\iota;t) - i h_{\times}(\varphi_0,\iota;t) = \sum_{\ell =2} \sum_{m=-\ell}^{m=+\ell} {}_{-2} Y_{\ell m}(\varphi_0,\iota)\,h_{\ell m}(t) \,.
\end{equation}
For nonprecessing binaries, the invariance of the system under reflection across the orbital plane (taken to be the $x\mbox{--}y$ plane) implies 
$h_{\ell m}= (-1)^\ell h_{\ell -m}^*$. The latter is a very convenient relationship --- for example it renders unnecessary to model modes 
with negative values of $m$. However, this relationship is no longer satisfied for precessing binaries. 

As investigated in previous NR studies~\cite{Pekowsky:2013ska,Boyle:2014ioa}, we expect the asymmetries 
between opposite-$m$ modes to be small as compared to the dominant $(2,2)$-mode emission (at least during the
inspiral) in a co-rotating frame that maximizes emission in the $(2,\pm 2)$ modes, also known as the \textit{maximum-radiation 
frame}~\cite{Boyle:2011gg,Boyle:2013nka}. However, while the asymmetries are expected to be small during the inspiral, 
the difference in phase and amplitude between positive and negative $m$-modes might become non-negligible at merger. 

\begin{figure}
	\includegraphics[width=\linewidth]{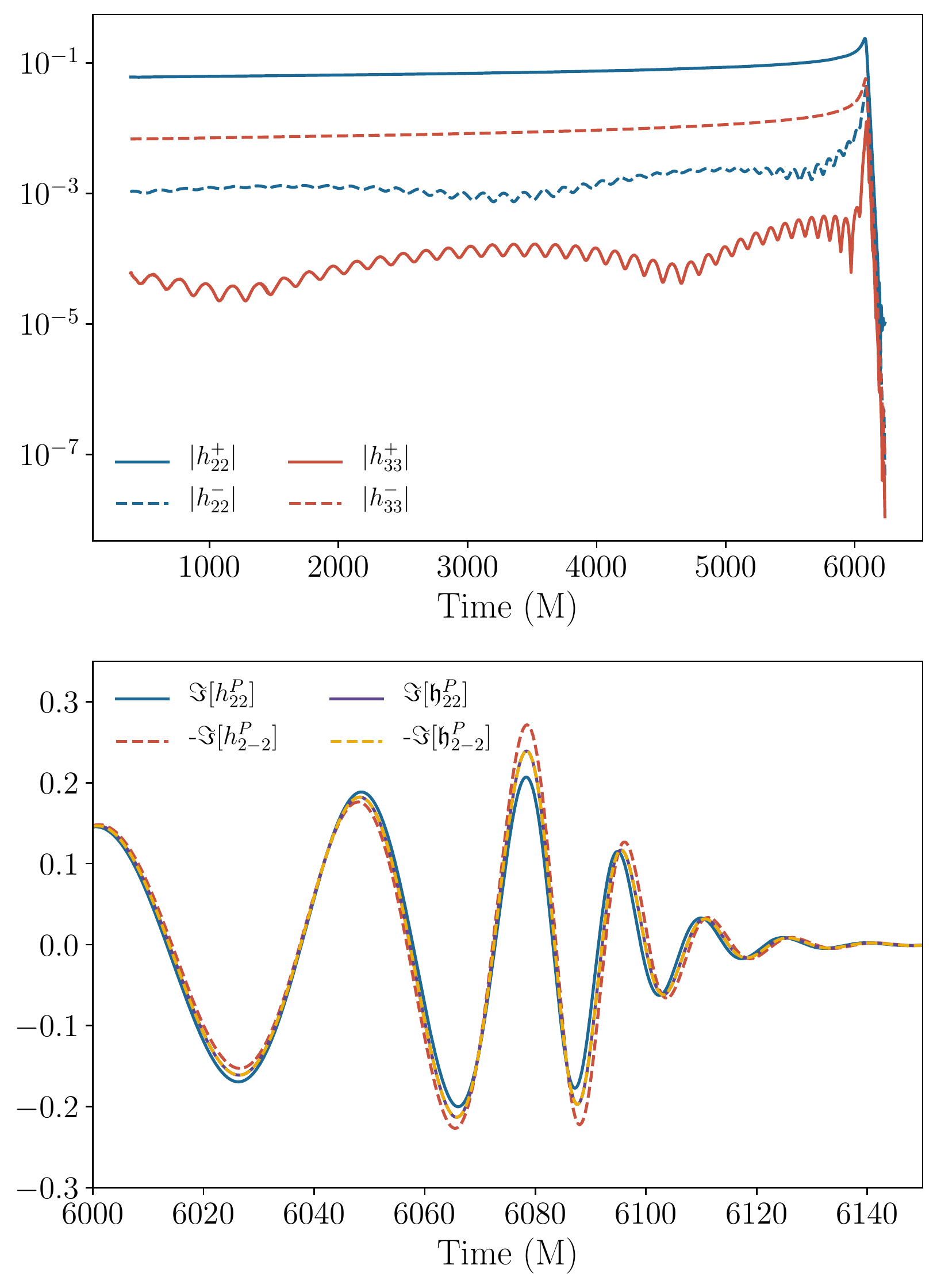}
        \caption{\emph{Top}: the behavior of $h_{\ell m}^{\pm}$ in the
          NR simulation {\texttt PrecBBH000078}. Note that especially during the inspiral,
          $|h_{22}^{+}|$ is much larger than $|h_{22}^{-}|$ while
          $|h_{33}^{-}|$ is much larger than $|h_{33}^{+}|$. \emph{Bottom}: an example of waveform symmetrization for the same NR
          case, shown in the co-precessing frame. The symmetrized
          waveform obeys the usual conjugation symmetry as expected,
          and represents a reasonable average to the behavior of the
          unsymmetrized modes.}
	\label{fig:example_symm_waveform}
\end{figure}

As we discuss in the next section, when building multipolar waveforms (\verb+SEOBNRv4PHM+) for precessing binaries by rotating 
modes from the co-precessing~\cite{Buonanno:2002fy,Schmidt:2010it,Boyle:2011gg,O'Shaughnessy:2011fx,Schmidt:2012rh} 
to the inertial frame of the observer, we shall neglect the mode asymmetries. To quantify the error introduced by this assumption, 
we proceed as follows. We first take NR waveforms in the co-precessing frame and construct \emph{symmetrized} waveforms. Specifically, we 
consider the combination of waveforms in the co-precessing frame defined by (e.g., see Ref.~\cite{Varma:2019csw})
\begin{equation}
h_{\ell m}^{\pm} = \frac{h_{\ell m}^{P}\pm h_{\ell -m}^{P*}}{2}\,.
\end{equation}
Note that if the assumption of conjugate symmetry holds (i.e., if
$h^P_{\ell -m} = (-1)^{\ell}h^{P*}_{\ell m}$), then for even (odd)
$\ell$ modes, $h_{\ell m}^{+}$ ($h_{\ell m}^{-}$) is non-zero while
the other component vanishes. If the assumption does not hold, it is
still true that at given $\ell$, one of the components is much larger
than the other, as shown in top panel of Fig.~\ref{fig:example_symm_waveform}. 
Motivated by this, we define the symmetrized modes (for $m>0$) as~\cite{Varma:2019csw} 
\begin{equation}
\mathfrak{h}_{\ell m}^{P} =
\begin{cases}
h_{\ell m}^{+} & \text{if}\ \ell\ \text{is even}\,, \\
h_{\ell m}^{-} & \text{if}\ \ell\ \text{is odd}\,. 
\end{cases} 
\end{equation}
The other modes are constructed as $\mathfrak{h}_{\ell -m}^{P}=\mathfrak{h}_{\ell m}^{P*}$ for $m<0$, 
and we set $m=0$ modes to zero. The bottom panel of Fig.~\ref{fig:example_symm_waveform} shows an example 
of asymmetrized waveform for the case \verb|PrecBBH000078| of the SXS catalogue, in the
co-precessing frame. It is obvious that the asymmetry between the
modes has been removed and that the symmetrized waveform does indeed
represent a reasonable ``average'' between the original modes. The symmetrized waveforms in the
inertial frame are obtained by rotating the co-precessing frames modes back to the inertial frame.

\begin{figure}
	\includegraphics[width=\linewidth]{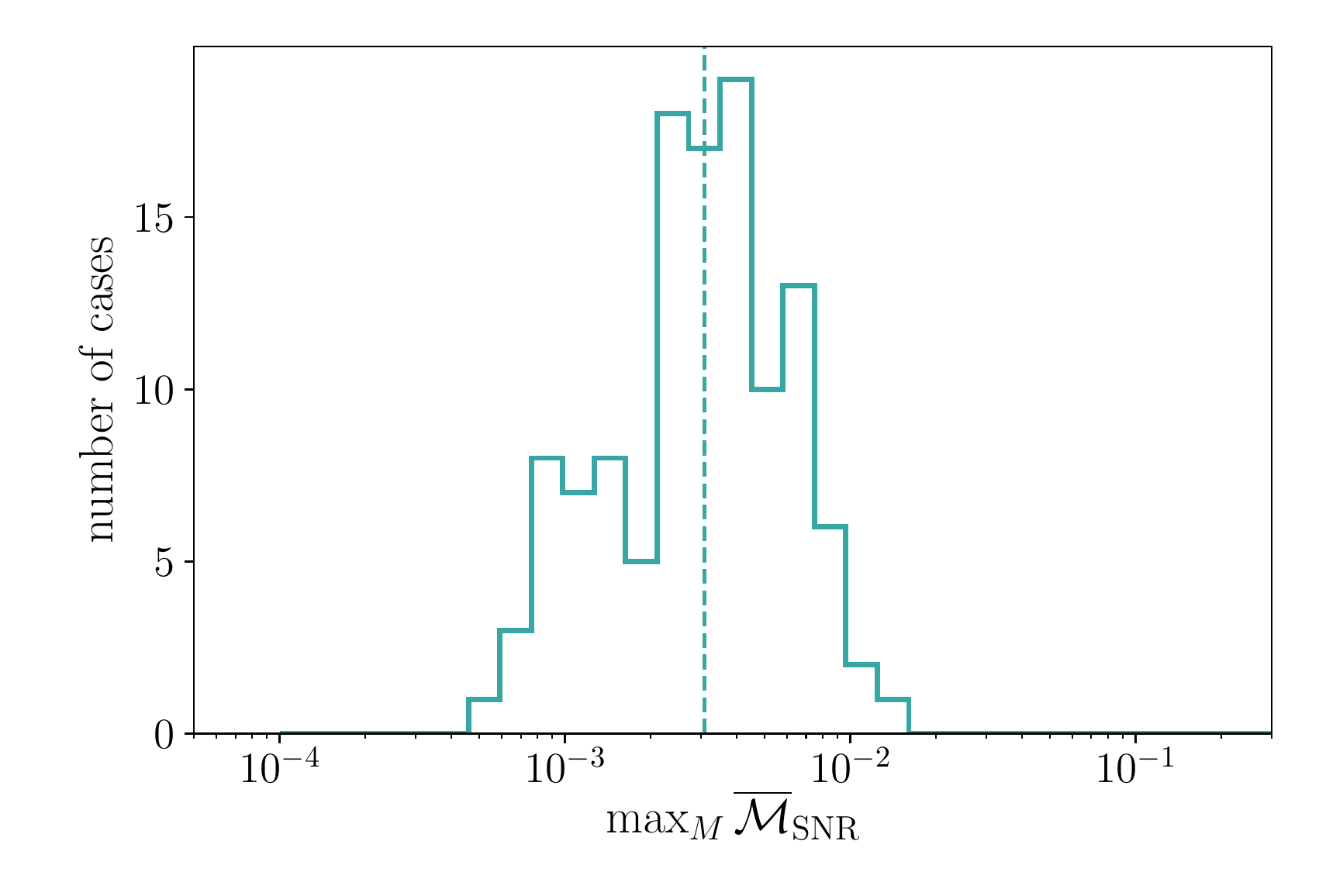}
	\caption{The sky-and-polarization-averaged unfaithfulness between NR and symmetrized 
NR waveforms, maximized over the total mass for the new 118 NR precessing waveforms. 
The inclination used is $\pi/3$. The vertical dashed line shows the median.}
	\label{fig:unfaith_NRNR_symm}
\end{figure}

In Fig.~\ref{fig:unfaith_NRNR_symm}, we show the sky-and-polarization 
averaged unfaithfulness between the NR waveforms and the
symmetrized waveforms described above, maximized over the total mass,
including all modes available in the NR simulation, that is up to $\ell=8$. 
For the vast majority of the cases, the unfaithfulness is $\sim0.5\%$, 
and all cases have unfaithfulness below $2\%$. This demonstrates that the effect of
neglecting the asymmetry is likely subdominant to other sources of
error such as the modeling of the waveform phasing, although the best 
way of quantifying the effect is to perform a Bayesian 
parameter-estimation study, which we leave to future work.  

\section{Multipolar EOB waveforms for spinning, precessing binary black holes}
\label{sec:multiEOB}

We briefly review the main ideas and building blocks of the EOB approach, and then describe 
an improved spinning, precessing EOBNR waveform model, which, for the first time, also contains 
multipole moments beyond the quadrupolar one. The model  is already
available in the LIGO Algorithm Library~\cite{LAL} under the name of \verb+SEOBNRv4PHM+.
We refer the reader to Refs.~\cite{Taracchini:2012ig,Taracchini:2013rva,Pan:2013rra,Babak:2016tgq,Cotesta:2018fcv}
for more details of the EOB framework and its most recent waveform models. 
Here we closely follow Ref.~\cite{Babak:2016tgq}, highlighting when needed differences with respect to the 
previous precessing waveform model developed in Ref.~\cite{Babak:2016tgq} (\verb+SEOBNRv3P+~\footnote{We note that 
whereas in LAL the name of this waveform approximant is {\tt SEOBNRv3}, here we 
add a ``P'' to indicate ``precession'', making the notation uniform with respect to 
the most recent developed model {\tt SEOBNRv4P}~\cite{Babak:2016tgq}.}). 

\subsection{Two-body dynamics}

The EOB formalism~\cite{Buonanno:1998gg, Buonanno:2000ef,Damour:2000we,Damour:2001tu} can describe
analytically the GW emission of the entire coalescence process, notably inspiral, merger and ringdown,  
and it can be made highly accurate by including information from NR. For the two-body conservative 
dynamics, the EOB approach relies on a Hamiltonian $H_{\textrm{EOB}}$, which is
constructed through: (i) the Hamiltonian $H_{\textrm{eff}}$
of a spinning particle of mass $\mu \equiv m_1 m_2/(m_1 + m_2)$ and
spin $\vS_* \equiv \vS_*(m_1,m_2,\vS_1,\vS_2)$ moving in an effective,
deformed Kerr spacetime of mass $M\equiv m_1 + m_2$ and spin
$\vS_{\textrm{Kerr}} \equiv \vS_1 +
\vS_2$~\cite{Barausse:2009aa,Barausse:2009xi,Barausse:2011ys}, and (ii) an
energy map between $H_{\textrm{eff}}$ and
$H_{\textrm{EOB}}$~\cite{Buonanno:1998gg}
\be
H_{\textrm{EOB}} \equiv
M\sqrt{1+2\nu\left(\frac{H_{\textrm{eff}}}{\mu} - 1\right)}-M\,,
\ee
where $\nu = \mu/M$ is the symmetric mass ratio. The deformation
of the effective Kerr metric is fixed by requiring that at any given PN order,
the PN-expanded Hamiltonian $H_{\textrm{EOB}}$ agrees with the PN Hamiltonian for
BBHs~\cite{Blanchet:2013haa}. In the EOB Hamiltonian used in this 
paper~\cite{Barausse:2009xi,Barausse:2011ys}, the spin-orbit (spin-spin) couplings are
included up to 3.5PN (2PN) order~\cite{Barausse:2009xi,Barausse:2011ys}, while 
the non-spinning dynamics is incorporated through 4PN order~\cite{Cotesta:2018fcv}. The dynamical variables
in the EOB model are the relative separation $\vR$ and its canonically conjugate momentum $\vP$, and the
spins $\vS_{1,2}$. The conservative EOB dynamics is completely general
and can naturally accommodate precession~\cite{Pan:2013rra,Babak:2016tgq} and 
eccentricity~\cite{Hinderer:2017jcs,Liu:2019jpg,Chiaramello:2020ehz}.

When BH spins have generic orientations, both the orbital plane and the
spins undergo precession about the total angular momentum of the
binary, defined as $\vJ \equiv \vL + \vS_1 + \vS_2$, where $\vL \equiv
\mu\, \vR \times \vP$. We also introduce the Newtonian orbital
angular momentum $\vL_N\equiv \mu\, \vR \times \dot{\vR}$, which at any
instant of time is perpendicular to the binary's orbital plane.
Black-hole spin precession is described by the following equations 
\be
\frac{\mathrm{d}\vS_{1,2}}{\mathrm{d}t} = \frac{\partial H_{\mathrm{EOB}}}{\partial \vS_{1,2}} \times \vS_{1,2}\,.
\ee

In the EOB approach, dissipative effects enter in the equations of motion through a nonconservative 
radiation-reaction force that is expressed in terms of the GW energy flux through the waveform 
multipole moments~\cite{Buonanno:2005xu,Damour:2007xr,Damour:2008gu,Pan:2010hz} as
\be
\label{RRforce}
\boldsymbol{\mathcal{F}} \equiv \frac{\Omega}{16\pi}
\frac{\vP}{|\vL|}\sum_{\ell=2}^{8}\sum_{m=-\ell}^{\ell} m^2\vert
d_{\textrm{L}} h_{\ell m}\vert^2\,,
\ee
where $\Omega \equiv |\vR \times \dot{\vR}|/|\vR|^2$ is the (angular) orbital frequency, $d_{\textrm L}$
is the luminosity distance of the BBH to the observer, and the
$h_{\ell m}$'s are the GW multipole modes. As discussed in 
Refs.~\cite{Cotesta:2018fcv,Bohe:2016gbl}, the $h_{\ell m}$ used in the 
energy flux are not the same as those used for building the 
gravitational polarizations in the inertial frame, since the latter  
include the nonquasi-circular corrections, which enforce that the 
SEOBNR waveforms at merger agree with the NR data, when available.

\begin{figure}
\includegraphics[angle=0,width=\linewidth]{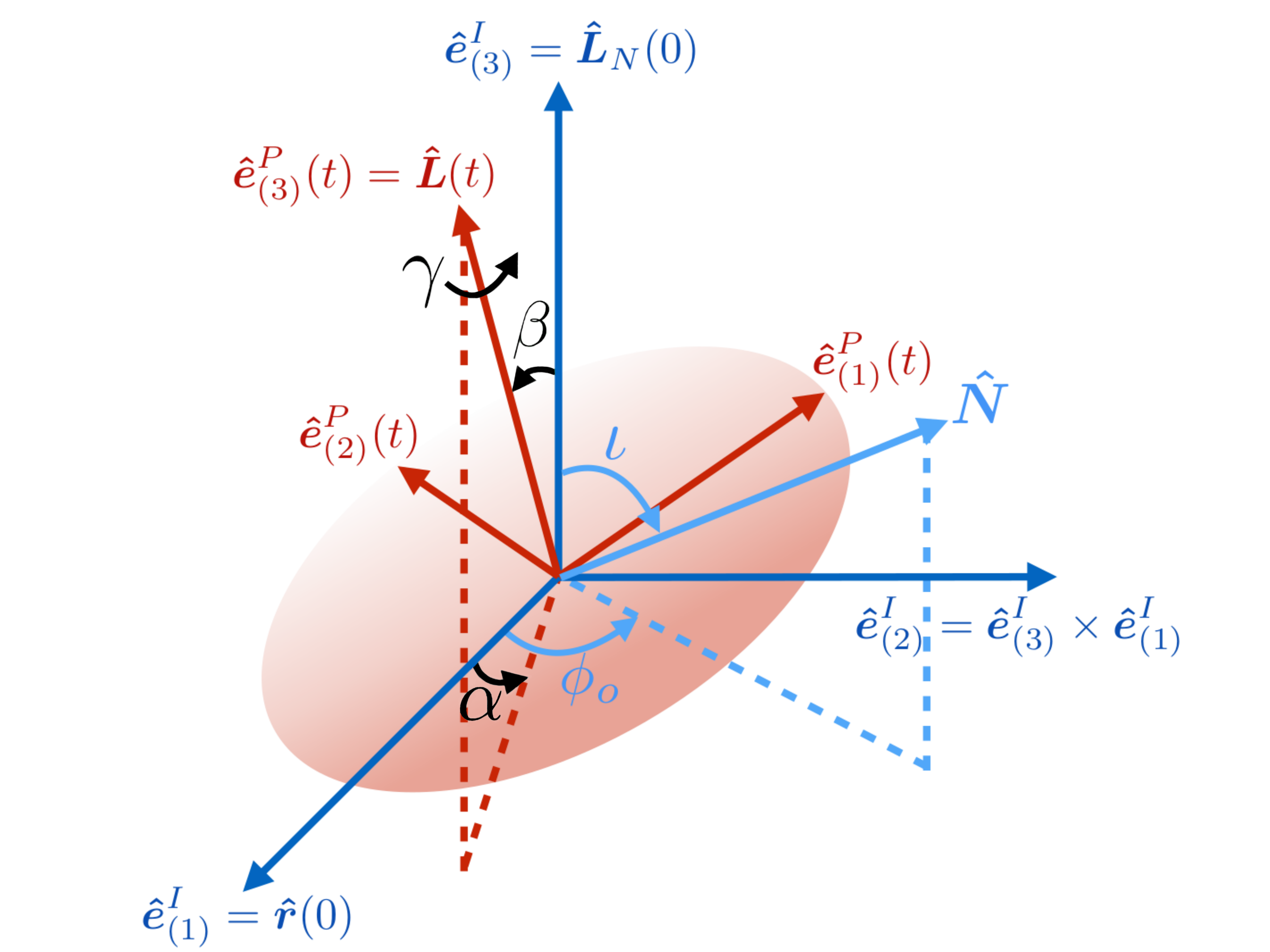}
\caption{Frames used in  the construction of the {\tt SEOBNRv4PHM} model: the observer's frame (blue), defined by the directions of the
  initial orbital angular momentum $\vLhat(0)$ and separation
  $\vR(0)$, and co-precessing frame (red), instantaneously aligned
  with $\vLhat(t)$ and described by the Euler angles $(\alpha, \beta,
  \gamma)$ (see text below for details). \label{F:Pframe}}
\end{figure}

\subsection{Inspiral-plunge waveforms}

For the inspiral-plunge waveform, the EOB approach uses a factorized, resummed 
version~\cite{Damour:2007xr,Damour:2008gu,Pan:2010hz,Cotesta:2018fcv} of the 
frequency-domain PN formulas of the modes~\cite{Arun:2008kb,Mishra:2016whh}.
As today, the factorized resummation has been developed only for quasicircular, 
nonprecessing BBHs~\cite{Damour:2008gu,Pan:2010hz}, and it has been shown to improve the accuracy of the PN expressions in the 
test-particle limit, where one can compare EOB predictions to numerical solutions of the
Regge-Wheeler-Zerilli and Teukolsky equations~\cite{Bernuzzi:2011aj,Barausse:2011kb,Taracchini:2013wfa,Harms:2015ixa}. 

The radiation-reaction force $\boldsymbol{\mathcal{F}}$ in Eq.~(\ref{RRforce}) 
depends on the amplitude of the individual GW modes $|h_{\ell m}|$, which, in the non-precessing
case, are functions of the constant aligned-spin magnitudes $\boldsymbol{\chi}_{1,2}\cdot \vLhat$.
In the precessing case, these modes depend on time, as  
$\boldsymbol{\chi}_{1,2}(t)\cdot \vLhat(t)$, and they depend on the generic, 
precessing orbital dynamics through the radial separation $r$ and orbital frequency $\Omega$, which carry 
modulations due to spin-spin couplings whenever precession is present. However, we stress that 
with this choice of the radiation-reaction force and waveform model, not all spin-precession effects are included, since 
the PN formulas of the modes~\cite{Arun:2008kb} also contain terms that depend 
on the in-plane spin components. 

For data-analysis purposes, we need to compute the GW polarizations in the inertial-frame
of the observer (or simply \textit{observer's frame}). We denote quantities 
in this frame with the superscript $I$. The observer's frame is defined  
by the triad $\{\vE^I_{(i)}\}$ ($i=1,2,3$), where
$\vE_{(1)}^I\equiv\boldsymbol{\hat{r}}(0)$, $\vE^I_{(3)} \equiv
\vLhat_N(0)$ and $\vE_{(2)}^I\equiv\vE_{(3)}^I \times \vE_{(1)}^I$. Moreover, in
this frame, the line of sight of the observer is parametrized as
$\boldsymbol{\hat{N}} \equiv
(\sin{\iota}\cos{\phi_o},\sin{\iota}\sin{\phi_o},\cos{\iota})$ (see
Fig.~\ref{F:Pframe}). We also introduce the observer's frame with the 
polarization basis $\{\vE_{(1)}^r,\vE_{(2)}^r\}$ such that $\vE_{(1)}^r \equiv
(\vE^I_{(3)}\times\boldsymbol{\hat{N}})/|\vE^I_{(3)}\times\boldsymbol{\hat{N}}|$
and $\vE_{(2)}^r \equiv \boldsymbol{\hat{N}} \times \vE_{(1)}^r$, 
which spans the plane orthogonal to $\boldsymbol{\hat{\vN}}$.

To compute the observer's-frame modes $h_{\ell m}^I$ during the
inspiral-plunge stage, it is convenient to introduce a non-inertial reference
frame that tracks the motion of the orbital plane, the so-called 
\textit{co-precessing frame} (superscript $P$), described 
by the triad $\{\vE^P_{(i)}\}$ ($i=1,2,3$). At each instant, its $z$-axis
is aligned with $\vLhat$: $\vE_{(3)}^P \equiv \vLhat(t)$~\footnote{Note that in 
Ref.~\cite{Babak:2016tgq}, the $z$-axis is aligned with $\vLhat_{\rm N}$ instead of $\vLhat$.}. In this frame, 
the BBH is viewed face-on at all times, and the GW radiation looks very much
nonprecessing~\cite{Buonanno:2002fy,Schmidt:2010it,Boyle:2011gg,O'Shaughnessy:2011fx,Schmidt:2012rh}. The
other two axes lie in the orbital plane and are defined such as they minimize
precessional effects in the precessing-frame modes $h_{\ell
  m}^P$~\cite{Buonanno:2002fy,Boyle:2011gg}. After introducing the
vector $\boldsymbol{\Omega}_e \equiv \vLhat\times {\textrm
  d}\vLhat/{\textrm d}t$, we enforce the minimum-rotation condition 
by requiring that ${\textrm d}\vE_{(1),(2)}^P/{\textrm d}t =
\boldsymbol{\Omega}_e\times \vE_{(1),(2)}^P$ and $\vE_{(1),(2)}^P(0) =
\vE^I_{(1),(2)}$ (see also Fig.~\ref{F:Pframe}). As usual, we parametrize the
rotation from the precessing to the observer's frame through 
time-dependent Euler angles $(\alpha(t),\beta(t),\gamma(t))$, which 
we compute using Eqs.~(A4)--(A6) in Appendix A of Ref.~\cite{Babak:2016tgq}. We notice that 
the minimum-rotation condition can also be expressed through the following differential 
equation for $\gamma$: $\dot{\gamma} = -\dot{\alpha}\cos{\beta}$ with $\gamma(0)=-\alpha(0) =
\pi/2$.

We compute the precessing-frame inspiral-plunge modes just like we do for the GW flux, 
namely by evaluating the factorized, resummed nonprecessing multipolar waveforms 
along the EOB precessing dynamics, and employing the time-dependent spin projections
$\boldsymbol{\chi}_{1,2}(t)\cdot \vLhat(t)$. Finally, the observer's-frame inspiral-plunge 
modes are obtained by rotating the precessing-frame inspiral-plunge modes using 
Eq.~(A13) in Appendix A of Ref.~\cite{Babak:2016tgq}. 

Following Ref.~\cite{Cotesta:2018fcv}, where an EOBNR  
nonprecessing multipolar  waveform  model was developed (\verb+SEOBNRv4HM+), 
here we include in the precessing frame of the \verb+SEOBNRv4PHM+ model the $(2,\pm 2), (2,\pm 1), (3,\pm 3), (4,\pm 4)$ 
and $(5,\pm 5)$ modes, 
and make the assumption $h^P_{l-m} = (-1)^{l}h^{P*}_{lm}$.  As shown in Sec.~\ref{sec:mode_asymm}, we expect that inaccuracies 
due to neglecting mode asymmetries should remain mild, or at most at the level of other modeling errors.

\subsection{Merger-ringdown waveforms}

The description of a BBH as a system composed of two individual
objects is of course valid only up to the merger. After that point,
the EOB model builds the GW emission (ringdown stage) via a 
phenomenological model of the quasinormal modes (QNMs) of the remnant BH, 
which forms after the coalescence of the progenitors. 
The QNM frequencies and decay times are known (tabulated) functions of the mass
$M_f$ and spin $\vS_f \equiv M_f^2 \boldsymbol{\chi}_f$ of the remnant
BH~\cite{Berti:2005ys}. Since the QNMs are defined with respect to the direction 
of the final spin, the specific form of the ringdown signal, as a linear
combination of QNMs, is formally valid only in an inertial frame whose
$z$-axis is parallel to $\boldsymbol{\chi}_f$. 

A novel feature of the \verb+SEOBNRv4PHM+ waveform model presented here is that we attach 
the merger-ringdown waveform (notably each multipole mode $h_{\ell m}^{\rm mergr-RD}$) 
directly in the co-precessing frame, instead of the observer's frame. As a consequence, 
we can employ here the merger-ringdown multipolar model developed for non-precessing BBHs 
(\verb+SEOBNRv4HM+) in Ref.~\cite{Cotesta:2018fcv} (see Sec.~IVE therein for details). 
By contrast, in the \verb+SEOBNRv3P+ waveform model~\cite{Babak:2016tgq}, the 
merger-ringdown waveform was built as a superposition of 
QNMs in an inertial frame aligned with the direction of the remnant spin.  
This construction was both more complicated to implement and more prone to numerical
instabilities.

To compute the waveform in the observer's frame, our approach requires 
a description of the co-precessing frame Euler angles $(\alpha, \beta, \gamma)$ that
extends beyond the merger. To prescribe this, we take advantage of
insights from NR simulations~\cite{OShaughnessy:2012iol}.  In
particular, it was shown that the co-precessing frame continues to
precess roughly around the direction of the final spin with a
precession frequency approximately equal to the differences between
the lowest overtone of the (2,2) and (2,1) QNM frequencies, while the
opening angle of the precession cone decreases somewhat at merger. We
find that this behavior is qualitatively correct for the NR waveforms used for 
comparison in this paper.

To keep our model generic for a wide range of mass ratios and spins,
we need an extension of the behavior noticed in Ref.~\cite{OShaughnessy:2012iol} 
to the retrograde case, where the remnant spin is negatively aligned with the orbital 
angular momentum at merger. Such configurations can occur for high mass-ratio binaries, when
the total angular momentum $\vJ$ is dominated by the spin of the
primary $\vS_{1}$ instead of the orbital angular momentum
$\vL$. This regime is not well explored by NR simulations, 
and includes in particular systems presenting transitional
precession~\cite{Apostolatos:1994mx}. In our model we keep imposing simple precession around the
direction of the remnant spin at a rate $\omega_{\rm prec} \geq 0$,
but we distinguish two cases depending on the direction of the final
spin $\bm{\chi}_{f}$ (approximated by the total angular momentum
$\vJ = \vL+ \vS_{1}+ \vS_{2}$ at merger) relative to the final orbital
angular momentum $\vL_{f}$:
\begin{equation}\label{eq:postmergerprec}
  \dot{\alpha} = \omega_{\rm prec}= \left\{\begin{aligned}
   &\omega^{\rm QNM}_{22}(\chi_{f})-\omega^{\rm QNM}_{21}(\chi_{f}) \quad \text{if} \quad \bm{\chi}_{f} \cdot \bm{L}_{f} > 0\\
   &\omega^{\rm QNM}_{2-1}(\chi_{f})-\omega^{\rm QNM}_{2-2}(\chi_{f}) \quad \text{if} \quad \bm{\chi}_{f} \cdot \bm{L}_{f} < 0
\end{aligned}\right.
\end{equation}
where $\chi_{f} = |\bm{\chi}_{f}|$, and the zero-overtone QNM frequencies for
negative $m$ are taken on the branch $\omega^{\rm QNM}_{lm} > 0$ that
continuously extends the $m>0$, $\omega^{\rm QNM}_{lm} > 0$
branch~\cite{Berti:2005ys} (the QNM refers to zero overtone). In both cases, $\dot{\alpha} \geq 0$. We
do not attempt to model the closing of the opening angle of the
precession cone and simply consider it to be constant during the post-merger phase,
$\beta = \mathrm{const}$. The third Euler angle $\gamma$ is then
constructed from the minimal rotation condition $\dot{\gamma} =
-\dot{\alpha} \cos\beta$. The integration constants are determined by
matching with the inspiral at merger. We find that the behavior
of  Eq.~\eqref{eq:postmergerprec} in the case $\bm{\chi}_{f} \cdot
\bm{L}_{f} < 0$ is qualitatively consistent with an NR simulation investigated 
by one of us~\cite{Ossokine:2020}. However, we stress that this
prescription for the retrograde case is much less tested than for the
prograde case.

\begin{figure}
    \includegraphics[width=\linewidth]{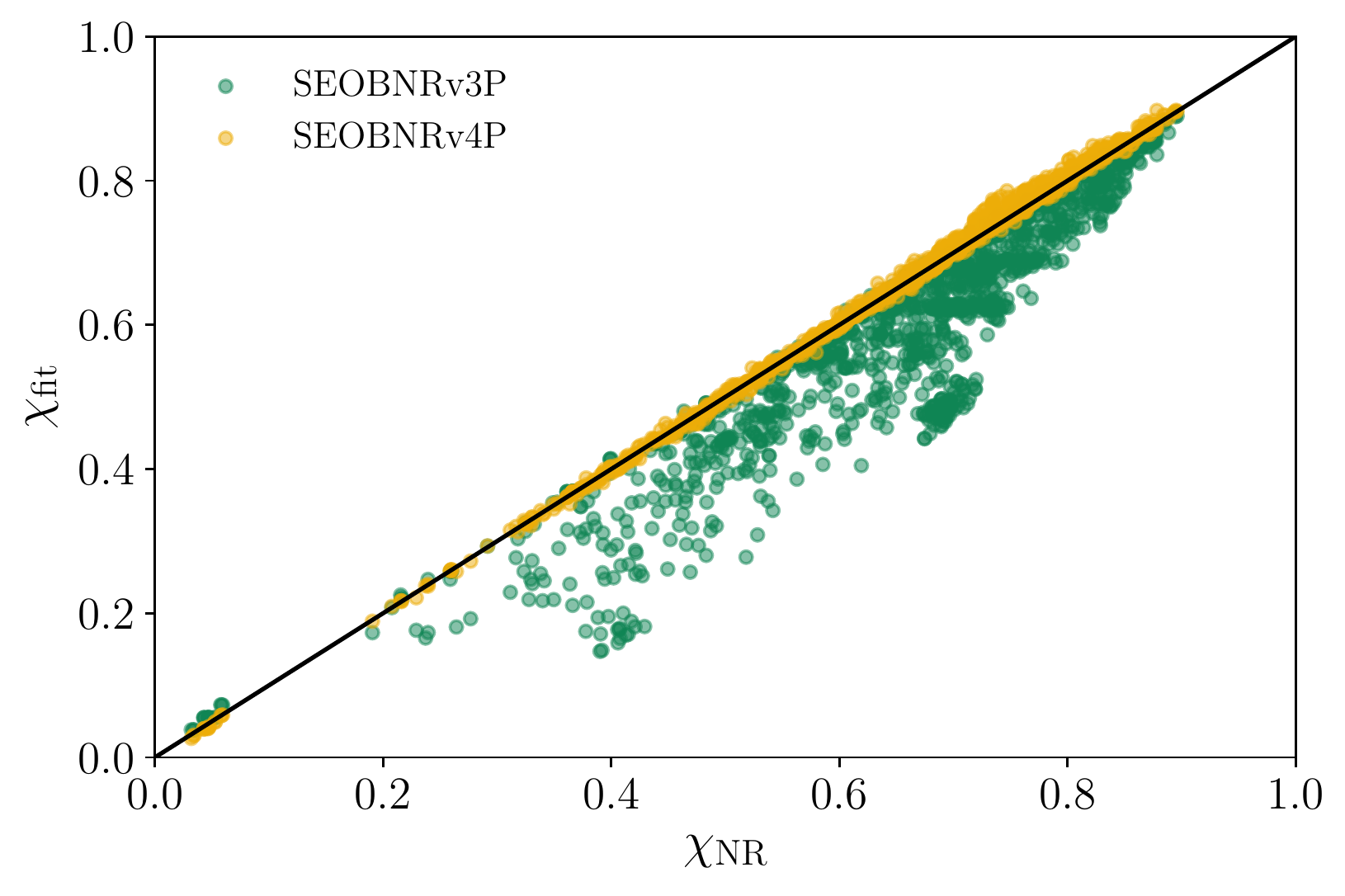}
    \caption{Comparison of the magnitude of the final spin  between the {\tt SEOBNRv3P} and
    {\tt SEOBNRv4P} models and NR results. For simplicity, the fits are evaluated 
using the NR data at the relaxed time. The black line is the identity. It is obvious that
    {\tt SEOBNRv4P} gives final-spin magnitudes much closer to the NR values.}
    \label{fig:final_spin}
\end{figure}

Furthermore, one crucial aspect of the above construction is the mapping from the
binary's component masses and spins to the final mass and spin, which is 
needed to compute the QNM frequencies of the merger remnant. Many groups have developed fitting formulae based
on a large number of NR simulations (e.g., see Ref.~\cite{Varma:2018aht} for an overview).
To improve the agreement of our EOB merger-ringdown model with NR, and to ensure agreement
in the aligned-spin limit with \verb+SEOBNRv4+~\cite{Bohe:2016gbl} and \verb+SEOBNRv4HM+~\cite{Cotesta:2018fcv},
we employ the fits from Hofmann et al.~\cite{Hofmann:2016yih}. In Fig.~\ref{fig:final_spin} we compare 
the performance of the fit used in the previous EOB precessing model \verb+SEOBNRv3P+~\cite{Pan:2013rra,Taracchini:2013rva,
Babak:2016tgq} to the fit from Hofmann et al. that we adopt for \verb+SEOBNRv4PHM+. It is clear that the new fit reproduces NR 
data much better. This in turn improves the correspondence between NR and EOB QNM frequencies.

For the final mass we employ the same fit as in previous EOB models, and we provide it here since it was not given 
explicitly anywhere before:

\bea
\frac{M_{f}}{M} &=& 1 - \left\{[1 - E_{\rm ISCO}(a)] \nu + 16\nu^{2} \left [ 0.00258 \right. \right. \nonumber \\
&& \left. \left. - \frac{0.0773 }{\left [a\,(1 +1/q)^{2}/(1 + 1/q^{2}) - 1.6939 \right ]} \right . \right .
\nonumber \\
&& \left. \left. - \frac{1}{4}(1 - E_{\rm ISCO}(a))\right ]\right\}\,,
\eea
where $a = \vLhat \cdot (\vchi_{1}+\vchi_{2}/q^{2})/(1+1/q)^{2}$, 
and $E_{\rm ISCO}(a)$ is the binding energy of the Kerr spacetime at the innermost stable circular orbit~\cite{Bardeen:1972fi}.

Finally, for precessing binaries, the individual components of the spins vary
with time. Therefore, in applying the fitting formulae to obtain final mass and spin, one must make a crucial choice in
selecting the time during the inspiral stage at which the spin directions are evaluated. In fact, even if one considers 
a given physical configuration, evaluating the final spin formulae with spin directions from
different times yields different final spins and consequently different
waveforms. We choose to evaluate the spins at a time corresponding to the
separation of $r=10M$. This choice is guided by two considerations: by the
empirical finding of good agreement with NR (e.g., performing better than
using the time at which the inspiral-plunge waveform is attached to the merger-ringdown 
waveform~\cite{Cotesta:2018fcv}), and by the restriction that the waveform must
start at $r>10.5 M$ in order to have small initial eccentricity~\cite{Babak:2016tgq}. 
Thus, our choice ensures that a given physical configuration always
produces the same waveform regardless of the initial starting frequency.

\begin{figure*}
	\includegraphics[width=\linewidth]{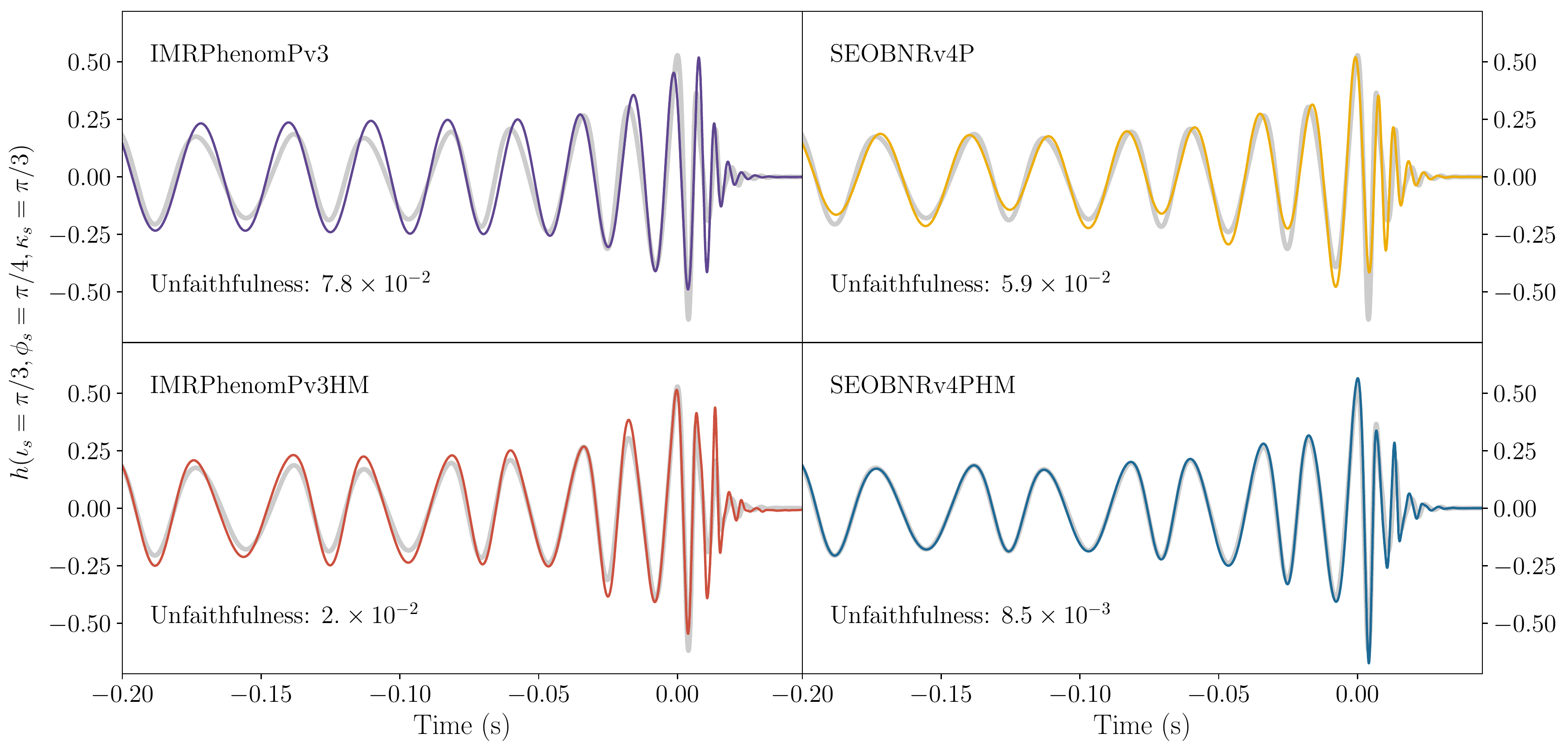}
	\caption{Time-domain comparison of state-of-the art waveform
          models to the NR waveform {\tt PrecBBH00078} with mass ratio
          $4$, BH's spins $0.7$ and total mass $M=70 M_\odot$. The
          source parameters are $\iota_{s}=\pi/3,\ \phi_{s}=\pi/4,\
          \kappa_{s}=\pi/4$. The NR waveform includes all modes up to
          and including $\ell=4$, and extends for 44 GW cycles before
          merger.  For models that include only $\ell=2$ modes, 
          the unfaithfulness are several
          percent 8\% for {\tt IMRPhenomPv3} and 6\% for {\tt
            SEOBNRv4P}.  Meanwhile, adding the higher mode content
          drastically improves the agreement, with mismatches going
          down to 2\% for {\tt IMRPhenomPv3HM} and 1\% for {\tt
            SEOBNRv4PHM}.  The agreement is particular good for {\tt
            SEOBNRv4PHM}, which reproduces the higher mode features at
          merger and ringdown faithfully.}
	\label{fig:time_domain_waveform}
\end{figure*}

To obtain the inspiral-merger-ringdown modes in the inertial frame, $h^{\rm I}_{\ell m}$, 
we rotate the inspiral-merger-ringdown modes $h^{\rm P}_{\ell m}$ from the co-precessing  frame to 
the observer's frame using the rotation formulas and Euler angles in Appendix A of Ref.~\cite{Babak:2016tgq}. 
The inertial frame polarizations then read
\begin{equation}
h^{\rm I}_{+}(\varphi_0,\iota;t) - i h^{\rm I}_{\times}(\varphi_0,\iota;t) = \sum_{\ell, m} {}_{-2} Y_{\ell m}(\varphi_0,\iota)\,h^{\rm I}_{\ell m}(t) \,.
\end{equation}

\subsection{On the fits of calibration parameters in presence of precession}

The \verb+SEOBNRv4PHM+ waveform model inherits the EOB Hamiltonian and GW energy flux from the aligned-spin model 
\verb+SEOBNRv4+~\cite{Bohe:2016gbl}, which features higher (yet unknown) PN-order terms in the dynamics 
calibrated to NR waveforms. These calibration parameters were denoted $K,d_{\rm SO}$ and $d_{\rm SS}$ in Ref.~\cite{Bohe:2016gbl}, 
and were fitted to NR and Teukolsky-equation--based waveforms
as polynomials in $\nu,\chi$ where $\chi \equiv {S_{\rm Kerr}^{z}}/({1-2\nu}$) with
${\vS}_{\rm Kerr}={\mathbf S}_{1}+{\mathbf S}_{2}$ the spin of the EOB background 
spacetime. In contrast to the \verb+SEOBNRv3P+ waveform model, which used the EOB Hamiltonian and GW energy flux 
from the aligned-spin model \verb+SEOBNRv2+\cite{Taracchini:2013rva}, the fits in 
Ref.~\cite{Bohe:2016gbl} include odd powers of $\chi$ and thus the sign of $\chi$ matters 
when the BHs precess. 

The most natural way to generalize these fits to the precessing case is to
project $\vS_{\rm Kerr}$ onto the orbital angular momentum $\hat{\vL}$ 
 in the usual spirit of reducing precessing 
quantities to corresponding aligned-spin ones. To test the impact of this prescription, we compute 
the sky-and-polarization-averaged unfaithfulness with the set of 118 NR simulations
described in Sec.~\ref{sec:NR}, and find that while the majority of the cases have low 
unfaithfulness ($\sim 1$\%), there are a handful of cases where it is significant($\sim10$\%), 
with many of them having large in-plane spins. 

To eliminate the high mismatches, we introduce the \emph{augmented spin} that includes contribution of the in-plane spins: 
\begin{equation}
  \tilde{\chi}=\frac{{\vS}_{\rm Kerr}\cdot {\vL}}{1-2\nu}+\alpha\frac{({\vS}_{1}^{\perp}+{\vS}_{2}^{\perp})\cdot 
{\vS}_{\rm Kerr}}{|{\vS}_{\rm Kerr}|(1-2\nu)}\,.
\label{eq:defn}
\end{equation}
Here ${\vS}_{i}^{\perp}\equiv {\vS}_{i}-({\vS}_{i}\cdot {\vL}) {\vL}$
and $\alpha$ is a \emph{positive} coefficient to be determined. Note that the
extra term in the definition of the augmented spin $\ge0$ for any combination
of the spins. We set $\tilde{\chi}=0$ when ${\vS}_{\rm Kerr}=0$. Fixing \(\alpha={1}/{2}\) 
insures that the augmented spin obeys the Kerr bound. Using the augmented spin eliminates all mismatches 
above $6\%$, and thus greatly improves the agreement of the model with NR data.

\section{Comparison of multipolar precessing models to numerical-relativity waveforms}
\label{sec:compEOBNR}

To assess the impact of the improvements incorporated in the \verb+SEOBNRv4PHM+ waveform model, we compare this model and other models publicly available 
in LAL (see Table~\ref{tbl:wf_models}) to the set of simulations described in Sec.~\ref{sec:NR}, as well as to all publicly available precessing {\tt SpEC} simulations~\footnote{The list of all
SXS simulations used can be found in \url{https://arxiv.org/src/1904.04831v2/anc/sxs_catalog.json}}.

We start by comparing in Fig.~\ref{fig:time_domain_waveform}, the precessing NR
waveform \verb+PrecBBH00078+ with mass ratio $4$, BH's spin magnitudes $0.7$, total mass $M=70 M_\odot$ 
and modes $\ell \leq 4$ from the new 118 SXS catalog (see Appendix~\ref{sec:NRparam}) to the precessing waveforms \verb+IMRPhenomPv3+ and \verb+SEOBNRv4P+ with modes $\ell = 2$ (upper panels), and to 
the precessing multipolar waveforms \verb+IMRPhenomPv3HM+ and \verb+SEOBNRv4PHM+
(lower panels). This NR waveform is the most ``extreme'' configuration from the
new set of waveforms and has about 44 GW cycles before merger, and the plot only shows 
the last $7$ cycles. More specifically, we plot the detector 
response function given in Eq.~(\ref{eq:det_strain_kappa}), but we leave out the overall constant amplitude. 
We indicate on the panels the unfaithfulness for the different cases. We note the improvement when including modes beyond the quadrupole. 
\verb+SEOBNRv4PHM+ agrees particularly well to this NR waveform, reproducing accurately the higher-mode features throughout merger and ringdown. 

\begin{figure}
  \includegraphics[width=\linewidth]{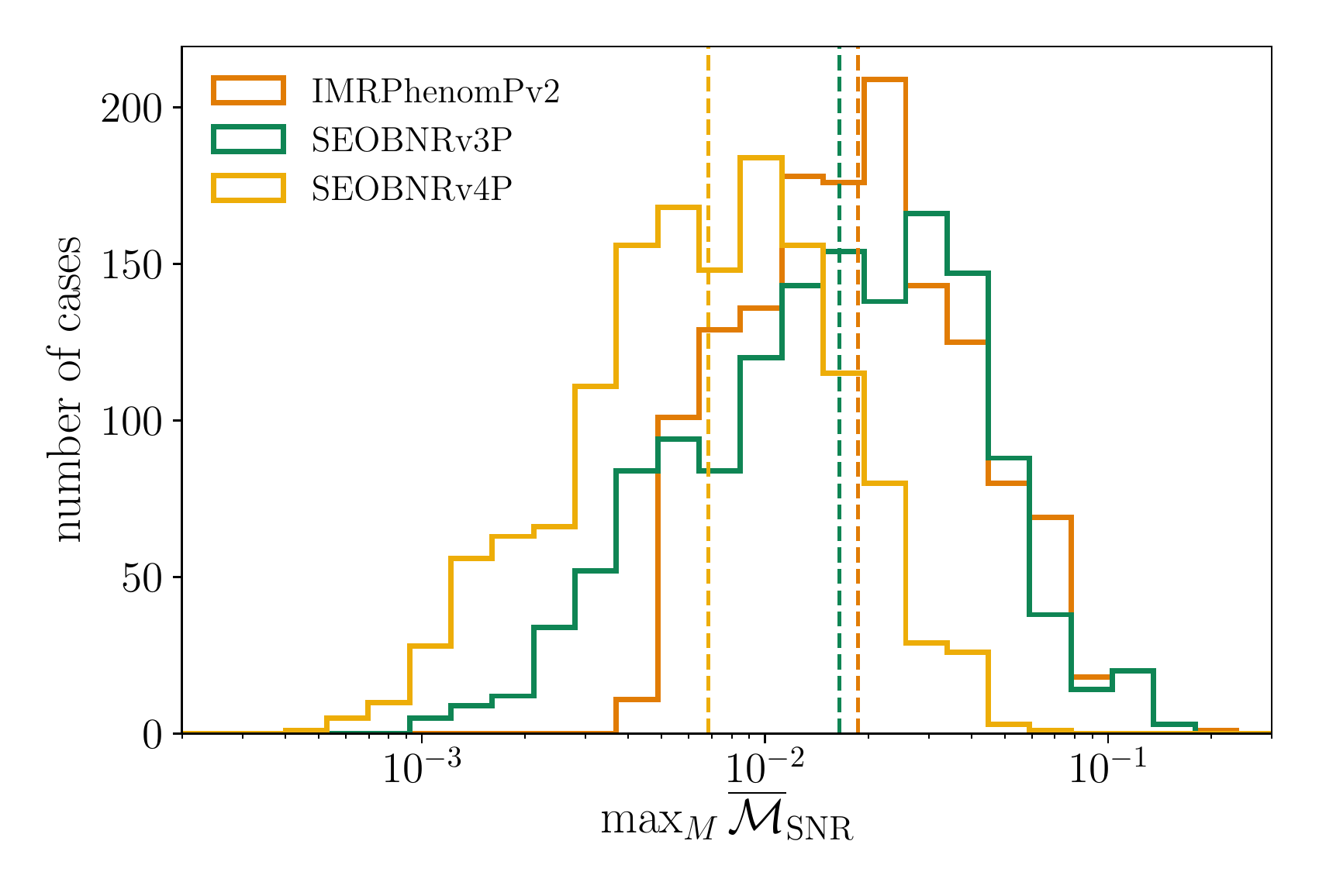}
  \caption{Sky-and-polarization averaged, SNR weighted unfaithfulness for
  an inclination $\iota=\pi/3$ between NR waveforms with $\ell =2$ and {\tt SEOBNRv4P}, and also 
{\tt SEOBNRv3P} and {\tt IMRPhenomPv2}, which were used in LIGO/Virgo publications.  The vertical dashed lines show the medians. 
It is evident the better performance of the newly developed precessing model {\tt SEOBNRv4P}.}
  \label{fig:all_approximants_ell2}
\end{figure}

\begin{figure}
  \includegraphics[width=\linewidth]{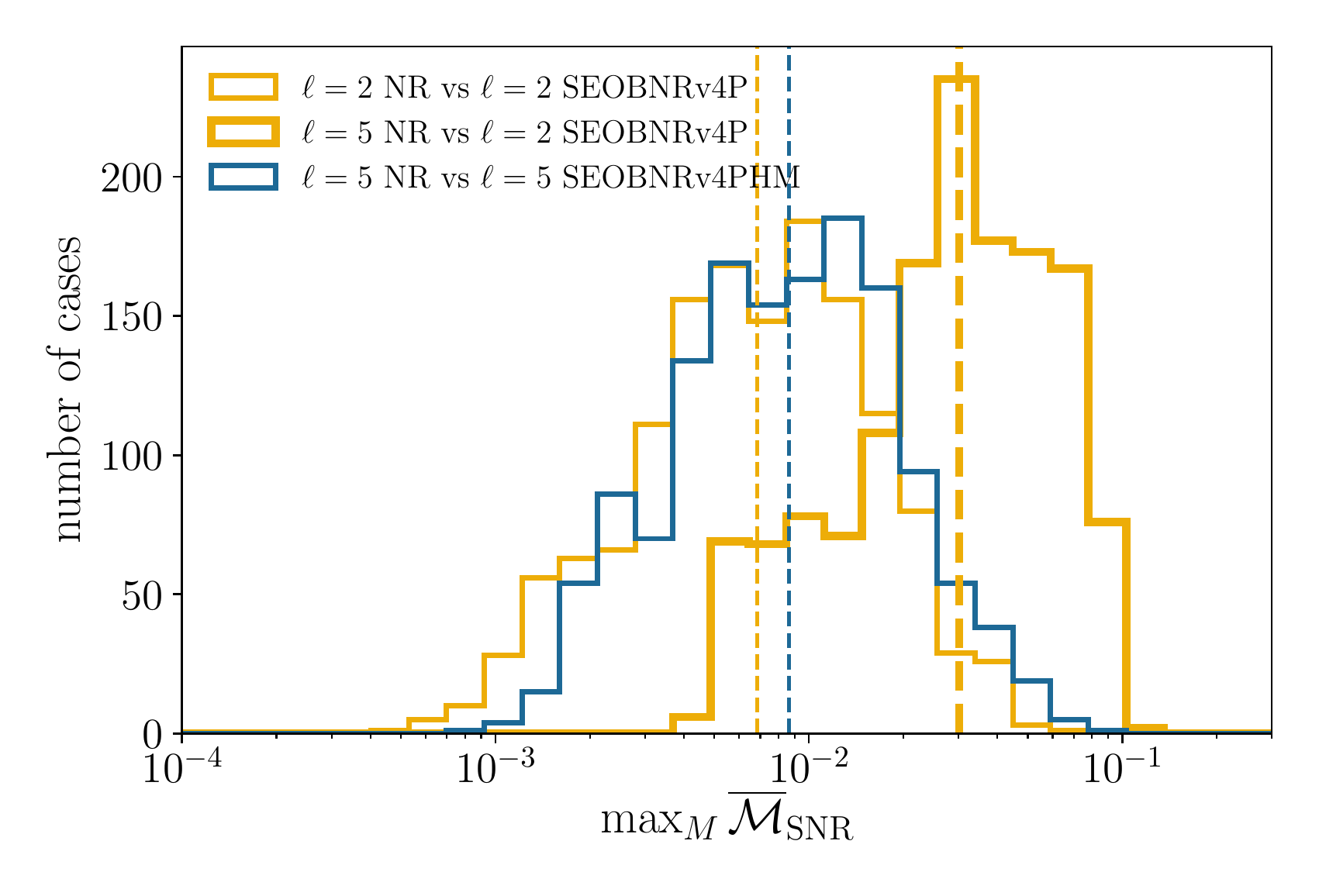}
  \caption{Sky-and-polarization averaged, SNR weighted unfaithfulness for
  an inclination $\iota=\pi/3$ between NR waveforms and {\tt SEOBNRv4PHM},
  including and omitting higher modes. The vertical dashed lines show the medians. Not including higher modes
  in the model results in high unfaithfulness. However, when they are 
included, the unfaithfulness between {\tt SEOBNRv4PHM} and NR is
  essentially at the same level as when only $\ell=2$ modes are compared (see Fig.~\ref{fig:all_approximants_ell2}).}
  \label{fig:higher_mode_effects}
\end{figure}

\begin{figure*}
	\includegraphics[width=\linewidth]{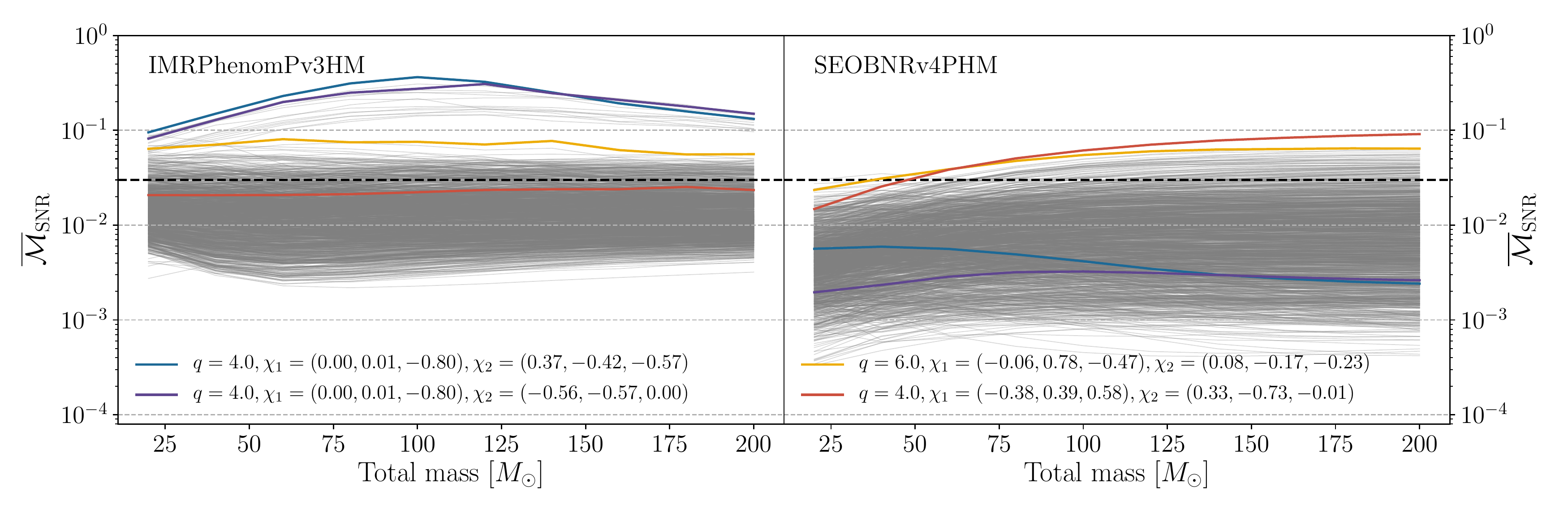}
	\caption{The sky-and-polarization averaged, SNR-weighted
          unfaithfulness as a function of binary's total mass for inclination
          $\iota=\pi/3$, between {\tt IMRPhenomPv3HM} and NR (left) and
          {\tt SEOBNRv4PHM} and NR (right) for 1404 quasi-circular
          precessing BBH simulations in the SXS public catalog. The
          colored lines highlight the cases with the worst maximum
          mismatches for both models. Note that for the majority of
          cases, both models have unfaithfulness below 5\%, but
          {\tt SEOBNRv4PHM} has no outliers beyond 10\% and many more cases
          at lower unfaithfulness.}
	 \label{fig:spaghetti_public}
\end{figure*}

We now turn to the public precessing SXS NR catalog of 1404 waveforms. First, to quantify the performance of the new precessing waveform model \verb+SEOBNRv4P+ 
with respect to previous precessing models used in LIGO and Virgo inference studies, we compute the unfaithfulness~\footnote{We always use the sky-and-polarization averaged, SNR-weighted faithfulness or 
unfaithfulness $\overline{\mathcal{M}}_{\rm SNR}$ unless otherwise stated.} against the precessing NR catalog, including only the dominant
$\ell=2$ multipoles in the co-precessing frame. Figure~\ref{fig:all_approximants_ell2} shows the histograms of the largest
mismatches when the binary total mass varies in the range $[20,200]M_\odot$. Here, we also consider the
precessing waveform models used in the first GW Transient Catalog~\cite{LIGOScientific:2018mvr} of the 
LIGO and Virgo collaboration (i.e., \verb+SEOBNRv3P+ and \verb+IMRPhenomPv2+).
Two trends are apparent: firstly, \verb+SEOBNRv3P+ and \verb+IMRPhenomPv2+ distributions are
broadly consistent, with both models having mismatches which extend beyond $10\%$ , although \verb+SEOBNRv3+ has more cases at lower unfaithfulness; secondly, \verb+SEOBNRv4P+ 
has  a distribution which is shifted to much lower values of the unfaithfulness and does not include outliers with the largest
unfaithfulness below $7\%$. 

Next, we examine the importance of higher modes. To do so, we use \verb+SEOBNRv4PHM+ 
with and without the higher modes, while always including all modes up to
$\ell=5$ in the NR waveforms. As can be seen in Fig.~\ref{fig:higher_mode_effects},
if higher modes are omitted, the unfaithfulness can be very large, with a
significant number of cases having unfaithfulness $>7\%$, as has been seen in many past studies. 
On the other hand, once higher modes are included in the model, the distribution of mismatches becomes
much narrower, with all mismatches below $9\%$. Furthermore, the distribution now
closely resembles the distribution of mismatches when only $\ell=2$ modes were
included in the  NR waveforms. Thus, we see that higher modes play an important role
and are accurately captured by \verb+SEOBNRv4PHM+ waveform model.

Moreover, in Fig.~\ref{fig:spaghetti_public} we display, for a specific choice of the inclination, the 
unfaithfulness versus the binary's total mass between the public precessing SXS NR catalog and 
\verb+SEOBNRv4PHM+ and \verb+IMRPhenomPv3HM+. We highlight with curves in color the NR configurations having worst maximum  
mismatches for the two classes of approximants. For the majority of cases, both models have unfaithfulness below 5\%, but 
\verb+SEOBNRv4PHM+ has no outliers beyond 10\% and many more cases at lower unfaithfulness ($<2\times 10^{-3}$). We find that
the large values of unfaithfulness above $10\%$ for \verb+IMRPhenomPv3HM+ come from simulations with $q\gtrsim4$ and large anti-aligned primary spin,
i.e. $\chi_{1}^{z}=-0.8$.  An examination of the waveforms  in this region reveals that unphysical features develop
in the waveforms, with unusual oscillations both in amplitude and phase. For lower spin magnitudes these features are milder, and disappear
for spin magnitudes $\lesssim0.65$. These features are present also in \verb+IMRPhenomPv3+ and are thus connected to the
precession dynamics, a region already known to potentially pose a challenge when modeling the 
precession dynamics as suggested in Ref.~\cite{Chatziioannou:2017tdw}, and adopted in Ref.~\cite{Khan:2019kot}.

\begin{figure*}
	\includegraphics[width=\linewidth]{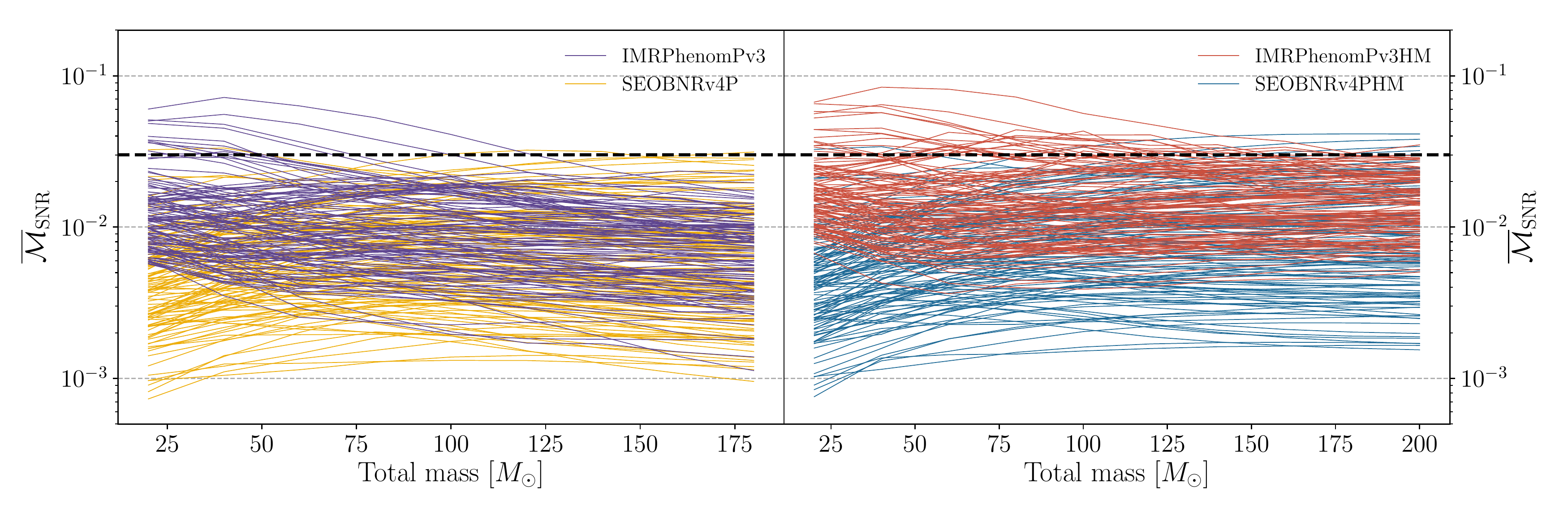}
	\caption{The sky-and-polarization averaged, SNR-weighted
		unfaithfulness as a function of binary's total mass for inclination
		$\iota=\pi/3$, between {\tt IMRPhenomPv3} and {\tt SEOBNRv4P} and NR (left),  and
		{\tt IMRPhenomPv3HM} and {\tt SEOBNRv4PHM} and NR (right) for the 118 SXS NR waveforms described in Appendix~\ref{sec:NRparam}. 
		The NR data has $\ell=2$ modes for the left panel, while all modes up to and including $\ell=4$ in the right panel. The unfaithfulness is low
	using both waveform families, however, {\tt SEOBNRv4P(HM)} has fewer cases above $3\%$, and the distribution is consistently shifted to lower values 
of unfaithfulness.}
	\label{fig:spaghetti_PrecBBH}
\end{figure*}

We now focus on the comparisons with the 118 SXS NR waveforms produced
in this paper. In Fig.~\ref{fig:spaghetti_PrecBBH} we show the
unfaithfulness for \verb+IMRPhenomPv3(HM)+ and \verb+SEOBNRv4P(HM)+ in
the left (right) panels. We compare waveforms without higher modes, to
NR data that has only the $\ell=2$ modes, and the other models to NR
data with $\ell\leq4$ modes. The performance of both waveform models
on this new NR data set is largely comparable to what was found for
the public catalog. Both families perform well on average, with most
cases having unfaithfulness below 2\% for models without higher modes
and 3\% for models with higher modes. However, for some configurations 
{\tt IMRPhenomPv3(HM)} reaches unfaithfulness values above $3\%$ for 
total masses below $125 M_\odot$. Once again, the overall
distribution is shifted to lower unfaithfulness values for
\verb+SEOBNRv4P(HM)+.

When studying the distribution of unfaithfulness for these 118 cases across parameter space, it is useful to introduce the widely used effective $\chi_{\rm eff}$~\cite{Damour:2001tu,Racine:2008qv,Santamaria:2010yb} and precessing $\chi_{p}$~\cite{Schmidt:2014iyl} spins. These capture the leading order aligned-spin and precession effects respectively, and are defined as
\begin{subequations}
\begin{align}
\chi_{\rm eff} &= \frac{(m_1\vchi_{1}+m_2\vchi_{2})}{m_1+m_2}\cdot\vLhat_N\,, \\
\chi_{\rm p}  &= \frac{1}{B_{1}m_{1}^{2}}\max(B_{1}m_{1}^{2}\chi_{1\perp},B_{2}m_{2}^{2}\chi_{2\perp})\,,
\end{align}
\end{subequations}
where with $B_{1} = 2+3{m_{2}}/{m_{1}}$, $B_{2}=2+3{m_1}/{m_2}$ and we indicate with $\chi_{i\perp}$ the projection of the spins  
on the orbital plane. We find that the unfaithfulness shows 2 general trends. First, it tends to increase with increasing $\chi_{\rm eff}$ and $\chi_{\rm p}$. 
Secondly, that cases with positive $\chi_{\rm eff}$ (i.e. aligned with Newtonian 
orbital angular momentum) tend to have larger unfaithfulness. This is likely driven 
by the fact that inspiral is longer for such cases and the binary merges at higher frequency. We do not find
any other significant trends based on spin directions.  It is interesting to note that the distribution of mismatches from the 118 cases is quite similar
to the distribution from the much larger public catalog. This suggests that the 118 cases do indeed explore many different regimes of precession. 

\begin{figure}
  \includegraphics[width=\linewidth]{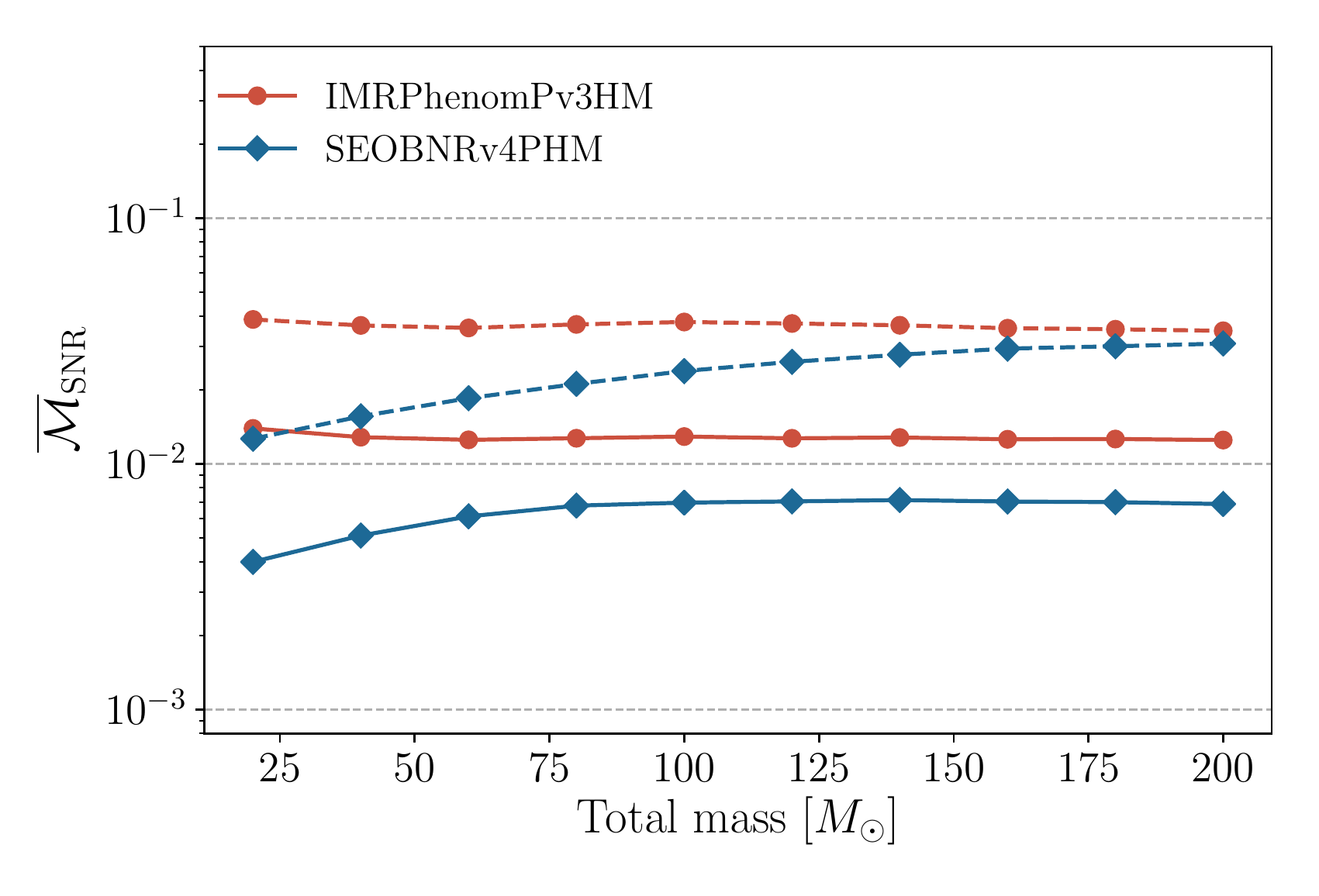}
  \caption{Summary of unfaithfulness as a function of the total
    mass, for all NR simulations considered as shown in Fig.~\ref{fig:spaghetti_public} and Fig~\ref{fig:spaghetti_PrecBBH}. The solid
    (dotted) line represents the median (95\%-percentile) of all 
    cases. For all total masses, we find that the median mismatch
    with {\tt SEOBNRv4PHM} is lower than 1\%, about a factor of 2
    lower than {\tt IMRPhenomPv3HM}. The 95th-percentile shows a
    stronger dependence on total mass for {\tt SEOBNRv4PHM}, with
    mismatches lower than {\tt IMRPhenomPv3HM} at low and medium total
    masses, becoming comparable at the highest total masses.}
\label{fig:all_runs_percentiles}
\end{figure}

\begin{figure}
	\includegraphics[width=\linewidth]{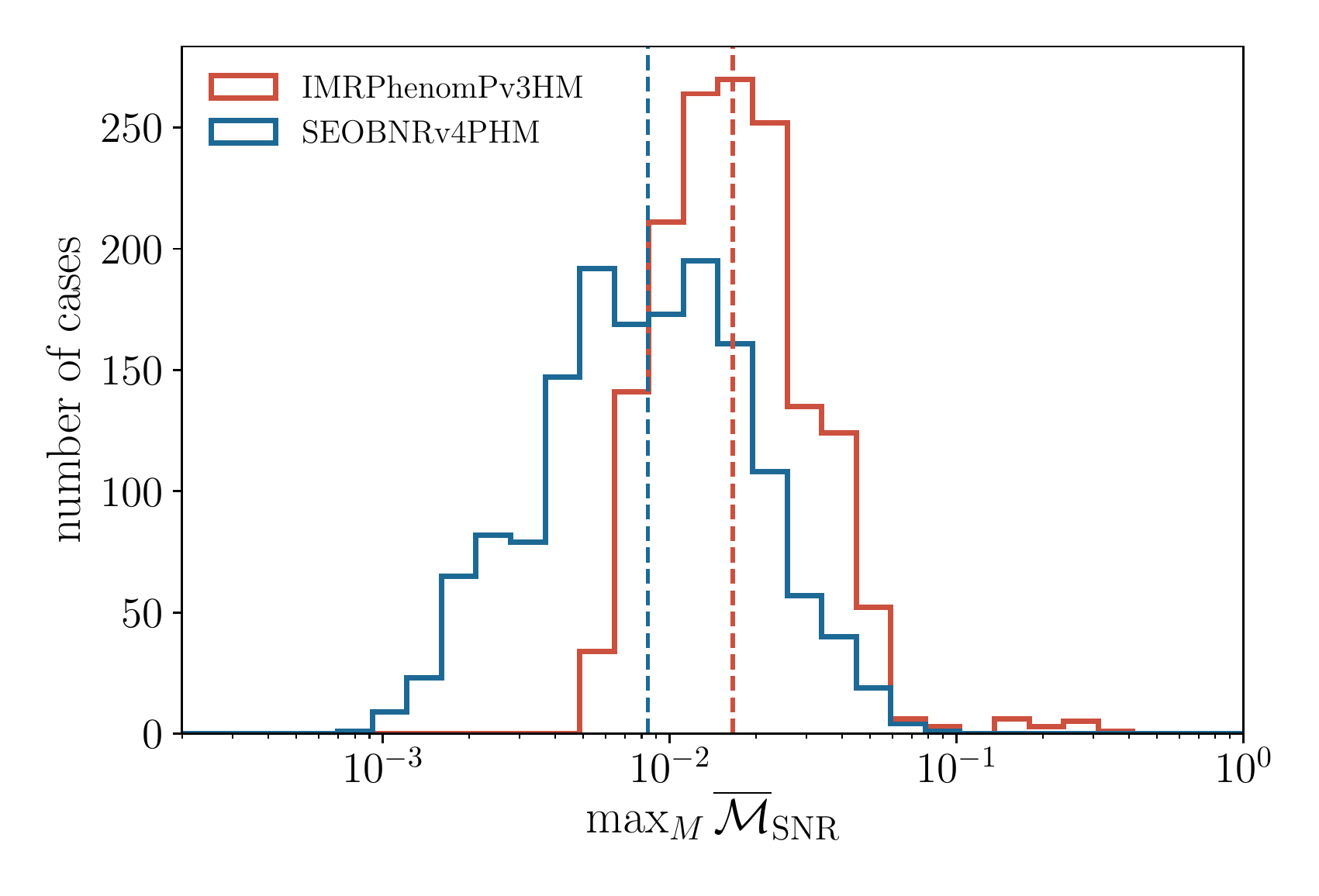}
	\caption{The highest unfaithfulness over total mass for all cases shown in Fig.~\ref{fig:all_runs_percentiles}. The median of unfaithfulness is 
		around 1\% for {\tt SEOBNRv4PHM} and 2\% for {\tt IMRPhenomPv3HM} (shown as dashed vertical lines). 
		Note that for {\tt SEOBNRv4PHM}, the worst unfaithfulness is below 10\% and the distribution is shifted to lower values.}
	\label{fig:all_runs_hist}
\end{figure}

To further quantify the results of the comparison between the precessing multipolar models \verb+SEOBNRv4PHM+ and \verb+IMRPhenomPv3HM+ and 
the NR waveforms, we show in Figs.~\ref{fig:all_runs_percentiles} and \ref{fig:all_runs_hist} the median and $95\%$-percentile of all cases,  
and the highest unfaithfulness as function of the total mass, respectively. These studies also demonstrate the better performance of 
\verb+SEOBNRv4PHM+ with respect to \verb+IMRPhenomPv3HM+. 

To summarize the performance against the entire SXS catalog (including the new 118 precessing waveforms) we find that 
for \verb+SEOBNRv4PHM+, out of a total of \aconfigs\ NR simulations we have considered, 864 cases (\EOBbone\ ) have a maximum unfaithfulness less than $1\%$,
and 1435 cases  (\EOBbthree\ ) have unfaithfulness less than $3\%$.  Meanwhile  for \verb+IMRPhenomPv3HM+ the numbers become 300 cases (\Phenombone\ ) below $1\%$, 1256 cases 
(\Phenombthree\ ) below $3\%$~\footnote{Due to technical details of the  {\tt IMRPhenomPv3HM} model, the total number of cases analyzed for this model is 1507 instead of 1523.}. 
The accuracy of the semi-analytical waveform models can be improved in the future by calibrating them to the precession sector of the SXS NR waveforms.

\begin{figure*}
	\includegraphics[width=\linewidth]{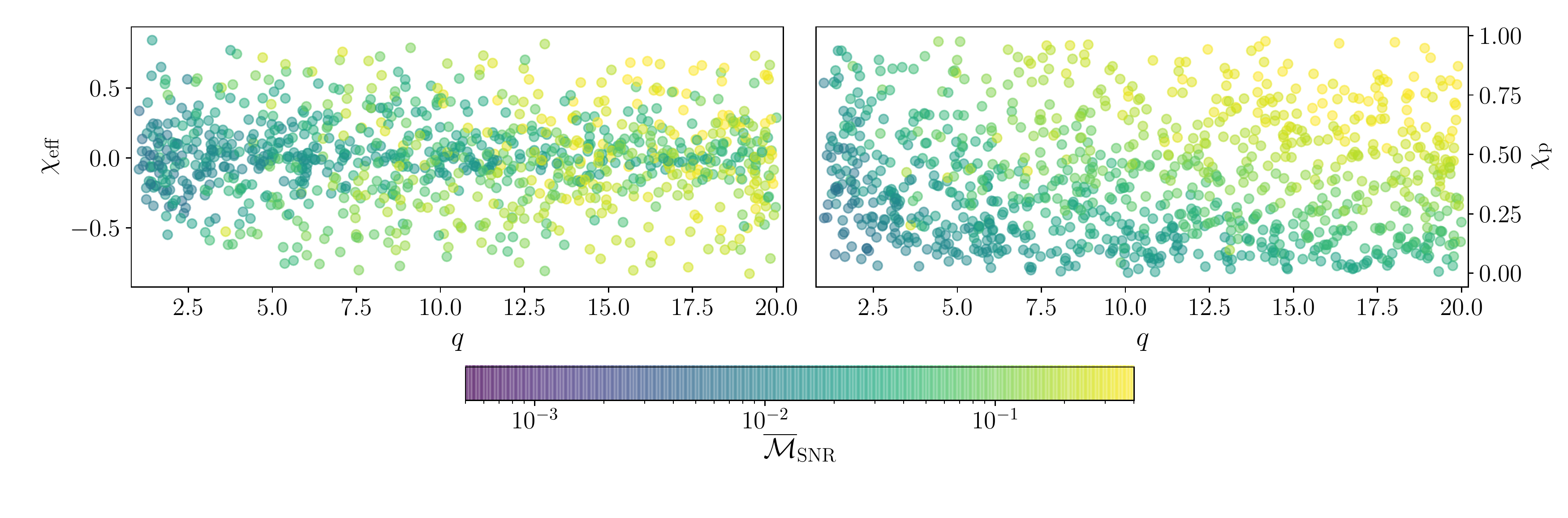}
	\caption{Sky-and polarization-averaged unfaithfulness between {\tt SEOBNRv4PHM} and {\tt IMRPhenomPv3HM} for 1000 random configurations.
	Notice that the unfaithfulness grows both with the mass ratio and the spin and can reach very large values for $q\approx20$ and high $\chi_{p}$. It's also
	clear that for cases with smaller spins the unfaithfulness remains much lower. }
	\label{fig:v4PHM_vs_Pv3HM}
\end{figure*}

An interesting question is to examine the behavior of the precessing models outside the region in which their underlying  aligned-spin waveforms were calibrated. To this effect we consider 1000 random cases
between mass ratios $q\in[1,20]$ and spin magnitudes $\chi_{1,2}\in[0,0.99]$ and compute $\overline{\mathcal{M}}_{\rm SNR}$ between
\verb+SEOBNRv4PHM+ and \verb+IMRPhenomPv3HM+. Figure~\ref{fig:v4PHM_vs_Pv3HM} shows the dependence of the unfaithfulness on the binary parameters,
in particular the mass ratio, and the effective and precessing spins. We find that for mass ratios $q< 8$, $50\%$ of cases have unfaithfulness below 2\% and 90\%
have unfaithfulness below 10\%. The unfaithfulness grows very fast with mass ratio and spin, with the highest unfaithfulness occurring at the highest mass ratio and 
precessing spin.  This effect is enhanced due to the fact that we choose to start all the waveforms at the same frequency and for higher mass ratios, the number of cycles in
band grows as $1/\nu$ where $\nu$ is the symmetric mass ratio. These results demonstrate the importance of producing long NR simulations for large mass ratios 
and spins, which can be used to validate waveform models in this more extreme region of the parameter space. To design more accurate semi-analytical models 
in this particular region, it will be relevant to incorporate in the models the information from gravitational self-force~\cite{Damour:2009sm,Bini:2018ylh,Antonelli:2019fmq}, 
and also test how the choice of the underlying EOB Hamiltonians with spin effects~\cite{Rettegno:2019tzh,Khalil:2020mmr} affects the accuracy.

Finally, in Appendix~\ref{sec:comparisonNRSurr} we quantify the agreement of the precessing multipolar waveform models 
\verb+SEOBNRv4PHM+ and \verb+IMRPhenomPv3HM+ against the NR surrogate model \verb+NRSur7dq4+~\cite{Varma:2019csw}, which 
was built for binaries with mass ratios $1\mbox{--}4$, BH's spins up to $0.8$ and binary's 
total masses larger than $\sim 60 M_\odot$. We find that the unfaithfulness between the semi-analytic models and the NR surrogate largely mirrors the results of the 
comparison in \cref{fig:all_runs_percentiles,fig:all_runs_hist}. Notably, as it can be seen in Fig.~\ref{fig:models_vs_NRsurr}, the unfaithfulness is generally below 3\% for both waveform 
families, but \verb+SEOBNRv4PHM+ outperforms \verb+IMRPhenomPv3HM+ with the former having a median at $3.3 \times 10^{-3}$, while the latter is at $1.5 \times 10^{-2}$.

\section{Bayesian analysis with multipolar precessing waveform models}
\label{sec:peEOBNR}

We now study how the accuracy of the waveform model \verb+SEOBNRv4PHM+ (and also \verb+IMRPhenomPv3HM+), which we have 
quantified in the previous section through the unfaithfulness, affects parameter inference when synthetic signal injections are performed. 
To this end, we employ two mock BBH signals and do not add any detector noise to them 
(i.e., we work in zero noise), which is equivalent to average over many different noise realizations. This choice avoids 
arbitrary biases introduced by a random-noise realization, and it is reasonable 
since the purpose of this analysis is to estimate possible biases in the 
binary's parameters due to inaccuracies in waveform models. 

We generate the first precessing-BBH mock signal with the
\verb+NRSur7dq4+ model. It has mass ratio $q = 3$ and a total
source-frame mass of $M = 70 M_\odot$. The spins of the two BHs are
defined at a frequency of $20$ Hz, and have components $\vchi_1 =
(0.30, 0.00, 0.50)$ and $\vchi_2 = (0.20,0.00,0.30)$. The masses and
spins'' magnitudes ($0.58$ and $0.36$) of this injection are compatible
with those of BBH systems observed so far with LIGO and Virgo
detectors~\cite{TheLIGOScientific:2016pea,LIGOScientific:2018mvr,Zackay:2019tzo,Venumadhav:2019lyq,Nitz:2019hdf}.
Although the binary's parameters are not extreme, we choose the
inclination with respect to the line of sight of the BBH to be
 $\iota = \pi/3$, to emphasize the effect of higher
modes. The coalescence and polarization phase, respectively $\phi$ and
$\psi$, are chosen to be 1.2 rad and 0.7 rad. The sky-position is
defined by its right ascension of 0.33 rad and its declination of -0.6
rad at a GPS-time of 1249852257 s. Finally, the distance to the source
is set by requesting a network-SNR of $50$ in the three detectors
(LIGO Hanford, LIGO Livingston and Virgo) when using the Advanced LIGO
and Advanced Virgo PSD at design sensitivity~\cite{Barsotti:2018}. The
resulting distance is $800$ Mpc. The unfaithfulness against this
injection is $0.2\%$ and $1\%$ for \verb+SEOBNRv4PHM+ and
\verb+IMRPhenomPv3HM+, respectively. Although the value of
the network-SNR is large for this synthetic signal, it is not
excluded that the Advanced LIGO and Virgo detectors at design
sensitivity could detect such loud BBH. With this study we
want to test how our waveform model performs on a system with moderate
precessional effect when detected with a large SNR value, 
considering that it has an unfaithfulness of $0.2\%$.

For the second precessing-BBH mock signal, we use a binary with larger mass ratio 
and spin magnitude for the primary BH. We employ the NR waveform \verb+SXS:BBH:0165+ from 
the public SXS catalog having mass ratio $q = 6$, and we choose the source-frame total mass 
$M = 76 M_\odot$. The BH's spins, defined at a frequency of $20$ Hz, have values 
$\vchi_1 = (-0.06, 0.78, -0.47)$ and $\vchi_2 =(0.08,-0.17,-0.23)$. The BBH system in this simulation has strong 
spin-precession effects.  We highlight that this NR waveform is one of the worst cases in term of unfaithfulness 
against \verb+SEOBNRv4PHM+, as it is clear from Fig.~\ref{fig:spaghetti_public}. 
For this injection we choose the binary's inclination to be 
edge-on at $20$ Hz to strongly emphasize higher modes. All 
the other binary parameters are the same of the previous injection, 
with the exception of the luminosity distance, which 
in this case is set to be $1.2$ Gpc to obtain a
network-SNR of $21$. The NR waveform used for
this mock signal has unfaithfulness of $4.4\%$ for \verb+SEOBNRv4PHM+ and $8.8\%$ for \verb+IMRPhenomPv3HM+, 
thus higher than in the first injection. 

For the parameter-estimation study we use the software \texttt{PyCBC}'s \texttt{pycbc\_generate\_hwinj}~\cite{alex_nitz_2020_3630601} to
prepare the mock signals, and we perform the Bayesian analysis with 
\texttt{parallel Bilby}~\cite{Smith:2019ucc}, a highly parallelized version of the
parameter-estimation software \texttt{Bilby}~\cite{Ashton:2018jfp}.  We choose a
uniform prior in component masses in the range $[5,150]
M_\odot$. Priors on the dimensionless spin magnitudes are uniform in
$[0,0.99]$, while for the spin directions we use prior isotropically
distributed on the unit sphere.  The priors on the other parameters
are the standard ones described in Appendix C.1 of 
Ref.~\cite{LIGOScientific:2018mvr}.

\begin{figure*}[hbt]
  \centering
  \includegraphics[width=0.45\textwidth]{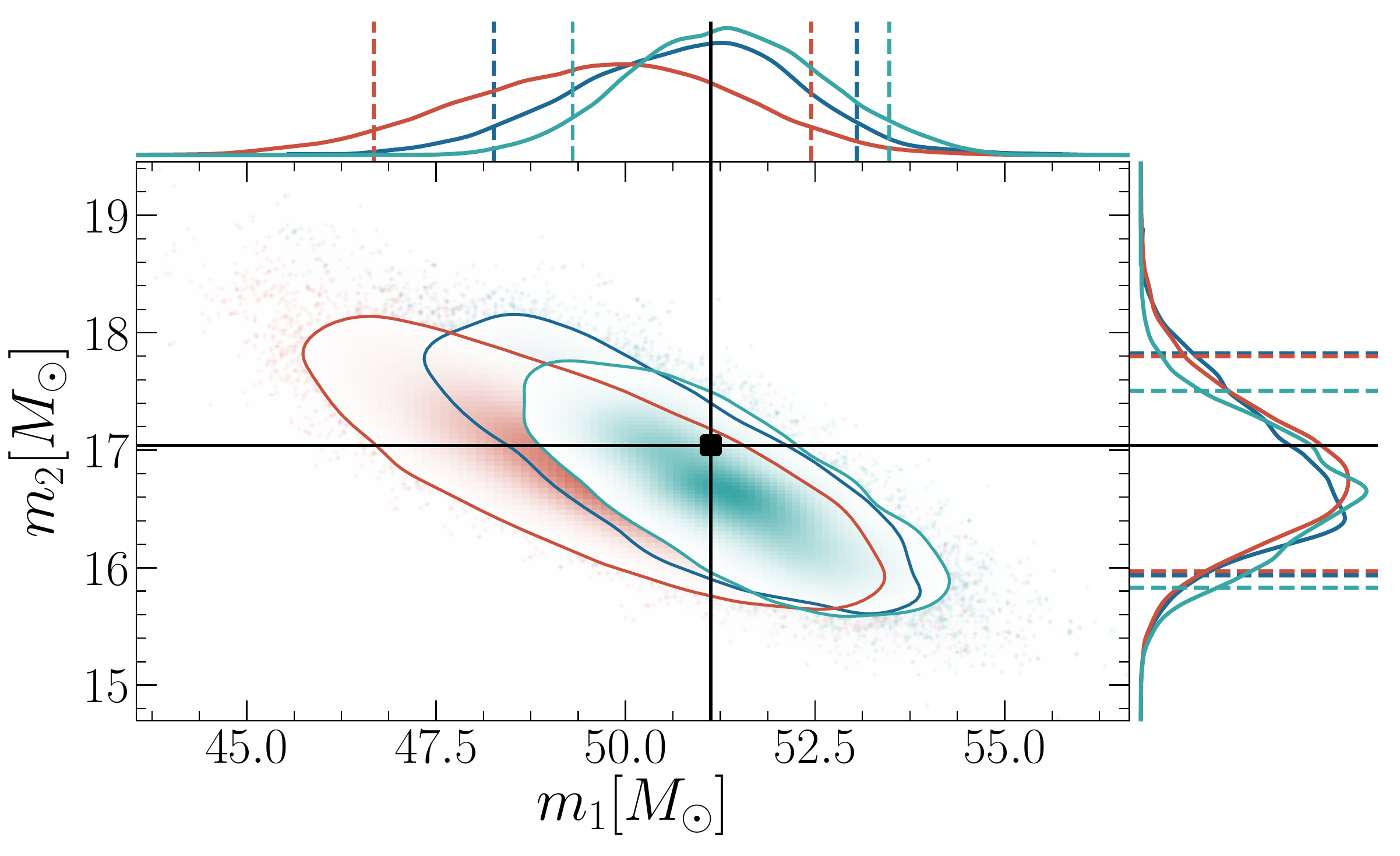}
  \includegraphics[width=0.45\textwidth]{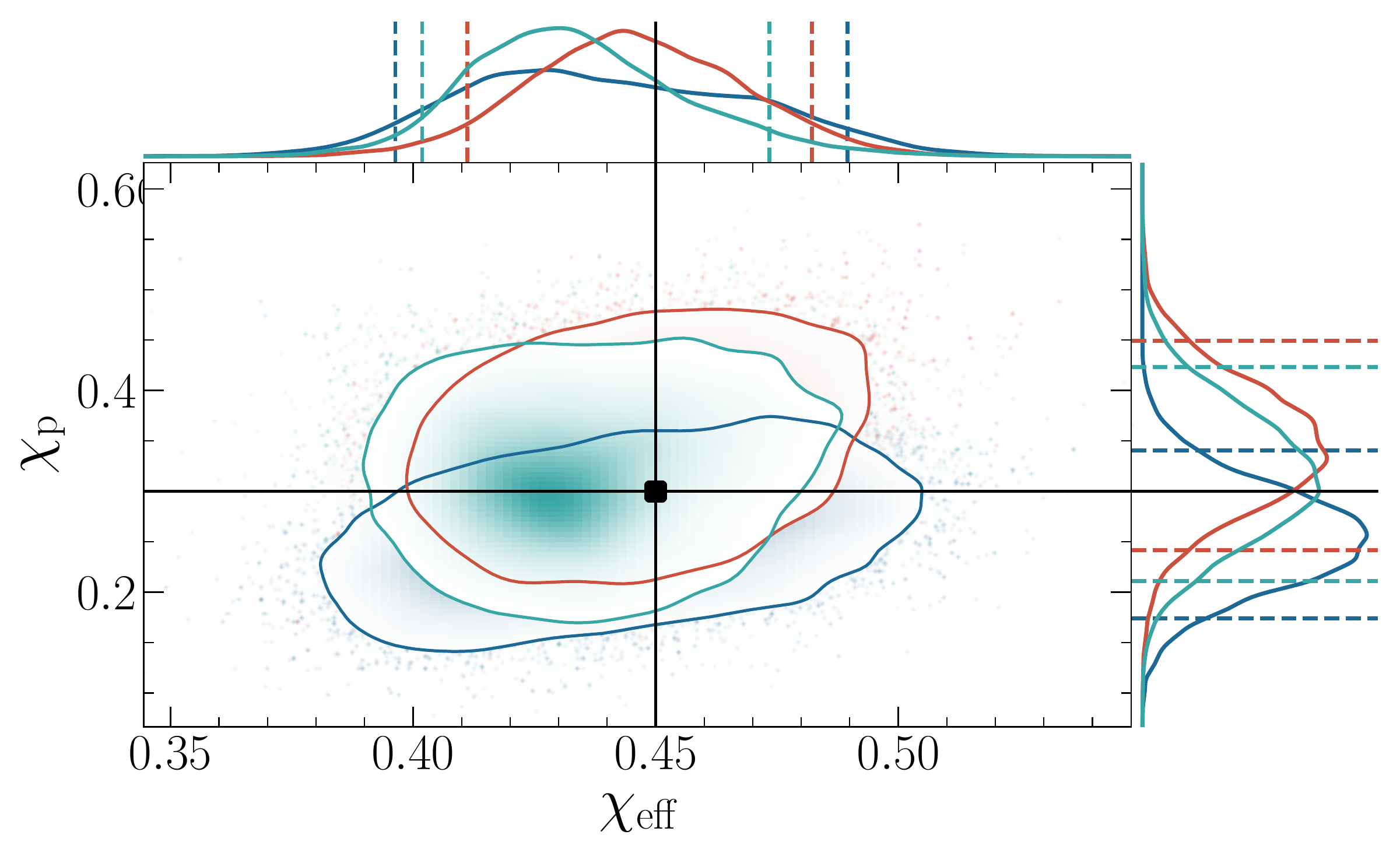}
  \includegraphics[width=0.45\textwidth]{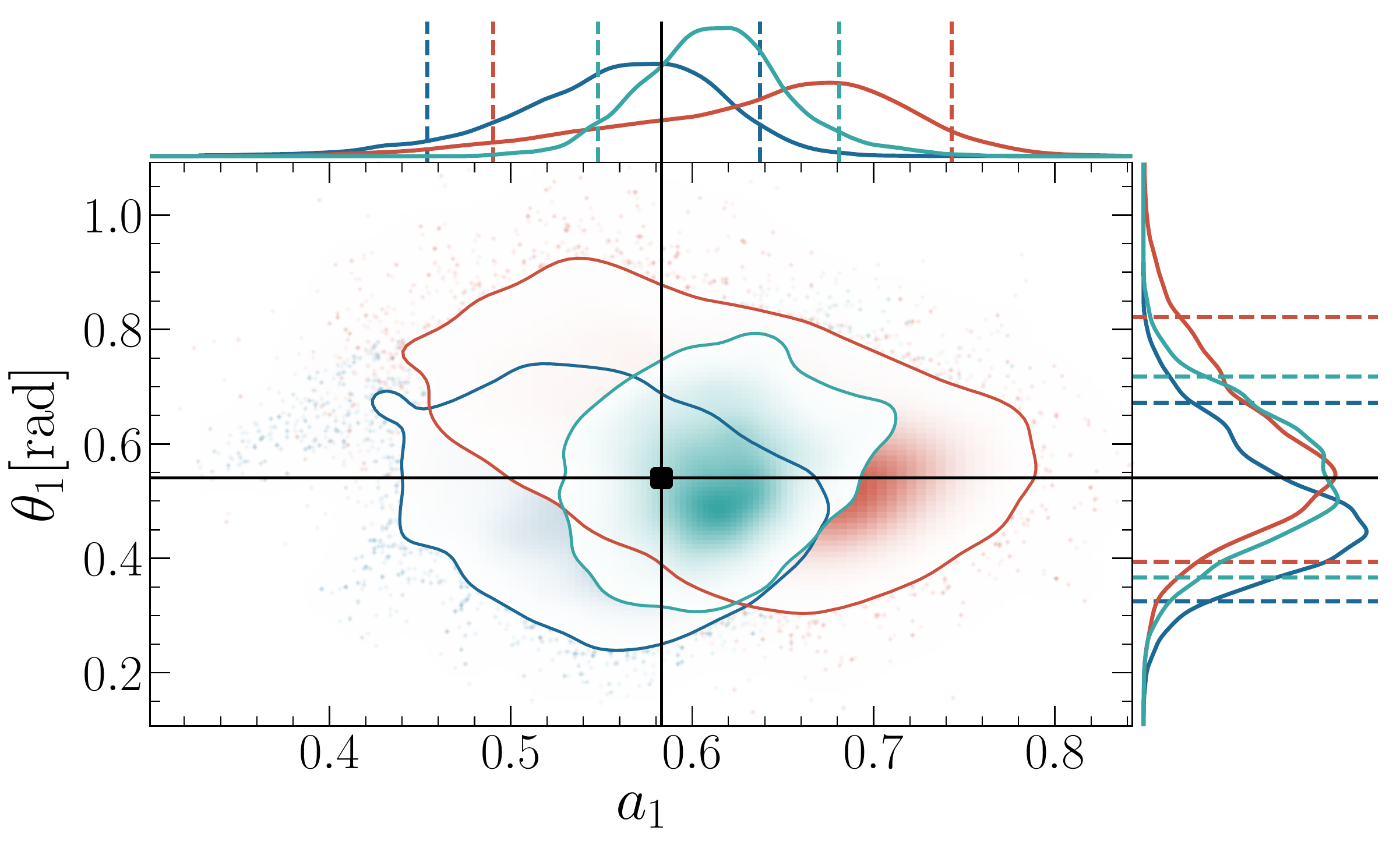}
  \includegraphics[width=0.45\textwidth]{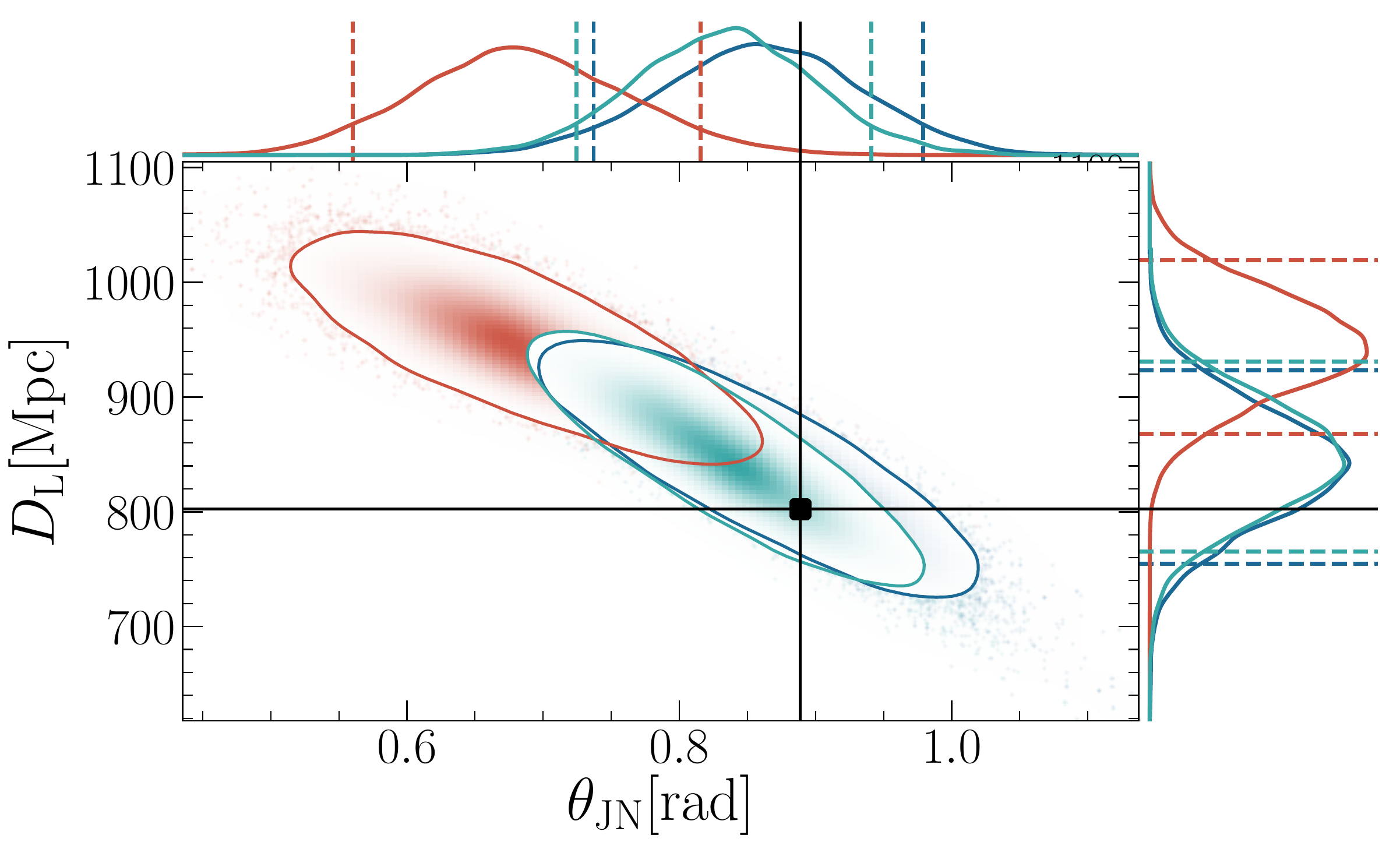}
  \includegraphics[clip, width=\textwidth]{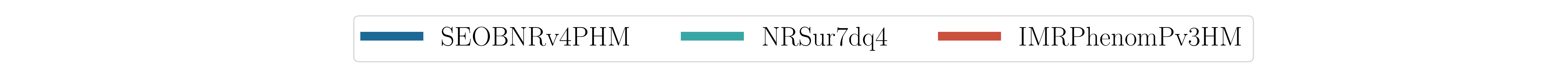}
  \caption{2D and 1D posterior distributions for some relevant
    parameters measured from the first synthetic BBH signal with mass ratio $q = 3$, total source-frame mass of $M = 70 M_\odot$,
spins of the two BHs $\vchi_1 = (0.30, 0.00, 0.50)$ and $\vchi_2 =
(0.20,0.00,0.30)$ defined at a frequency of $20$ Hz . The inclination with respect to the line of
sight of the BBH is $\iota = \pi/3$. The other parameters are specified in the text. The signal waveform is generated with the
    waveform model \texttt{NRSur7dq4}. In the 2D posteriors, solid
    contours represent $90\%$ credible intervals and black dots show
    the value of the parameter used in the synthetic signal. In the 1D
    posteriors they are represented respectively by dashed lines and
    black solid lines. The parameter estimation is performed with the
    waveform models \texttt{SEOBNRv4PHM} (blue), \texttt{NRSur7dq4}
    (cyan) and \texttt{IMRPhenomPv3HM} (red).  \emph{Top left:}
    component masses in the source frame, \emph{Top right:}
    $\chi_{\mathrm{eff}}$ and $\chi_{\mathrm{p}}$, \emph{Bottom left:}
    magnitude and tilt angle of the primary spin, \emph{Bottom right:}
    $\theta_{\mathrm{JN}}$ and luminosity distance.}
  \label{fig:PE_q_3}
\end{figure*}

We summarize in Fig.~\ref{fig:PE_q_3} the results of the parameter
estimation for the first mock signal for \verb+SEOBNRv4PHM+ (blue),
\verb+IMRPhenomPv3HM+ (red) and \verb+NRSur7dq4+ (cyan).  We report
the marginalized 2D and 1D posteriors for the component masses $m_1$
and $m_2$ in the source frame (top left), the effective spin
parameters $\chi_{\mathrm{eff}}$ and $\chi_\mathrm{p}$ (top right),
the spin magnitude of the more massive BH $a_1$ and its tilt angle
$\theta_1$ (bottom left) and finally the angle $\theta_{\mathrm{JN}}$
and the luminosity distance (bottom right). In the 2D posteriors, solid
contours represent $90\%$ credible intervals and black dots show the
value of the parameter used in the synthetic signal. In the 1D
posteriors, they are represented respectively by dashed lines and black
solid lines. As it is clear from Fig.~\ref{fig:PE_q_3}, when using the
waveform models \verb+SEOBNRv4PHM+ and \verb+NRSur7dq4+, all the
parameters of the synthetic signal are correctly measured within the
statistical uncertainty.  Moreover, the shape of the posterior
distributions obtained when using \verb+SEOBNRv4PHM+ are similar to
those recovered with \verb+NRSur7dq4+ (the model used to create the
synthetic signal). This means that the systematic error due to a non
perfect modeling of the waveforms is negligible in this case.

For the model \verb+IMRPhenomPv3HM+ while masses and spins
are correctly measured within the statistical uncertainty, the
luminosity distance $D_{\rm L}$ and the angle $\theta_{\rm JN}$ are
biased.  This is consistent with the prediction obtained using
Lindblom's criterion in Refs.~\cite{Flanagan:1997kp,Lindblom:2008cm,McWilliams:2010eq,Chatziioannou:2017tdw}~\footnote{The
  criterion says that a sufficient, but not necessary condition for
  two waveforms to become distinguishable is that the unfaithfulness $
  \geq (N_{\rm intr}-1)/(2{\rm SNR}^2)$, where $N_{\rm intr}$ is the
  number of binary's intrinsic parameters, which we take to be 8 for a
  precessing-BBH system. Note, however, that in practice this factor
  can be much larger, see discussion in
  Ref.~\cite{Purrer:2019jcp}.}. In fact, according to this criterion,
an unfaithfulness of $1\%$ for {\tt IMRPhenomPv3HM} would be
sufficient to produce biased results at a network-SNR of $19$. Thus,
it is expected to observe biases when using \verb+IMRPhenomPv3HM+ at
the network-SNR of the injection, which is $50$. In the case of
\verb+SEOBNRv4PHM+ the unfaithfulness against the signal waveform is
$0.2\%$ and according to Lindblom's criterion we should also expect
biases for network-SNRs larger than $42$, but in practice we do not
observe them. We remind that Lindblom's criterion is only approximate
and it has been shown in Ref.~\cite{Purrer:2019jcp} to be too
conservative, therefore the lack of bias that we observe  is not
surprising. 

\begin{figure*}[hbt]
  \centering
  \includegraphics[width=0.45\textwidth]{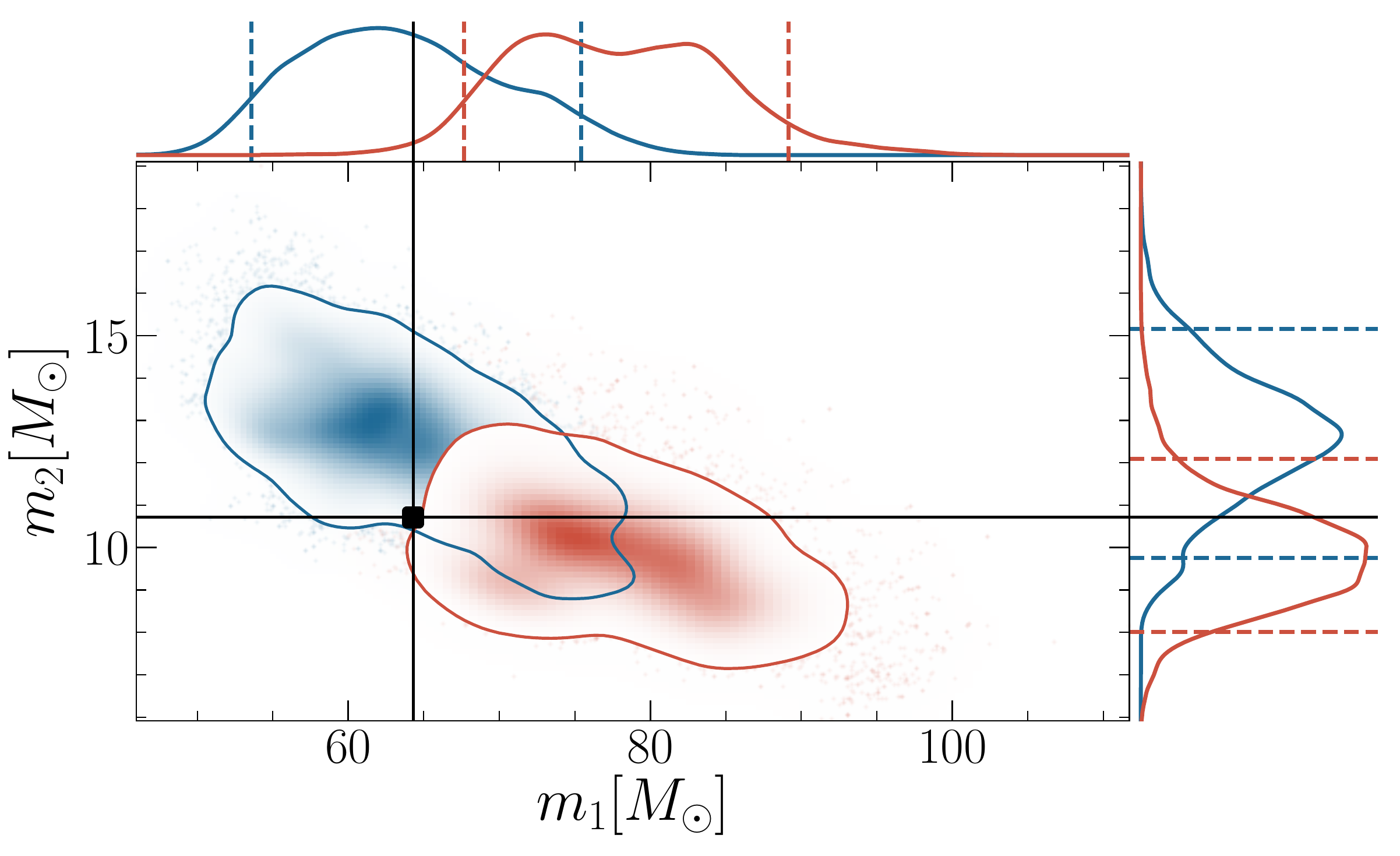}
  \includegraphics[width=0.45\textwidth]{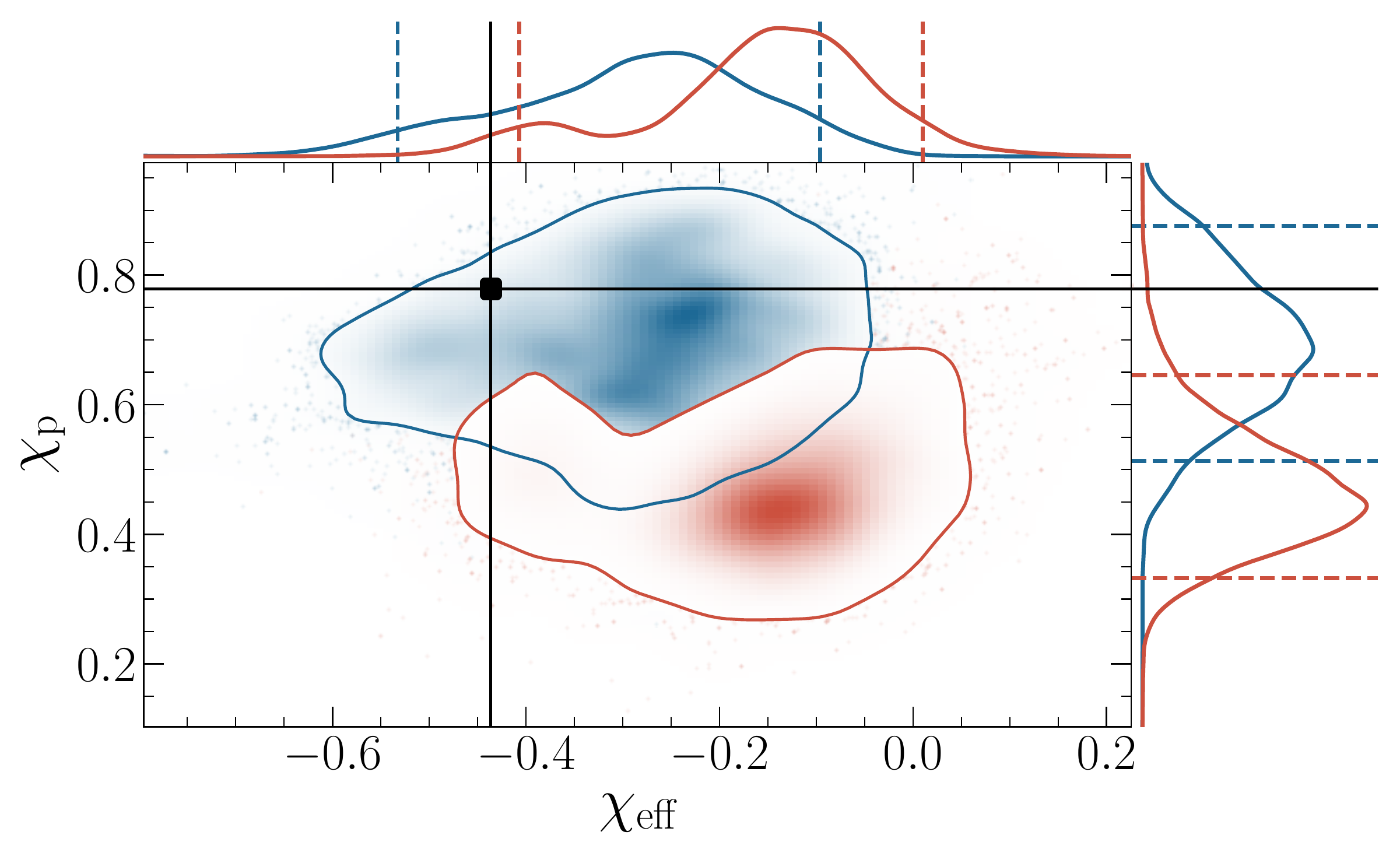}
  \includegraphics[width=0.45\textwidth]{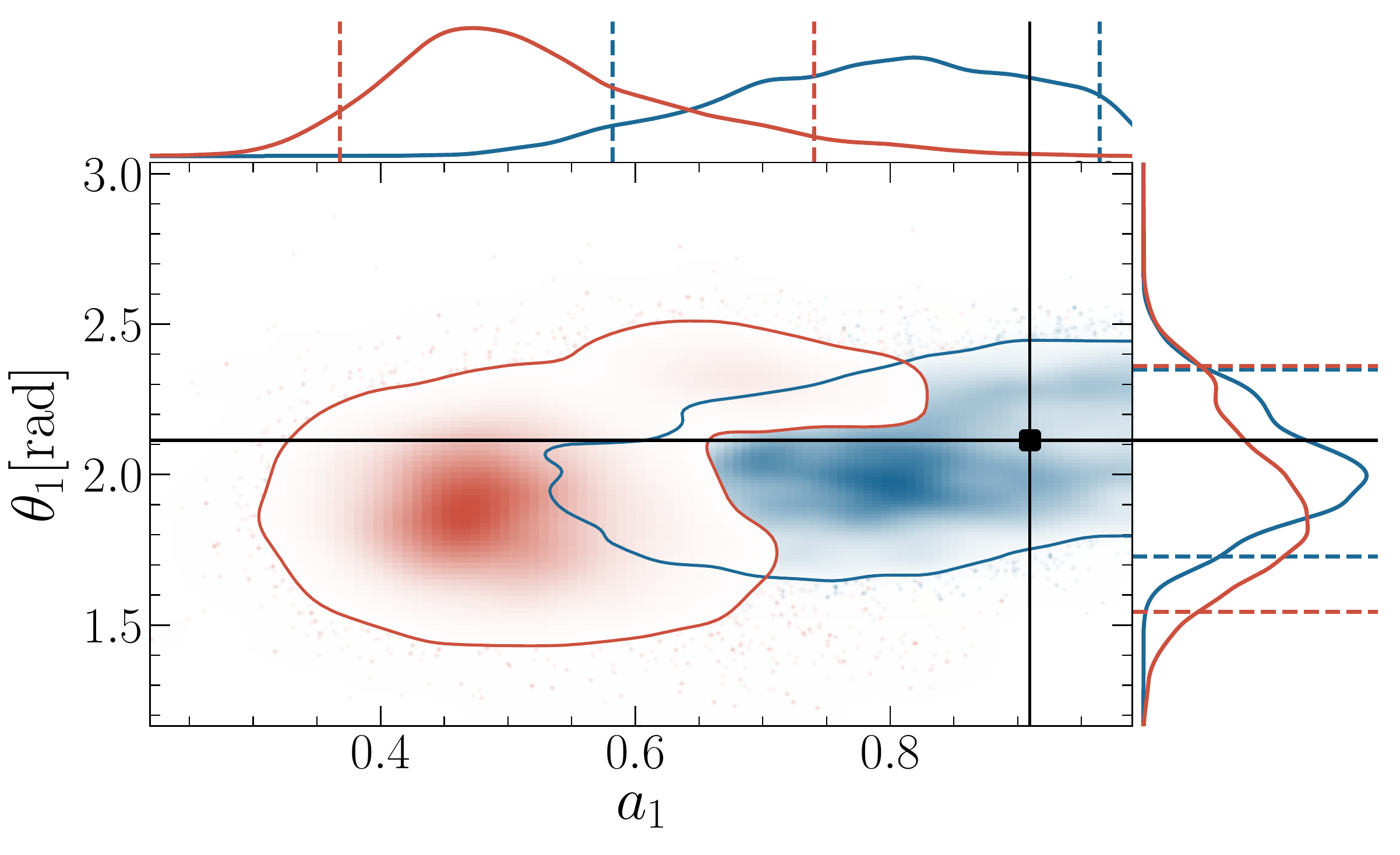}
  \includegraphics[width=0.45\textwidth]{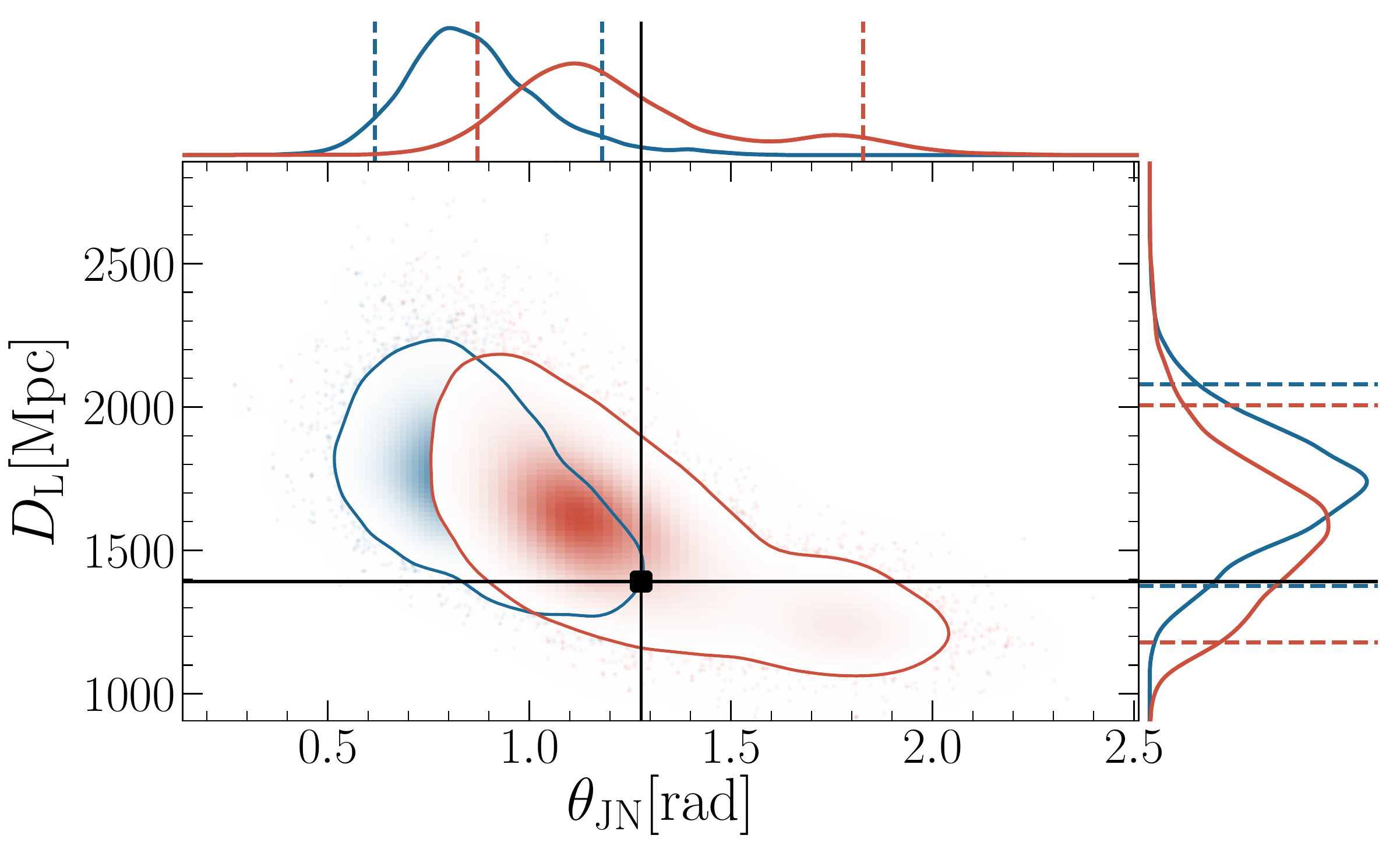}
  \includegraphics[ clip, width=0.5\textwidth]{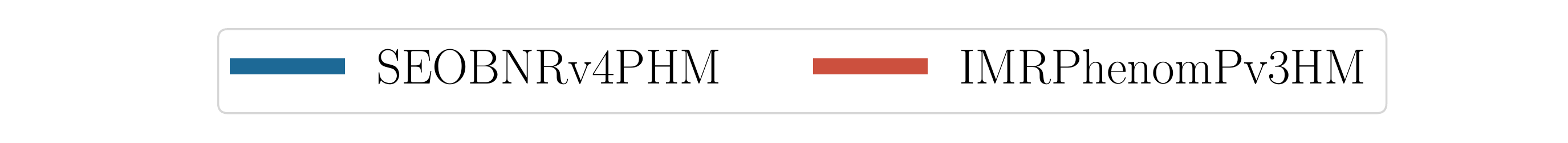}
  \caption{2D and 1D posterior distributions for some relevant
    parameters measured from the first synthetic BBH signal with mass ratio $q = 6$, total source-frame mass of $M = 76 M_\odot$,
spins of the two BHs $\vchi_1 = (-0.06, 0.78, -0.47)$ and $\vchi_2 =(0.08,-0.17,-0.23)$ defined at a frequency of $20$ Hz . The inclination with respect to the line of
sight of the BBH is edge-on, i.e., $\iota = \pi/2$. The other parameters are specified in the text. The signal waveform is generated using the NR waveform from the SXS public catalog \texttt{SXS:BBH:0165}. In the 2D posteriors solid
    contours represent $90\%$ credible intervals and black dots show
    the value of the parameter used in the synthetic signal. In the 1D
    posteriors they are represented respectively by dashed lines and
    black solid lines. The parameter estimation is performed with the
    waveform models \texttt{SEOBNRv4PHM} (blue) and \texttt{IMRPhenomPv3HM} (red).  \emph{Top left:}
    component masses in the source frame, \emph{Top right:}
    $\chi_{\mathrm{eff}}$ and $\chi_{\mathrm{p}}$, \emph{Bottom left:}
    magnitude and tilt angle of the primary spin, \emph{Bottom right:}
    $\theta_{\mathrm{JN}}$ and luminosity distance.}
  \label{fig:PE_q_6}
\end{figure*}

In Fig.~\ref{fig:PE_q_6} we summarize the results of 
the second mock-signal injection. The plots are the same as in Fig.~\ref{fig:PE_q_3} 
with the only exception that we do not have results for the \verb+NRSur7dq4+ 
model since it is not available in this region of the parameter space.  In
this case the unfaithfulness between \verb+SEOBNRv4PHM+
(\verb+IMRPhenomPv3HM+) and the NR waveform used for the mock signal
is $4.4\%$ ($8.8\%$). According to Lindblom's criterion, at the
network-SNR of this mock signal we should expect the bias due to
non-perfect waveform modeling to be dominant over the statistical
uncertainty for an unfaithfulness $\gtrsim
1\%$. Therefore we might expect some biases in inferring parameters 
for both models. Lindblom's criterion does not say which parameters 
are biased and by how much. The results in Fig.~\ref{fig:PE_q_6} clearly show 
that both models have biases in the measurement of some
parameters, but unfaithfulness of $4.4\%$ and $8.8\%$ induce different 
amount of biases and also on different set of parameters (intrinsic and extrinsic).

In particular for the component masses (top left panel of
Fig.~\ref{fig:PE_q_6}), the 2D posterior distribution obtained with
\verb+SEOBNRv4PHM+ barely include the value used for the mock signal
in the $90\%$ credible region. This measurement looks better when
focusing on the 1D posterior distributions for the individual masses
for which the injection values are well within the $90\%$ credible intervals. 
The situation is worst for the \verb+IMRPhenomPv3HM+ model, for which the 2D posterior
distribution barely excludes the injection value at 
$90\%$ credible level. In this case also the true value of $m_1$
is excluded from the $90\%$ credible interval of the marginalized 1D
posterior distribution. Furthermore, $\chi_{\rm eff}$ and $\chi_{\rm p}$ (top right
panel of Fig.~\ref{fig:PE_q_6}) are correctly measured with
\verb+SEOBNRv4PHM+ while the measurement with \verb+IMRPhenomPv3HM+
excludes the true value from the 2D $90\%$ credible region. From the
1D posterior distributions it is clear that the source of this inaccuracy
is the incorrect measurement of $\chi_{\rm p}$, while $\chi_{\rm eff}$
is correctly recovered within the $90\%$ credible interval. A similar
situation is observed in the measurement of $a_1$ the spin magnitude
of the heavier BH and $\theta_1$ its tilt angle (bottom left panel of
Fig.~\ref{fig:PE_q_6}). Also in this case \verb+SEOBNRv4PHM+ correctly
measures the parameters used in the mock signal, while
\verb+IMRPhenomPv3HM+ yields an incorrect measurement due to a bias in
the estimation of $a_1$. Finally, we focus on the measurement of the
angle $\theta_{\rm JN}$ and the luminosity distance $D_L$ (bottom left
panel of Fig.~\ref{fig:PE_q_6}). In this case the value of these
parameters used in the synthetic signal is just slightly measured
within the $90\%$ credible region of the 2D posterior distribution
obtained with \verb+SEOBNRv4PHM+. As a consequence the luminosity
distance is also barely measured within the $90\%$ credible interval
from the marginalized 1D posterior distribution and the measured value
of $\theta_{\rm JN}$ results outside the $90\%$ credible interval of
the 1D posterior distribution. The posterior distributions obtained
using \verb+IMRPhenomPv3HM+ are instead correctly measuring the
parameters of the mock signal. We can conclude that even with an
unfaithfulness of $4.4\%$ against the NR waveform used for the mock
signal the \verb+SEOBNRv4PHM+ model is able to correctly measure most
of the binary parameters, notably the intrinsic ones, such as masses 
and spins.

\section{Conclusions}
\label{sec:concl}

In this paper we have developed and validated the first inspiral-merger-ringdown precessing 
waveform model in the EOB approach, \verb+SEOBNRv4PHM+, that includes multipoles beyond the dominant quadrupole. 

Following previous precessing SEOBNR models~\cite{Pan:2013rra,Taracchini:2013rva,Babak:2016tgq}, 
we have built such a model twisting up the aligned-spin waveforms of \verb+SEOBNRv4HM+~\cite{Bohe:2016gbl,Cotesta:2018fcv} 
from the co-precessing~\cite{Buonanno:2002fy,Schmidt:2010it,Boyle:2011gg,O'Shaughnessy:2011fx,Schmidt:2012rh} 
to the inertial frame, through the EOB equations of motion for the spins and orbital angular momentum. 
With respect to the previous precessing SEOBNR model, \verb+SEOBNRv3P+~\cite{Babak:2016tgq}, 
which has been used in LIGO/Virgo data analysis~\cite{Abbott:2016izl,Abbott:2017vtc,LIGOScientific:2018mvr}, the new model 
(i) employs a more accurate aligned-spin two-body dynamics, since, in the non-precessing limit, it reduces to 
\verb+SEOBNRv4HM+, which was calibrated to 157 SXS NR simulations~\cite{Mroue:2013xna,Chu:2015kft}, and 13 waveforms~\cite{Barausse:2011kb} 
from BH perturbation theory,  (ii) includes in the co-precessing frame the modes $(2,\pm 2), (2,\pm 1), (3,\pm 3), (4,\pm 4)$ 
and $(5,\pm 5)$, instead of only $(2,\pm 2), (2,\pm 1)$, (iii) incorporates the merger-ringdown 
signal in the co-precessing frame instead of the inertial frame, (iv) describes the merger-ringdown stage through a phenomenological 
fit to NR waveforms~\cite{Bohe:2016gbl,Cotesta:2018fcv}, and (v) uses more accurate NR fits  for the final spin of the remnant BH.

\begin{figure*}
	\includegraphics[width=0.49\linewidth]{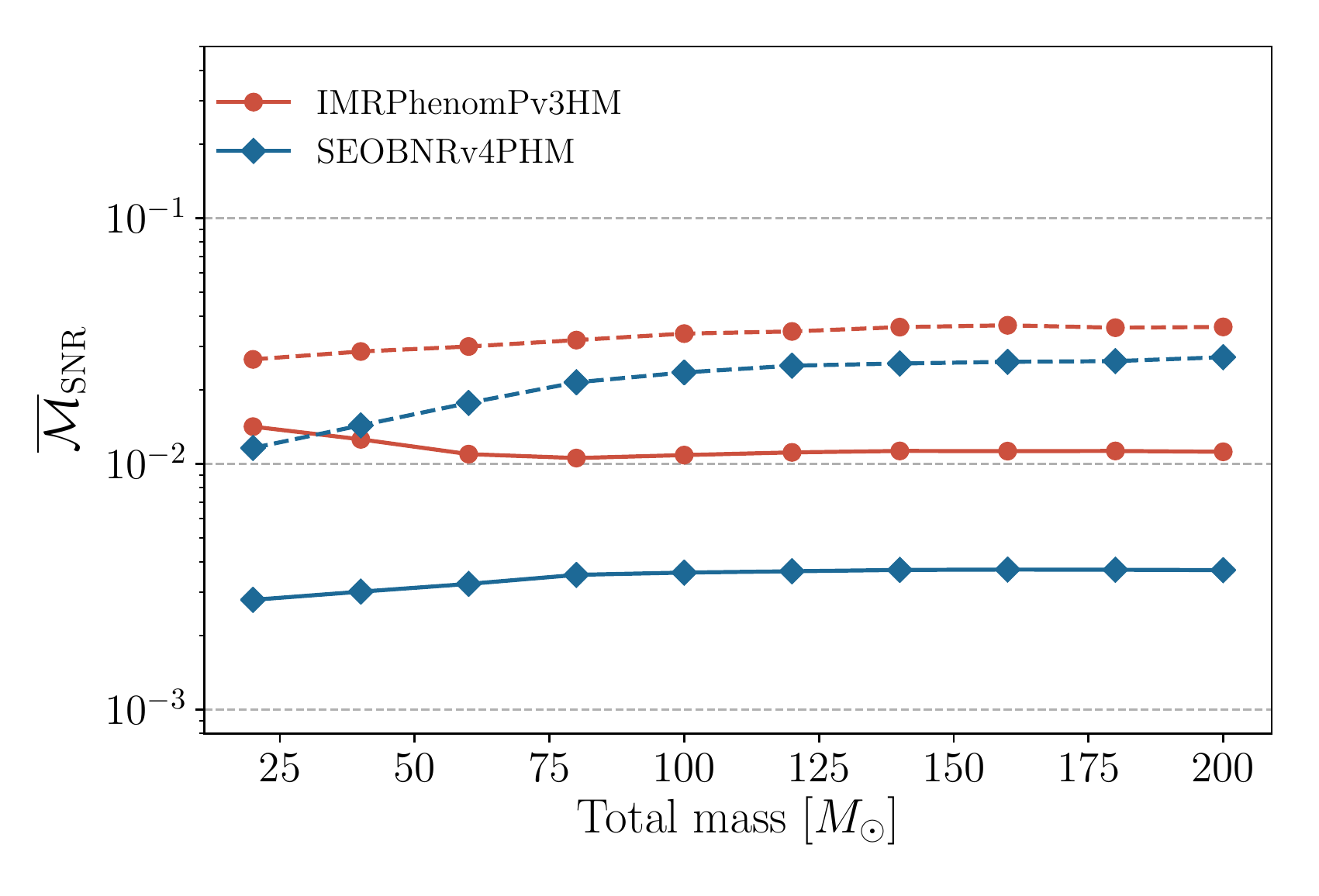}
		\includegraphics[width=0.49\linewidth]{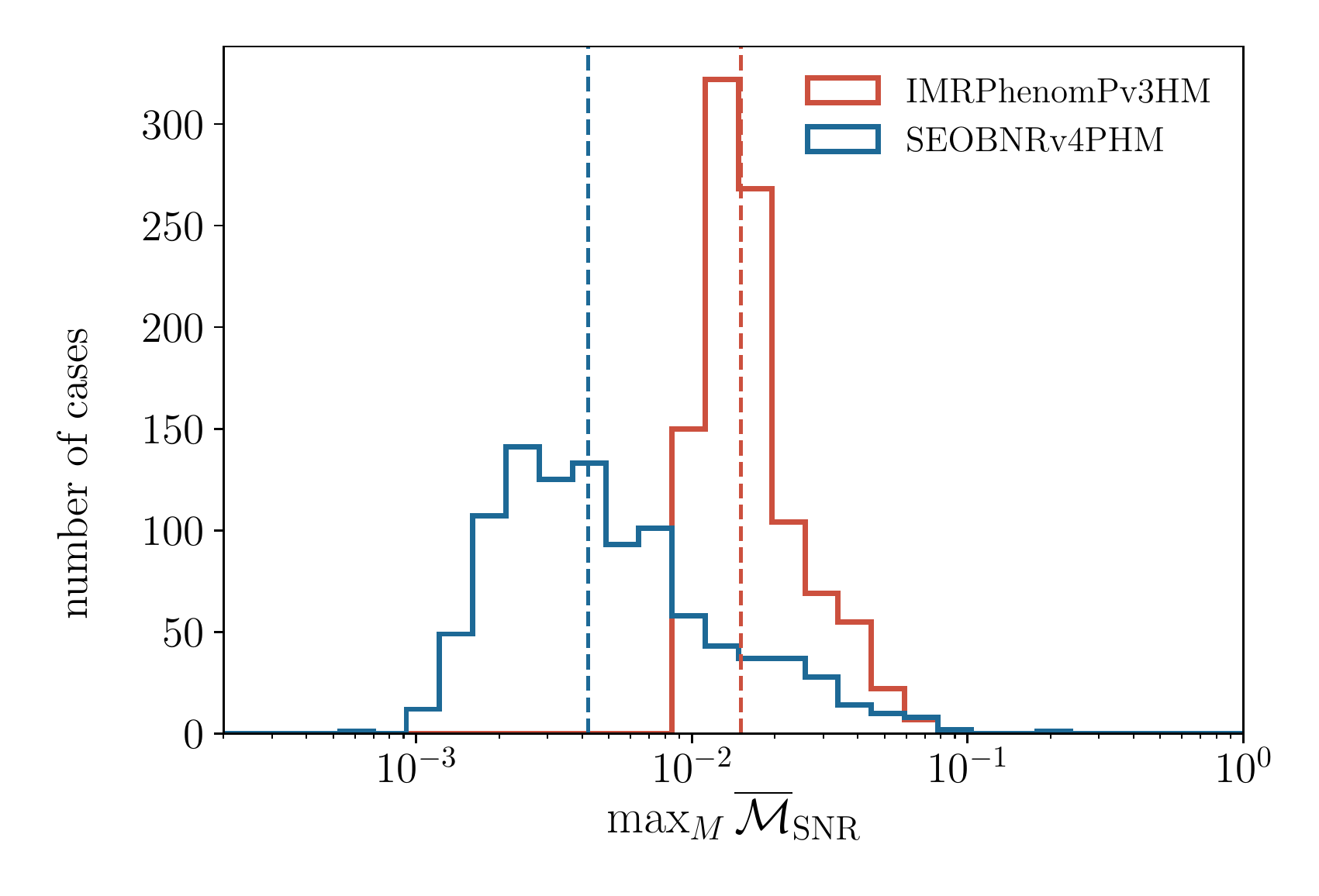}
	\caption{The summary of the sky-and-polarization averaged, SNR-weighted
		unfaithfulness as a function of binary's total mass for inclination
		$\iota=\pi/3$, among the {\tt NRSur7dq4} model and the {\tt IMRPhenomPv3HM} and {\tt SEOBNRv4PHM} models. 
\emph{Left}: The solid (dashed) lines show the median (95th percentile) as a function of total mass,
		cf Fig.~\ref{fig:all_runs_percentiles}. \emph{Right}: maximum unfaithfulness over all total masses, cf. Fig.~\ref{fig:all_runs_hist}.
		The unfaithfulness is low 
		using both waveform families, however, {\tt SEOBNRv4P(HM)} has lower median unfaithfulness by a factor of ~4.3.}
	\label{fig:models_vs_NRsurr}
\end{figure*}

\begin{figure*}
	\includegraphics[width=\linewidth]{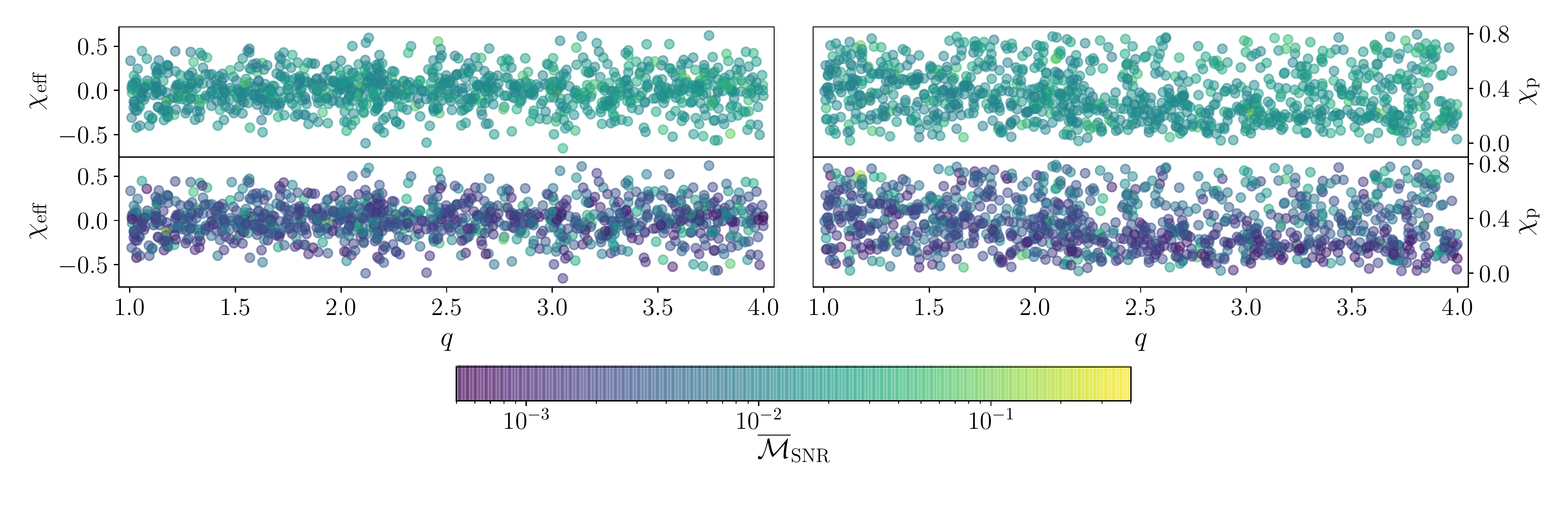}
	\caption{The maximum  sky-and-polarization averaged, SNR-weighted
		unfaithfulness as a function of binary's total mass for inclination
		$\iota=\pi/3$, between the models {\tt IMRPhenomPv3HM} (\emph{top}) and {\tt SEOBNRv4PHM} (\emph{bottom}),
		and the NR surrogate, cf. Fig.~\ref{fig:v4PHM_vs_Pv3HM}.
		The unfaithfulness is strongly dependent on the intrinsic parameters, especially the spins.
	}
	\label{fig:2d_models_vs_NRsurr}
\end{figure*}

The improvement in accuracy between \verb+SEOBNRv4+ and  \verb+SEOBNRv3P+ (i.e., the models with only the $\ell = 2$ modes)
is evident from Fig.~\ref{fig:all_approximants_ell2}, where we have compared those models to the public SXS catalog of 1405 precessing
NR waveforms, and the new 118 SXS NR waveforms produced for this work. The impact of including higher modes in semi-analytical models to achieve higher accuracy to multipolar NR waveforms is demonstrated in Fig.~\ref{fig:higher_mode_effects}. Figures~\ref{fig:spaghetti_public}, \ref{fig:spaghetti_PrecBBH}, \ref{fig:all_runs_percentiles} and \ref{fig:v4PHM_vs_Pv3HM} quantify the comparison of the
multipolar precessing \verb+SEOBNRv4PHM+ and  \verb+IMRPhenomPv3HM+ to all SXS NR precessing waveforms at our disposal. We have found that for the \verb+SEOBNRv4PHM+ model, \EOBbthree\ (\EOBbone\ ) of the cases have maximum unfaithfulness value, in the
total mass range $20\mbox{--}200 M_\odot$, below $3\%$ ($1\%$). Those numbers change to \Phenombthree\ (\Phenombone\ ) when using the \verb+IMRPhenomPv3HM+. The better accuracy of
\verb+SEOBNRv4PHM+ with respect to \verb+IMRPhenomPv3HM+ is also confirmed by
the comparisons with the NR surrogate model \verb+NRSur7dq4+, as shown in Fig.~\ref{fig:models_vs_NRsurr}. We have investigated
in which region of the parameter space the unfaithfulness against NR waveforms and {\tt NRSur7dq4} lies, and have found, not surprisingly,
that it occurs where both mass ratios and spins are large (see Fig.~\ref{fig:2d_models_vs_NRsurr}). When comparing {\tt SEOBNRv4PHM} and {\tt IMRPhenomPv3HM} outside
the region in which their corresponding  aligned-spin underlying models were calibrated, we have also found that the largest differences reside when mass ratios are larger than 4 and spins
larger than 0.8 (see Fig.~\ref{fig:v4PHM_vs_Pv3HM}).
To improve the accuracy of the models in those more challenging regions, we would need NR simulations, but also more information from analytical
 methods, such as the gravitational self-force~\cite{Damour:2009sm,Bini:2018ylh,Antonelli:2019fmq}, and resummed EOB Hamiltonians
with spins~\cite{Rettegno:2019tzh,Khalil:2020mmr}.

To quantify how the modeling inaccuracy, estimated by the unfaithfulness, impacts the inference of binary's parameters, we
have perfomed two parameter-estimation studies using Bayesian analysis. Working with
the Advanced LIGO and Virgo network at design sensitivity, we have injected in zero noise two precessing-BBH
mock signals with mass ratio 3 and 6, having SNR of 50 and 21, with inclination of  $\pi/3$ and $\pi/2$ with respect to the line of sight respectively,  and recovered them with {\tt SEOBNRv4PHM} and {\tt IMRPhenomPv3HM}.
The unfaithfulness values of those models against the synthetic signals considered (i.e., {\tt NRSurd7q4} and {\tt SXS:BBH:0165})
range from $0.2\%$ to $8.8\%$. The results are summarized in Figs.~\ref{fig:PE_q_3} and \ref{fig:PE_q_6}. Overall, we have found that Lindblom's criterion
\cite{Flanagan:1997kp,Lindblom:2008cm,McWilliams:2010eq,Chatziioannou:2017tdw,Purrer:2019jcp} is too conservative
and predicts visible biases at SNRs lower than what we have obtained through the Bayesian analysis. In particular, we 
have found, when doing inference with {\tt SEOBNRv4PHM}, that an unfaithfulness of $0.2\%$ may produce no biases up to SNR of 50, while 
an unfaithfulness of $2.2\%$ can produce biases only for some extrinsic parameters, such as distance and inclination, but not for binary's masses and 
spins at SNR of 21. A more comprehensive Bayesian study will be needed to quantify, in a more realistic manner, the modeling systematics of
{\tt SEOBNRv4PHM}, if this model were used during the fourth observation run of Avanced LIGO and Virgo in 2022 (i.e., the run at design sensitivity).

The improvement in accuracy between \verb+SEOBNRv4+ and
\verb+SEOBNRv3P+ (i.e., the models with only the $\ell = 2$ modes) is
evident from Fig.~\ref{fig:all_approximants_ell2}, where we have
compared those models to the public SXS catalog of 1405 precessing NR
waveforms, and the new 118 SXS NR waveforms produced for this
work. The impact of including higher modes in semi-analytical models
to achieve higher accuracy to multipolar NR waveforms is demonstrated
in
Fig.~\ref{fig:higher_mode_effects}. Figures~\ref{fig:spaghetti_public},
\ref{fig:spaghetti_PrecBBH}, \ref{fig:all_runs_percentiles} and
\ref{fig:v4PHM_vs_Pv3HM} quantify the comparison of the multipolar
precessing \verb+SEOBNRv4PHM+ and \verb+IMRPhenomPv3HM+ to all SXS NR
precessing waveforms at our disposal. We have found that for the
\verb+SEOBNRv4PHM+ model, \EOBbthree\ (\EOBbone\ ) of the cases have
maximum unfaithfulness value, in the total mass range
$20\mbox{--}200 M_\odot$, below $3\%$ ($1\%$). Those numbers change to
\Phenombthree\ (\Phenombone\ ) when using the
\verb+IMRPhenomPv3HM+. We have found several cases with large
unfaithfulness ($>10\%$) for {\tt IMRPhenomPv3HM}, coming from a
region of parameter space with $q\gtrsim 4$ and large ($\simeq0.8$)
spins anti-aligned with the orbital angular momentum, which
appear to be connected to unphysical features in the underlying
precession model, and cause unusual oscillations in the 
waveform's amplitude and phase. The better accuracy of \verb+SEOBNRv4PHM+ with
respect to \verb+IMRPhenomPv3HM+ is also confirmed by the comparisons
with the NR surrogate model \verb+NRSur7dq4+, as shown in
Fig.~\ref{fig:models_vs_NRsurr}. We have investigated in which region
of the parameter space the unfaithfulness against NR waveforms and
{\tt NRSur7dq4} lies, and have found, not surprisingly, that it occurs
where both mass ratios and spins are large (see
Fig.~\ref{fig:2d_models_vs_NRsurr}). When comparing {\tt SEOBNRv4PHM}
and {\tt IMRPhenomPv3HM} outside the region in which the aligned-spin
underlying model was calibrated, we have also found that the largest
differences reside when mass ratios are larger than 4 and spins larger
than 0.8 (see Fig.~\ref{fig:v4PHM_vs_Pv3HM}).  To improve the accuracy
of the models in those more challenging regions, we would need NR
simulations, but also more information from analytical methods, such
as the gravitational
self-force~\cite{Damour:2009sm,Bini:2018ylh,Antonelli:2019fmq}, and
resummed EOB Hamiltonians with
spins~\cite{Rettegno:2019tzh,Khalil:2020mmr}.

The newly produced 118 SXS NR waveforms extend the coverage of binary's parameter space, spanning  
mass ratios $q=1\mbox{--}4$, (dimensionless) spins $\chi_{1,2}=0.3\mbox{--}0.9$, and different orientations 
to maximize the number of precessional cycles. As we have emphasized, the waveform model \verb+SEOBNRv4HM+ 
is not calibrated to NR waveforms in the precessing sector, only the aligned-spin sector was calibrated 
in Refs.~\cite{Bohe:2016gbl,Cotesta:2018fcv}. Despite this, 
the accuracy of the model is very good, and it can be further improved in the future if we calibrate  
the model to the $1404$ plus $118$ SXS NR precessing waveforms at our disposal. This will be an important 
goal for the upcoming LIGO and Virgo O4 run in early 2022. Furthermore, \verb+SEOBNRv4HM+ assumes the following 
symmetry among modes $h_{\ell m}= (-1)^\ell h_{\ell -m}^*$ in the co-precessing frame, which however 
no longer holds in presence of precession. As discussed in Sec.~\ref{sec:mode_asymm}, forcing 
this assumption causes unfaithfulness on the order of a few percent. Thus, to achieve better accuracy, 
when calibrating the model to NR waveforms, the mode-symmetry would need to be relaxed.

Finally, \verb+SEOBNRv4HM+ uses PN-resumed factorized modes that were developed for aligned-spin BBHs~\cite{Damour:2008gu,Pan:2010hz}, 
thus they neglect the projection of the spins on the orbital plane. To obtain high-precision waveform models, it will 
be relevant to extend the factorized modes to precession. Considering the variety of GW signals 
that the improved sensitivity of LIGO and Virgo detectors is allowing to observe, it will also be important 
to include in the multipolar SEOBNR waveform models the more challenging $(3,2)$ and $(4,3)$ modes, which are characterized 
my mode mixing~\cite{Buonanno:2006ui,Berti:2014fga,Mehta:2019wxm}. Their contribution is no longer negligible for 
high total-mass and/or large mass-ratio binaries, especially if observed away from face-on (face-off). 

Lastly, being a time-domain waveform model generated by solving ordinary differential equations, \verb+SEOBNRv4HM+ is not a fast 
waveform model, especially for low total-mass binaries. To speed up the waveform generation, a reduced-order modeling version 
has been recently developed~\cite{Gadre:2020a}. Alternative methods that employ a fast evolution of the EOB Hamilton equations 
in the post-adiabatic approximation during the long inspiral phase have been suggested~\cite{Nagar:2018gnk}, and we are currently implementing 
them in the simpler nonprecessing limit in LAL.

\section*{Acknowledgments}
It is our pleasure to thank Andrew Matas for providing us with the scripts to make 
the parameter-estimation plots, and Sebastian Khan for useful
discussions on the faithfulness calculation.  We would also like to thank the SXS collaboration for
help and support with the {\tt SpEC} code in producing the new NR
simulations presented in this paper, and for making the large catalog
of BBH simulations publicly available.  The new 118 SXS NR simulations
were produced using the high-performance compute (HPC) cluster {\tt
  Minerva} at the Max Planck Institute for Gravitational Physics in
Potsdam, on the {\tt Hydra} cluster at the Max Planck Society at the 
Garching facility, and on the {\tt
  SciNet} cluster at the University of Toronto.  The data-analysis
studies were obtained with the HPC clusters {\tt Hypatia} and {\tt
  Minerva} at the Max Planck Institute for Gravitational Physics. The
transformation and manipulation of waveforms were done using the {\tt
  GWFrames} package~\cite{GWFrames,Boyle:2013nka}. 

\appendix
\section{Comparison of multipolar precessing models to numerical-relativity surrogate waveforms}
\label{sec:comparisonNRSurr}

In this appendix we compare directly \verb+SEOBNRv4PHM+ and \verb+IMRPhenomPv3HM+ to
the NR surrogate model \verb+NRSur7dq4+.  We choose a starting frequency corresponding to 20 Hz at 70 $M_{\odot}$ (this is  essentially the limit of the length for NR
surrogate waveforms). We generate 1000 random configurations, uniform
in mass ratio $q\in[1,4]$ and in spin magnitudes $\in[0,0.8]$,
and with random directions uniform on the unit sphere. The left panel  of  Fig.~\ref{fig:models_vs_NRsurr}
shows the summary of the unfaithfulness as a function of total mass for all the cases
considered, for \verb+IMRPhenomPv3HM+ and \verb+SEOBNRv4PHM+.  We see that the 
median and 95th percentile values for both models are close to the values in Fig.~\ref{fig:all_runs_percentiles}, with \verb+SEOBNRv4PHM+
having a median unfaithfulness below 1\% and \verb+IMRPhenomPv3HM+ about a factor of 3 larger. The right panel of Fig.~\ref{fig:models_vs_NRsurr} 
shows the maximum unfaithfulness distribution and the same trends are also observed. \verb+SEOBNRv4PHM+ outperforms 
\verb+IMRPhenomPv3HM+, with the median of the former being 4 times smaller than the one of the latter. Finally, to gain further insight 
into the behavior of the waveform models across the parameter space, we show in Fig.~\ref{fig:2d_models_vs_NRsurr} the maximum unfaithfulness 
as a function of mass ratio and the effective spin.

\clearpage
\onecolumngrid

\section{Parameters of the new 118 NR simulations}
\label{sec:NRparam}

\begin{longtable}{llllll}
            ID &     $q$ &                $\chi_{1}$ &                $\chi_{2}$ &  $M\Omega$ &  \# of  orbits \\
\hline
 PrecBBH000001 &  1.2499 &   (-0.272, -0.628, 0.414) &    (-0.212, -0.653, 0.41) &    0.01632 &        21 \\
 PrecBBH000002 &  1.2500 &    (-0.629, 0.202, 0.451) &    (-0.13, -0.708, 0.348) &    0.01645 &        20 \\
 PrecBBH000003 &  1.2499 &     (0.68, -0.104, 0.408) &    (0.71, -0.153, -0.335) &    0.01616 &        19 \\
 PrecBBH000004 &  1.2501 &    (0.309, -0.593, 0.439) &    (0.611, 0.325, -0.401) &    0.01627 &        18 \\
 PrecBBH000005 &  1.2500 &   (0.269, -0.684, -0.317) &       (0.393, -0.57, 0.4) &    0.01626 &        18 \\
 PrecBBH000006 &  1.2500 &   (0.561, -0.488, -0.293) &      (0.37, 0.611, 0.361) &    0.01623 &        18 \\
 PrecBBH000007 &  1.2499 &   (-0.671, 0.287, -0.328) &   (-0.694, 0.205, -0.339) &    0.01651 &        16 \\
 PrecBBH000008 &  1.2501 &     (-0.7, 0.269, -0.277) &  (-0.133, -0.669, -0.418) &    0.01653 &        16 \\
 PrecBBH000009 &  2.4998 &     (0.279, 0.579, 0.387) &     (0.138, 0.631, 0.381) &    0.01604 &        24 \\
 PrecBBH000010 &  2.5000 &     (-0.577, 0.26, 0.403) &   (-0.021, -0.679, 0.317) &    0.01633 &        24 \\
 PrecBBH000011 &  2.4999 &     (-0.604, 0.23, 0.381) &   (-0.608, 0.096, -0.428) &    0.01631 &        23 \\
 PrecBBH000012 &  2.4998 &    (-0.587, 0.238, 0.402) &   (-0.014, -0.576, -0.48) &    0.01630 &        23 \\
 PrecBBH000013 &  2.4998 &   (-0.531, 0.349, -0.399) &    (-0.65, -0.043, 0.371) &    0.01636 &        19 \\
 PrecBBH000014 &  2.4998 &   (-0.554, 0.332, -0.382) &    (0.012, -0.683, 0.309) &    0.01638 &        19 \\
 PrecBBH000015 &  2.4998 &    (0.052, 0.633, -0.399) &    (-0.096, 0.62, -0.411) &    0.01605 &        18 \\
 PrecBBH000016 &  2.4997 &    (0.615, 0.166, -0.396) &   (-0.326, 0.497, -0.457) &    0.01606 &        18 \\
 PrecBBH000017 &  3.4997 &     (0.421, 0.298, 0.306) &     (0.301, 0.417, 0.309) &    0.01598 &        27 \\
 PrecBBH000018 &  3.4992 &     (0.464, 0.218, 0.312) &    (-0.348, 0.402, 0.277) &    0.01599 &        27 \\
 PrecBBH000019 &  3.4996 &     (0.242, 0.455, 0.307) &    (0.127, 0.471, -0.349) &    0.01598 &        26 \\
 PrecBBH000020 &  3.4999 &     (0.514, -0.006, 0.31) &   (-0.139, 0.451, -0.371) &    0.01602 &        26 \\
 PrecBBH000021 &  3.4993 &     (-0.4, 0.297, -0.335) &   (-0.518, -0.054, 0.298) &    0.01631 &        22 \\
 PrecBBH000022 &  3.4995 &     (0.464, 0.18, -0.335) &    (-0.358, 0.395, 0.275) &    0.01605 &        22 \\
 PrecBBH000023 &  3.4991 &   (0.414, -0.273, -0.338) &    (0.472, -0.09, -0.358) &    0.01606 &        21 \\
 PrecBBH000024 &  3.4997 &   (0.256, -0.431, -0.329) &    (0.225, 0.401, -0.385) &    0.01609 &        21 \\
 PrecBBH000025 &  1.2501 &    (-0.661, 0.193, 0.407) &          (0.0, -0.0, 0.0) &    0.01645 &        19 \\
 PrecBBH000026 &  1.2501 &    (-0.466, -0.618, -0.2) &          (0.0, -0.0, 0.0) &    0.01638 &        17 \\
 PrecBBH000027 &  2.4999 &     (0.099, 0.637, 0.383) &          (0.0, -0.0, 0.0) &    0.01601 &        23 \\
 PrecBBH000028 &  2.5003 &    (0.557, 0.357, -0.354) &          (0.0, 0.0, -0.0) &    0.01609 &        19 \\
 PrecBBH000029 &  3.5006 &    (0.458, -0.242, 0.302) &           (0.0, 0.0, 0.0) &    0.01603 &        27 \\
 PrecBBH000030 &  3.4996 &   (-0.397, -0.32, -0.316) &          (0.0, -0.0, 0.0) &    0.01619 &        22 \\
 PrecBBH000031 &  1.0001 &    (-0.752, 0.179, 0.461) &         (-0.0, -0.0, 0.0) &    0.01646 &        19 \\
 PrecBBH000032 &  1.0002 &   (-0.836, 0.259, -0.206) &         (-0.0, -0.0, 0.0) &    0.01649 &        17 \\
 PrecBBH000033 &  2.0000 &    (-0.709, 0.269, 0.445) &         (-0.0, -0.0, 0.0) &    0.01638 &        22 \\
 PrecBBH000034 &  2.0004 &    (0.027, 0.793, -0.379) &         (-0.0, 0.0, -0.0) &    0.01605 &        18 \\
 PrecBBH000035 &  3.2002 &     (0.681, 0.112, 0.405) &           (0.0, 0.0, 0.0) &    0.01599 &        26 \\
 PrecBBH000036 &  3.1995 &    (0.162, 0.597, -0.507) &         (-0.0, -0.0, 0.0) &    0.01600 &        20 \\
 PrecBBH000037 &  3.9994 &    (0.596, -0.106, 0.352) &        (-0.0, -0.0, -0.0) &    0.01598 &        29 \\
 PrecBBH000038 &  4.0003 &  (-0.146, -0.481, -0.487) &         (-0.0, 0.0, -0.0) &    0.01613 &        22 \\
 PrecBBH000039 &  1.0000 &    (-0.542, 0.137, 0.332) &        (-0.0, -0.0, -0.0) &    0.01646 &        19 \\
 PrecBBH000040 &  1.0000 &   (-0.614, 0.183, -0.108) &        (-0.0, -0.0, -0.0) &    0.01649 &        17 \\
 PrecBBH000041 &  2.0001 &     (-0.48, 0.195, 0.303) &          (0.0, -0.0, 0.0) &    0.01639 &        21 \\
 PrecBBH000042 &  2.0003 &   (-0.509, 0.261, -0.181) &         (0.0, -0.0, -0.0) &    0.01644 &        19 \\
 PrecBBH000043 &  2.5002 &     (0.349, 0.252, 0.254) &          (0.0, 0.0, -0.0) &    0.01606 &        22 \\
 PrecBBH000044 &  2.5000 &     (0.456, 0.13, -0.158) &        (-0.0, -0.0, -0.0) &    0.01607 &        20 \\
 PrecBBH000045 &  3.9999 &    (-0.265, 0.146, 0.176) &          (0.0, 0.0, -0.0) &    0.01621 &        27 \\
 PrecBBH000046 &  4.0003 &    (0.25, -0.213, -0.122) &        (-0.0, -0.0, -0.0) &    0.01603 &        25 \\
 PrecBBH000047 &  2.9997 &     (0.249, 0.072, 0.152) &         (0.0, -0.0, -0.0) &    0.01604 &        23 \\
 PrecBBH000048 &  3.0000 &   (0.228, -0.183, -0.067) &        (-0.0, -0.0, -0.0) &    0.01610 &        22 \\
 PrecBBH000050 &  1.0001 &    (-0.709, 0.187, 0.522) &     (-0.18, -0.79, 0.391) &    0.01644 &        21 \\
 PrecBBH000051 &  1.0000 &    (-0.768, 0.118, 0.453) &   (-0.747, 0.299, -0.402) &    0.01646 &        18 \\
 PrecBBH000053 &  1.0000 &   (-0.747, 0.299, -0.402) &    (-0.768, 0.117, 0.453) &    0.01646 &        18 \\
 PrecBBH000054 &  1.0001 &    (-0.79, 0.265, -0.339) &   (-0.161, -0.801, 0.377) &    0.01648 &        18 \\
 PrecBBH000055 &  1.0000 &    (-0.748, 0.286, -0.41) &    (-0.748, 0.286, -0.41) &    0.01651 &        16 \\
 PrecBBH000056 &  1.0001 &   (-0.791, 0.266, -0.335) &   (-0.207, -0.71, -0.514) &    0.01655 &        15 \\
 PrecBBH000057 &  1.9997 &    (-0.715, 0.242, 0.452) &    (-0.757, 0.023, 0.448) &    0.01634 &        23 \\
 PrecBBH000058 &  2.0000 &    (-0.681, 0.276, 0.484) &   (-0.061, -0.797, 0.368) &    0.01636 &        23 \\
 PrecBBH000059 &  1.9997 &    (-0.725, 0.222, 0.447) &    (-0.706, 0.16, -0.499) &    0.01634 &        21 \\
 PrecBBH000060 &  2.0001 &    (-0.695, 0.242, 0.482) &  (-0.059, -0.655, -0.584) &    0.01636 &        21 \\
 PrecBBH000061 &  1.9998 &    (0.674, 0.294, -0.483) &     (0.529, 0.539, 0.451) &    0.01611 &        18 \\
 PrecBBH000062 &  1.9999 &  (-0.441, -0.618, -0.444) &    (0.762, -0.218, 0.381) &    0.01632 &        18 \\
 PrecBBH000063 &  1.9998 &   (-0.628, 0.392, -0.475) &     (-0.7, 0.137, -0.514) &    0.01643 &        16 \\
 PrecBBH000064 &  2.0000 &   (-0.188, 0.727, -0.458) &    (-0.608, -0.3, -0.561) &    0.01604 &        16 \\
 PrecBBH000065 &  3.1997 &    (-0.633, 0.268, 0.409) &   (-0.689, -0.046, 0.403) &    0.01622 &        27 \\
 PrecBBH000066 &  3.1998 &    (-0.611, 0.292, 0.426) &    (0.012, -0.728, 0.331) &    0.01623 &        27 \\
 PrecBBH000067 &  3.1996 &     (0.606, 0.327, 0.408) &    (0.436, 0.335, -0.581) &    0.01598 &        26 \\
 PrecBBH000068 &  3.1998 &     (-0.624, 0.27, 0.421) &   (0.018, -0.487, -0.634) &    0.01623 &        25 \\
 PrecBBH000069 &  3.1995 &   (-0.444, 0.373, -0.551) &   (-0.692, -0.085, 0.391) &    0.01634 &        20 \\
 PrecBBH000070 &  3.1992 &    (-0.51, -0.29, -0.544) &     (0.632, -0.342, 0.35) &    0.01627 &        20 \\
 PrecBBH000071 &  3.1991 &      (0.4, 0.409, -0.559) &    (0.228, 0.504, -0.577) &    0.01611 &        18 \\
 PrecBBH000072 &  3.1994 &    (0.245, 0.527, -0.549) &    (-0.51, 0.053, -0.613) &    0.01600 &        18 \\
 PrecBBH000073 &  4.0002 &     (0.604, 0.004, 0.354) &     (0.559, 0.214, 0.363) &    0.01597 &        30 \\
 PrecBBH000074 &  3.9992 &   (-0.004, -0.595, 0.369) &     (0.573, 0.241, 0.322) &    0.01607 &        30 \\
 PrecBBH000075 &  4.0000 &    (-0.549, 0.252, 0.354) &   (-0.441, 0.048, -0.542) &    0.01616 &        28 \\
 PrecBBH000076 &  3.9993 &    (-0.538, 0.262, 0.363) &   (0.034, -0.402, -0.572) &    0.01618 &        28 \\
 PrecBBH000077 &  4.0003 &   (-0.361, 0.309, -0.513) &     (-0.6, -0.101, 0.345) &    0.01623 &        22 \\
 PrecBBH000078 &  4.0001 &    (0.466, 0.089, -0.515) &    (-0.366, 0.503, 0.321) &    0.01604 &        22 \\
 PrecBBH000079 &  4.0003 &   (-0.435, 0.179, -0.518) &   (-0.416, -0.118, -0.55) &    0.01624 &        21 \\
 PrecBBH000080 &  4.0000 &    (0.139, 0.456, -0.513) &   (-0.422, -0.03, -0.557) &    0.01599 &        21 \\
 PrecBBH000081 &  1.0000 &     (-0.545, 0.12, 0.333) &     (-0.545, 0.12, 0.333) &    0.01643 &        20 \\
 PrecBBH000082 &  1.0000 &    (-0.519, 0.141, 0.365) &   (-0.132, -0.565, 0.293) &    0.01645 &        20 \\
 PrecBBH000083 &  1.0000 &     (-0.549, 0.107, 0.33) &   (-0.581, 0.198, -0.213) &    0.01646 &        18 \\
 PrecBBH000084 &  1.0000 &     (-0.52, 0.126, 0.369) &  (-0.163, -0.574, -0.258) &    0.01648 &        18 \\
 PrecBBH000085 &  1.0000 &   (-0.582, 0.197, -0.213) &      (-0.55, 0.106, 0.33) &    0.01646 &        18 \\
 PrecBBH000086 &  1.0000 &   (-0.596, 0.188, -0.176) &    (-0.123, -0.57, 0.286) &    0.01648 &        18 \\
 PrecBBH000087 &  1.0000 &   (-0.582, 0.192, -0.215) &   (-0.582, 0.192, -0.215) &    0.01650 &        17 \\
 PrecBBH000088 &  1.0000 &     (-0.6, 0.181, -0.172) &  (-0.151, -0.573, -0.266) &    0.01651 &        16 \\
 PrecBBH000089 &  1.9999 &    (0.513, -0.058, 0.305) &     (0.511, 0.046, 0.311) &    0.01615 &        22 \\
 PrecBBH000090 &  2.0003 &    (-0.467, 0.197, 0.322) &   (-0.039, -0.537, 0.264) &    0.01638 &        22 \\
 PrecBBH000091 &  2.0001 &     (0.278, 0.433, 0.308) &      (0.238, 0.5, -0.231) &    0.01604 &        21 \\
 PrecBBH000092 &  1.9999 &     (-0.47, 0.186, 0.323) &  (-0.041, -0.533, -0.273) &    0.01638 &        21 \\
 PrecBBH000093 &  1.9999 &   (-0.495, 0.257, -0.221) &    (-0.518, 0.005, 0.302) &    0.01639 &        19 \\
 PrecBBH000094 &  1.9999 &   (0.553, -0.092, -0.214) &    (-0.063, 0.531, 0.272) &    0.01612 &        19 \\
 PrecBBH000095 &  1.9999 &   (-0.494, 0.258, -0.221) &   (-0.544, 0.074, -0.242) &    0.01641 &        18 \\
 PrecBBH000096 &  1.9998 &  (-0.532, -0.185, -0.206) &   (0.332, -0.419, -0.273) &    0.01637 &        18 \\
 PrecBBH000097 &  2.4999 &      (0.003, 0.43, 0.255) &    (-0.085, 0.423, 0.252) &    0.01603 &        23 \\
 PrecBBH000098 &  2.5000 &      (0.075, 0.421, 0.26) &    (-0.435, 0.003, 0.246) &    0.01604 &        23 \\
 PrecBBH000099 &  2.5000 &      (0.128, 0.41, 0.256) &    (0.057, 0.461, -0.184) &    0.01601 &        22 \\
 PrecBBH000100 &  2.4999 &     (-0.06, 0.424, 0.259) &  (-0.435, -0.135, -0.206) &    0.01601 &        22 \\
 PrecBBH000101 &  2.5002 &   (0.195, -0.426, -0.176) &     (0.318, -0.294, 0.25) &    0.01616 &        20 \\
 PrecBBH000102 &  2.5001 &   (0.468, -0.022, -0.175) &    (-0.121, 0.428, 0.229) &    0.01610 &        20 \\
 PrecBBH000103 &  2.5002 &   (0.456, -0.095, -0.182) &    (0.459, 0.038, -0.194) &    0.01610 &        20 \\
 PrecBBH000104 &  2.5000 &    (0.293, 0.366, -0.173) &    (-0.41, 0.199, -0.206) &    0.01610 &        19 \\
 PrecBBH000105 &  4.0011 &    (-0.302, 0.006, 0.177) &    (-0.256, -0.156, 0.18) &    0.01616 &        28 \\
 PrecBBH000106 &  4.0001 &     (0.251, 0.165, 0.179) &    (-0.236, 0.197, 0.168) &    0.01596 &        28 \\
 PrecBBH000107 &  4.0008 &     (0.252, 0.166, 0.177) &    (0.198, 0.253, -0.138) &    0.01598 &        27 \\
 PrecBBH000108 &  3.9994 &     (0.038, -0.298, 0.18) &      (0.267, 0.17, -0.15) &    0.01608 &        27 \\
 PrecBBH000109 &  4.0002 &    (0.171, 0.277, -0.129) &     (0.048, 0.298, 0.177) &    0.01592 &        25 \\
 PrecBBH000110 &  4.0005 &  (-0.122, -0.303, -0.125) &     (0.305, 0.038, 0.168) &    0.01611 &        25 \\
 PrecBBH000111 &  3.9999 &     (0.278, 0.168, -0.13) &    (0.193, 0.257, -0.138) &    0.01597 &        24 \\
 PrecBBH000112 &  4.0001 &     (0.249, 0.21, -0.128) &   (-0.274, 0.165, -0.143) &    0.01598 &        24 \\
 PrecBBH000113 &  3.0002 &    (-0.232, 0.115, 0.152) &   (-0.257, -0.004, 0.154) &    0.01627 &        24 \\
 PrecBBH000114 &  3.0001 &    (0.001, -0.257, 0.154) &     (0.243, 0.096, 0.148) &    0.01614 &        24 \\
 PrecBBH000115 &  3.0000 &     (0.123, 0.227, 0.153) &    (0.069, 0.281, -0.078) &    0.01601 &        23 \\
 PrecBBH000116 &  2.9997 &     (0.236, 0.102, 0.154) &   (-0.179, 0.226, -0.083) &    0.01615 &        23 \\
 PrecBBH000117 &  2.9999 &   (-0.252, 0.144, -0.074) &   (-0.258, -0.009, 0.153) &    0.01629 &        22 \\
 PrecBBH000118 &  3.0001 &   (0.163, -0.242, -0.071) &      (0.14, 0.222, 0.144) &    0.01611 &        22 \\
 PrecBBH000119 &  3.0002 &     (0.018, 0.29, -0.073) &   (-0.055, 0.285, -0.076) &    0.01600 &        22 \\
 PrecBBH000120 &  3.0001 &   (-0.253, 0.144, -0.071) &    (0.011, -0.286, -0.09) &    0.01631 &        22 \\
\hline
\caption{The parameters of the runs in the new precessing catalog. Note that all the parameters are provided at the relaxed time and in the LAL source frame\cite{Schmidt:2017btt}.}

\end{longtable}

\twocolumngrid

\bibliography{references}

\end{document}